\documentclass[preprints,article,accept,pdftex,moreauthors]{Definitions/mdpi}
\usepackage{journals}

\firstpage{1} 
\pubvolume{1}
\issuenum{1}
\articlenumber{0}
\pubyear{2025}
\copyrightyear{2025}
\datereceived{ } 
\daterevised{ } % Comment out if no revised date
\dateaccepted{ } 
\datepublished{ } 
\hreflink{https://doi.org/} % If needed use \linebreak
\Title{The Edge-on Galaxies in the DESI survey (EGIDE): sample building and photometry}

\TitleCitation{Boxy/peanut bulges in EGIPS galaxies and TNG50 models}

\Author{Alexander A. Marchuk $^{1, 2}$*\orcidA{}
,  Sergey S. Savchenko$^{1, 2}$\orcidB{} 
,  Dmitry I. Makarov$^{1, 3}$\orcidE{}
, Vladimir P. Reshetnikov$^{1, 2}$\orcidD{}
, Ilia V. Chugunov$^{1, 4}$\orcidC{}
, Matvey D. Kozlov$^{1, 2}$\orcidH{}
, Aleksandra V. Antipova$^{1, 3}$\orcidF{}
, Anastasia M. Sypkova$^{1, 2}$\orcidG{}
, Evgenii V. Rubtsov$^{4}$\orcidI{}
, Dmitry V. Bizyaev$^{4, 5}$\orcidJ{}
}

\AuthorNames{Alexander Marchuk, Sergey Savchenko, Dmitry Makarov, Vladimir Reshetnikov, Ilia Chugunov,  Matvey Kozlov, Aleksandra Antipova, Anastasia Sypkova, Evgenii Rubtsov, Dmitry Bizyaev}

\AuthorCitation{Marchuk, A.; Savchenko, S.; Makarov, D.; Reshetnikov, V.; Chugunov, I.; Kozlov, M.; Antipova, A.; Sypkova, A.; Rubtsov, E.; Bizyaev D.}

\address{%
$^{1}$ \quad Central (Pulkovo) Astronomical Observatory of RAS, Pulkovskoye Chaussee 65/1, 196140 St. Petersburg, Russia\\
$^{2}$ \quad Saint Petersburg State University, 7/9 Universitetskaya nab., St.Petersburg 199034, Russia \\
$^{3}$ \quad Special Astrophysical Observatory, Russian Academy of Sciences, Nizhnii Arkhyz 369167, Russia\\
$^{4}$ \quad Sternberg Astronomical Institute, Lomonosov Moscow State University, Universitetsky Pr. 13, 119234 Moscow, Russia \\
$^{5}$ \quad Apache Point Observatory and New Mexico State
University, P.O. Box 59, Sunspot, NM, 88349-0059, USA
}

\corres{Correspondence: aamarchuk+astro@gmail.com}

\abstract{We present the EGIDE (The Edge-on Galaxies in the DESI survey) project - a catalogue of 149,215 edge-on galaxy candidates created using the data of the DESI Legacy Imaging Survey DR10 images. The catalogue size is ten times bigger than its predecessor and covers more than half of the sky. It is constructed in an automatic way utilizing the full power of manual annotations from the GalaxyZoo volunteers, implemented in the \textsc{Zoobot} neural model, which was fine-tuned to search for edge-on galaxies specifically. To ensure the credibility of the dataset, subsequent manual supervision was done. The EGIDE catalogue provides homogeneous \textsc{SExtractor} photometry in $griz$ bands, total stellar mass estimation, redshift values for 98\% of the sample, star formation rates and other information. All of this is publicly available at The Edge-on Galaxy Database site. The preliminary analysis focused on differences between edge-on galaxies in the so-called blue sequence and red cloud populations. These galaxies demonstrate distinct properties: the number of redder galaxies drops with increasing $a/b$ ratio faster than for the bluer galaxies; galaxy thickness varies with galaxy colour: red sequence galaxies are thicker than blue cloud galaxies; the flattening ratio $q=b/a$ increases with total stellar mass $M_{\star}$ for the high-mass end. It is an intriguing result,  that the same trend of $q$ increasing is detected from both the statistical models of figures of revolution and direct observations of edge-on galaxies in EGIDE independently. The full extent of the validity of this relationship can only be determined after correctly accounting for the contributions of the bulge and the PSF.
}


\begin{document}
\section{Introduction}

%Edge-on disc galaxies, which are by definition inclined at nearly $90^\circ$ to the line-of-sight, are a unique laboratory for studying the physical processes of the internal and external evolution of galaxies and larger structures in the Universe. This orientation allows one to obtain in observations a lot of information about galaxies that is not available at other angles of inclination, which results in many available results and ongoing interest in these objects since early pioneering works \citep{1978ApJ...223L..63K,1979ApJ...234..829B,1981A&A....95..105V}. We list below here some of them (see also a recent review in \citep{2026PhR..1163....1J}).

Edge-on disk galaxies, which are by definition inclined at nearly $90^\circ$ to the line of sight, are a unique laboratory for studying the physical processes governing the internal and external evolution of galaxies and larger structures in the Universe. This orientation allows one to obtain, from observations, a lot of information about galaxies that is not available at other inclination angles. Consequently, there is a rich body of results and ongoing interest in these objects dating back to early pioneering works \citep{1978ApJ...223L..63K,1979ApJ...234..829B,1981A&A....95..105V}. We list some of them below (see also a recent review in \citep{2026PhR..1163....1J}).

 %thick and thin disk, heating mechanisms
%First of all, for edge-on galaxies we can directly get information on the thickness of the stellar disk and the vertical distribution of stellar matter in it. The presence of so-called thick and thin disks, which are characterized by different heights of stellar population $h_z$, are clearly established in our Galaxy using very different methods  (for example 5 methods applied for APOGEE DR17 in \citep{2025arXiv251110092A}). The exponential scale in the thin disk is larger (2-3 kpc in MW) than in the thick disk (2 kpc), and the former comprises younger populations. These observations are also held true for other galaxies \citep{2025arXiv251119632M}, which raises the question if thick discs are heat up during evolution or born thick already. Different possible causes of vertical heating are proposed, such as bar, tidal interaction or major/minor merging, spiral arms, radial migration, adiabatic heating, clumps/GMC scattering \citep{2026arXiv260221171S,2016MNRAS.459..199G}. 

First of all, for edge-on galaxies we can directly obtain information on the thickness of the stellar disk and the vertical distribution of stellar matter within it. In the same way we can estimate thickness and properties of gaseous disk \citep{2022MNRAS.515.1598B}. The presence of so-called thick and thin disks, which are characterized by different scale heights of the stellar population $h_z$, is clearly established in our Galaxy using a variety of methods (for example, five methods applied to APOGEE DR17 in \citep{2025arXiv251110092A}). In other galaxies, these two disks were first discovered in the edge-on galaxy NGC~4565 by \citep{1981A&A....95..105V}. The exponential scale height of the thick disk is larger than that of the thin disk ($\sim$1~kpc and $\sim$0.3~kpc in the Milky Way, \citep{2016ARA&A..54..529B}), and the latter comprises younger populations. These observations also hold true for other galaxies \citep{2025arXiv251119632M}, which raises the question of whether thick disks are heated up during evolution or are born thick. Different possible causes of vertical heating have been proposed, such as bars, tidal interactions, major or minor mergers, spiral arms, radial migration, adiabatic heating, and scattering by clumps or giant molecular clouds  \citep{2026arXiv260221171S,2016MNRAS.459..199G}.

%This discussion was heated up again after James Webb Space Telescope (JWST) observations started. With the ability to see edge-on regular disk galaxies up to $z\approx 5$ \citep{2024ApJ...975...44Y} and even further, it can now potentially directly observe all epochs when discs have been formed. The data from Hubble \citep{2023ApJ...956..147H} and early based on JWST work \citep{2024ApJ...960L..10L} suggests that disc thickness is almost constant across early epochs. Separation by thin and thick discs in \citep{2025MNRAS.540.3493T} reveal, that thick disk forms first, and the transition to two disks varies across mass bins of the total stellar mass $M_{\star}$. The results however, are not settled yet, when various JWST samples gives different conclusions: \citep{2026arXiv260103339V} found that disks are more likely to heat up during evolution than to be born thick, and \citep{2026arXiv260104988Y} suggest that discs have intermediate thickness at birth and then heat up, but are definitely not born thin. Edge-on orientation of special galaxies may also be a possible explanation of Little Red Dots nature \citep{2026arXiv260301668P}.

This discussion has been reignited with the beginning of observations with James Webb Space Telescope (JWST). With the ability to observe edge-on regular disk galaxies up to $z\approx 5$ \citep{2024ApJ...975...44Y} and even beyond, JWST can now potentially directly observe all epochs during which disks formed. Data from Hubble Space Telescope \citep{2023ApJ...956..147H} and early JWST-based work suggest that disk thickness is almost constant across early epochs. The separation into thin and thick disks in \citep{2025MNRAS.540.3493T} reveals that the thick disk forms first, and the transition epoch to two disks varies in time depending on total stellar mass $M_{\star}$. The results, however, are not yet settled, as various JWST samples lead to different numbers: \citep{2026arXiv260104988Y} suggest that disks have intermediate thickness at birth and then thicken due to the combined effects of decreasing surface mass density and violent gravitational instabilities, but they are definitely not born as thin as found in \citep{2026arXiv260103339V}.

Quantitatively, the thickness of disks may be characterized by the intrinsic flattening $q_{int}$ (also called oblateness or aspect ratio), which is defined as the ratio of disk height to its radius. For observed galaxies we see the apparent flattening $q_{vis} = b/a$, where $b$ and $a$ are the disk minor and major axes. Due to projection effects, as well as the influence of the bulge and dust, we find $q_{vis} \geq q_{int}$, but the true intrinsic galaxy shapes can still be inferred from these measurements in a statistical sense by rotating various shapes and measuring the resulting projection's $b$ and $a$. This study was initiated by Edwin Hubble in the 1930s, and since \citep{1970ApJ...160..831S} we already know that $q_{int}\approx 0.25$, although individual galaxies have a fairly wide range of thicknesses, including ultra-thin galaxies, for which $q\approx 0.1$ \citep{1979BAAS...11..668G,1981ApJ...250...79G}. Statistical inference of $q_{int}$ for large samples such as GAMA DR3 \citep{2018MNRAS.474.3875B} and SDSS \citep{2008MNRAS.388.1321P,2020ApJS..249....3A}, along with comparisons with cosmological simulations (Illustris, IllustrisTNG, EAGLE) in \citep{2022ApJ...925..183H}, reveal that the high fraction of thin disks observed in real galaxies does not match the simulations (see also \citep{2019MNRAS.484..869V,2025arXiv250305612K}). There is an ongoing debate as to whether this effect is real or relates to an incorrect accounting of the spherical component \citep{2024ApJ...974...88X} or perhaps to the fact that the apparent $q_{vis}$ does not properly reflect the true three-dimensional shape \citep{2025arXiv250305612K}. Recent estimates of oblateness over a wide $M_{\star}$ range in \citep{2025arXiv251211035B} show that for 40\% of galaxies $q_{int}<0.2$, which results in an extremely large number of "flat" dwarfs, producing an even greater inconsistency with simulations. Since edge-on galaxies have the advantage that we can measure flattening directly rather than statistically, larger samples of galaxies with inclination $i\approx 90^{\circ}$ are needed to resolve this discrepancy in $q_{vis}$ versus $q_{int}$.

To further complicate the situation, there are indications from JWST observations that high-redshift galaxies may not exhibit oblate (disky) geometry but instead contain a significant number of prolate objects \citep{2024ApJ...974...48G,2024ApJ...963...54P}, which could be misidentified as edge-on systems. The edge-on orientation of these prolate galaxies may also be a possible explanation for the nature of Little Red Dots \citep{2026arXiv260301668P}. Potentially, such objects also can help to distinguish between different models of dark matter origin  \citep{2025NatAs.tmp..246P}. Moreover, knowing the disk thickness in edge-on galaxies allows us to understand the properties of the dark halo. For example, the vertical force gradient can be measured as in UGC~7321, and the full halo shape can be recovered \citep{2010A&A...515A..63O}. Another approach is that, under plausible assumptions about the Toomre parameter $Q$, we can derive the proportionality $h_z/h_R \propto M_{disk}/M_{tot}$, where $h_R$ is the radial disk scale and $M_{tot}$ is the total mass within a certain radius \citep{2006AstL...32..649S,2010AN....331..731K,2025PPulO.237...33M,rodionov_sotnikova2013,1991SvAL...17..374Z,2004ApJ...613..886B}. In other words, this imposes an independent constraint on the dark halo mass. Finally, the existence of superthin galaxies with observed $q < 0.1$ (e.g. IC~5249) and with weak or no galactic bulges (cf. \citep{edgeon1}) may imply the presence of an unusually massive dark halo with a compact core \citep{1991SvAL...17..374Z,2017MNRAS.466.3753B,2018MNRAS.479.5686K,2021ApJ...914..104B}. An alternative explanation derived from the TNG100 simulation in \citep{2024RAA....24g5019H} links superthin galaxies to disk radial growth driven by specific mergers. In their picture, the existence of superthin galaxies is explained by subhalos that fall in at small angles and in the direction of rotation, i.e. when angular momentum continuously pumped into the system since $z\approx 1$.

Many disk features are best seen when the disk is viewed edge-on and can be reliably studied only in this orientation. These include disk warps \citep{2025A&A...697L...1R,2016MNRAS.461.4233R,1998A&A...337....9R}, flaring \citep{2024ApJ...977...66R,2025arXiv251119632M}, central X-shaped structures from boxy/peanut-shaped (B/PS) bulges \citep{marchuk2022,2026Galax..14....4S,smirnov_savchenko2020}, polar rings \citep{2011MNRAS.418..244M,2024A&A...681L..15M,2022MNRAS.516.3692S,2015MNRAS.447.2287R}, and some low-surface-brightness structures surrounding galaxies \citep{2026arXiv260102579B}. Photometric breaks in radial disk profiles and truncations are also more visible in edge-on galaxies \citep{2026arXiv260200626C,2025arXiv251119632M}, as for the lower surface brightnesses (Savchenko et al., in prep.). Additionally, significantly lower uncertainty in the measured orientation parameters (inclination $i$ and position angle) for edge-on galaxies is important when conducting research where precise knowledge of these parameters is crucial. This applies to studies related to the rotation velocity of the galaxy, such as Tully-Fisher measurements \citep{2018MNRAS.479.3373M}, or research that depends on the mutual orientation with cosmic filaments \citep{2025PASA...42...84A,2025ApJ...993..205Z} or satellites \citep{2005A&A...431..517K,2024MNRAS.528.2805K}.

% which model to use, volumetric SF, gravitational instability, DIG

%Last to be mentioned but not least important advantage of galaxies visible from edge-on is the ability to study processes where knowledge about height above disk plane is essential. For example, extra-planar diffuse ionized gas (DIG) contaminate photometry in H$\alpha$ filter when viewed face-on and need to be separated from the disc emission \citep{2023AstL...49..151P,2025MNRAS.544.1861S}. Another example is a presence of extraplanar cold gas \citep{2013A&A...554A.125G,2026arXiv260306913J}, which may be essential for lenticular S0 galaxies formation \citep{2019ApJS..244....6S}. Disc vertical scale $h_z$ is also crucial when volumetric information is needed, like in studies of large-scale star formation (cite K-S), disc gravitational stability (cite me, Romeo), secular bar formation \citep{2017MNRAS.470.3685A} and many other.

Last but not least, an important advantage of galaxies viewed edge-on is the ability to study processes for which knowledge of the height above the disk plane is essential. For example, extraplanar diffuse ionized gas (DIG) contaminates photometry in the H$\alpha$ filter when viewed face-on and must be separated from the disk emission \citep{2023AstL...49..151P,2025MNRAS.544.1861S}. Another example is the presence of extraplanar cold gas \citep{2013A&A...554A.125G,2026arXiv260306913J}, which may be essential for the formation of lenticular S0 galaxies \citep{2019ApJS..244....6S}. The disk vertical scale height $h_z$ is also crucial when volumetric information is needed, as in studies of large-scale star formation \citep{2012ApJ...745...69K,2019A&A...622A..64B}, disk gravitational stability \citep{2018MNRAS.476.3591M,2018MNRAS.475.4891M,2011MNRAS.416.1191R,2013MNRAS.433.1389R,2025A&A...698L...6K}, secular bar formation \citep{2017MNRAS.470.3685A}, and many others.

For all the reasons listed above, edge-on disk galaxies are therefore of particular interest for various areas of extragalactic astrophysics. Although disk thickness can now be easily studied in many available simulations, real observations are still preferred due to the possible presence of unphysical conditions in simulations (e.g., \citep{2025A&A...703L..16M}). The importance of edge-on galaxies is emphasized by the efforts spent on compiling catalogues (and surveys, such as GECKOS \citep{2024IAUS..377...27V}) devoted to them. The classic Flat Galaxies Catalogue (FGC) \citep{1993AN....314...97K} and the Revised Flat Galaxy Catalogue (RFGC) \citep{1999BSAO...47....5K} together contain around 4000 large ($a\geq40$ arcsec) galaxies. A sample one and a half times larger was found in SDSS and presented in the $gri$ EGIS (Edge-on Galaxies In SDSS) catalogue \citep{bizyaev_etal2014}. Subsequent efforts allowed the construction of an even larger sample of 16,551 edge-on galaxies in the Pan-STARRS survey (EGIPS\footnote{\url{https://www.sao.ru/edgeon/catalogs.php?cat=EGIPS}}, \citep{egips,2024A&C....4600771S}), where edge-on galaxies were visually inspected after automatic selection from the Pan-STARRS DR2 survey. There are also specialized samples, such as the 2MASS-selected Flat Galaxy Catalog (2MFGC) \citep{2010MNRAS.401..559M} or \citep{2009ApJ...702.1567B,2002A&A...389..795B} from the same datasource, the B/PS-related subsample in \citep{2015MNRAS.446.3749Y}, a catalogue of low-surface-brightness edge-on galaxies \citep{2023ApJS..269...59X}, and others.

The number of disk galaxies with inclination between $i_1$ and $i_2$ varies as $\cos i_1 - \cos i_2$, which gives $\approx 1.75\%$ of total population for $i = 89^{\circ}$-$90^{\circ}$. Therefore, in the forthcoming era of contemporary wide-field surveys, such as Rubin, Roman, Euclid, 4MOST, and others, it is anticipated that samples of approximately $10^5$ highly inclined galaxies will be identified. The manual detection of such sizable samples is well beyond the reasonable capabilities of any individual or group. For example, the largest relevant project, Galaxy Zoo, has already collected 104 million annotations for a million galaxies \citep{2025arXiv251223691W}, but only around 15,000 edge-on galaxies with reliable annotations were found (see their figure~3). The detection of the axis ratio $q$ from photometric data, which is another method for finding edge-on galaxies, is susceptible to various influences, including projection effects, significant errors, and challenges in automatic decomposition. At the same time, the simplicity of form is a distinguishing characteristic of edge-on galaxies, which exhibit geometry and features (e.g., dust lanes) that are relatively straightforward to comprehend. In a previous study, an approach was proposed for EGIPS \citep{egips} that involved the implementation of an automatic detection system utilizing a neural network, which yielded satisfactory results. 

In this study, we propose an expanded methodology, whereby a pre-trained model from an external dataset is utilized and subsequently fine-tuned through the application of transfer learning. This approach is employed for the identification of edge-on galaxies within the wide-field DESI Legacy Imaging DR10 survey. The resulting sample is referred to as EGIDE\footnote{\url{https://www.sao.ru/edgeon/catalogs.php?cat=EGIDE}} (The Edge-on Galaxies in the DESI Survey). The EGIDE catalog, which we present here, contains more than 145,000 edge-on galaxies. The methodology employed here can be successfully applied to forthcoming Rubin, Roman, and Euclid data, and the EGIDE sample size is analogous to the expected sizes of future samples from these telescopes.

The present work is structured as follows. In Section~\ref{sec:CandidateSelection}, we describe the DESI DR10 data and outline the procedure used to detect edge-on galaxies. Section~\ref{sec:Photometry} presents an overview of the photometry extraction, its precision and the obtained results. In Section~\ref{sec:redshifts}, we list information about distances to the galaxies in EGIDE. Section~\ref{sec:completeness} discusses the completeness of our sample. In Section~\ref{sec:results}, we discuss our results, specifically the galaxy colour-magnitude relation in Section~\ref{sec:GalaxyCMD}, the colour-flattening relation in Section~\ref{sec:color_flatness}, the mass-disk thickness relation in Section~\ref{sec:mass}, and the number of galaxies versus flattening in Section~\ref{sec:countdrop}. Finally, Section~\ref{sec:conclusions} summarizes our conclusions. In Appendixes we provide additional details about neural network training, data cross-validation, and used methods.

%In Appendixes~\ref{ap:database_structure}-\ref{ap:gallery}, we provide additional details about the catalog, neural network training, data cross-validation, the methods used, and examples of galaxies from EGIDE.

Throughout this paper, we assume a standard flat $\Lambda$CDM cosmology with $\Omega_m = 0.3$, $\Omega_{\Lambda} = 0.7$ and $\mathrm{H}_0$ = 70 km/s/Mpc, and use the AB magnitude system.

\section{EGIDE sample building}
\label{sec:CandidateSelection}

%As a source of observational images we use in this work DESI Legacy Imaging Surveys. The DESI Legacy Imaging Surveys (DESI Legacy) represent a comprehensive astrophysical campaign designed to provide the essential imaging dataset for target selection of the Dark Energy Spectroscopic Instrument (DESI; cite smth). Initially DESI Legacy delivers a uniform imaging coverage of approximately 14,000 square degrees of the sky in the optical bands $grz$, combining data from three individual telescopes: the Dark Energy Camera Legacy Survey (DECaLS; Flaugher et al. 2015; Dark Energy Survey Collaboration et al. 2016), the Beijing-Arizona Sky Survey (BASS; Williams et al. 2004), and the Mayall $z$-band Legacy Survey (MzLS; Dey et al. 2019). These observations are complemented by photometry at infrared wavelengths using data from the Wide-field Infrared Survey Explorer (WISE) mission, thereby providing spectral energy distribution coverage from 0.4 to 4.5 $\upmu $m microns. Additional data provided in catalogs obtained using probabilistic astronomical source detection algorithm called TRACTOR (Lang et al. 2016). Another significant source of information is recently released DESI Data Release 1 (DR1), which presents spectroscopic redshift measurements for millions of galaxies \citep{2025arXiv250314745D}. 

As a source of observational images, we use the DESI Legacy Imaging Surveys in this work. The DESI Legacy Imaging Surveys (DESI Legacy) represent a comprehensive astrophysical campaign designed to provide the essential imaging dataset for target selection of the Dark Energy Spectroscopic Instrument (DESI; \citep{dey2018}). Initially, DESI Legacy delivers a uniform imaging coverage of approximately 14,000 square degrees of the sky in the optical $grz$ bands, combining data from three individual telescopes: the Dark Energy Camera Legacy Survey (DECaLS; \citep{2015AJ....150..150F}), the Beijing-Arizona Sky Survey (BASS; \citep{2017PASP..129f4101Z}), and the Mayall $z$-band Legacy Survey (MzLS; \citep{dey2018}). These observations are complemented by photometry at infrared wavelengths using data from the Wide-field Infrared Survey Explorer (WISE) mission, thereby providing spectral energy distribution coverage from 0.4~$\mu$m to 4.5~$\mu$m. Additional data are provided in catalogs obtained using the \textsc{TRACTOR} probabilistic astronomical source detection algorithm \citep{tractor}. Another significant source of information is the recently released DESI Data Release 1 (DESI DR1), which presents spectroscopic redshift measurements for millions of galaxies \citep{2025arXiv250314745D}.

%Throughout this work we use the images from the latest version of DESI LIS available, which is DR10\footnote{ For details about release see  \url{https://www.legacysurvey.org/dr10/description/}.}. The two most notable changes in this release are, first, inclusion  for the first time of the $i$-band observations and, second, prolongation of the southern $\delta \leq 32.375^{\circ}$ part of the survey (Drlica-Wagner et al. 2021; Zenteno et al. 2025). The good representation of the survey footprint presented in Figure~\ref{fig:skymap}. It is easy to see, that the full DESI DR10 cover significant fractions of both northern and southern celestial hemispheres, and contains more than 20000 square degrees of the sky. Besides large coverage, DESI Legacy DR10 another advantage is a significant depth of the images, which reaches averagely 28.5-29 mag arcsec$^2$ within a 10~arcsec square box at the 3$\sigma$ level in the $g$-band \citep{2026arXiv260102579B}.

\begin{figure}
\centering
\includegraphics[width=0.85\textwidth]{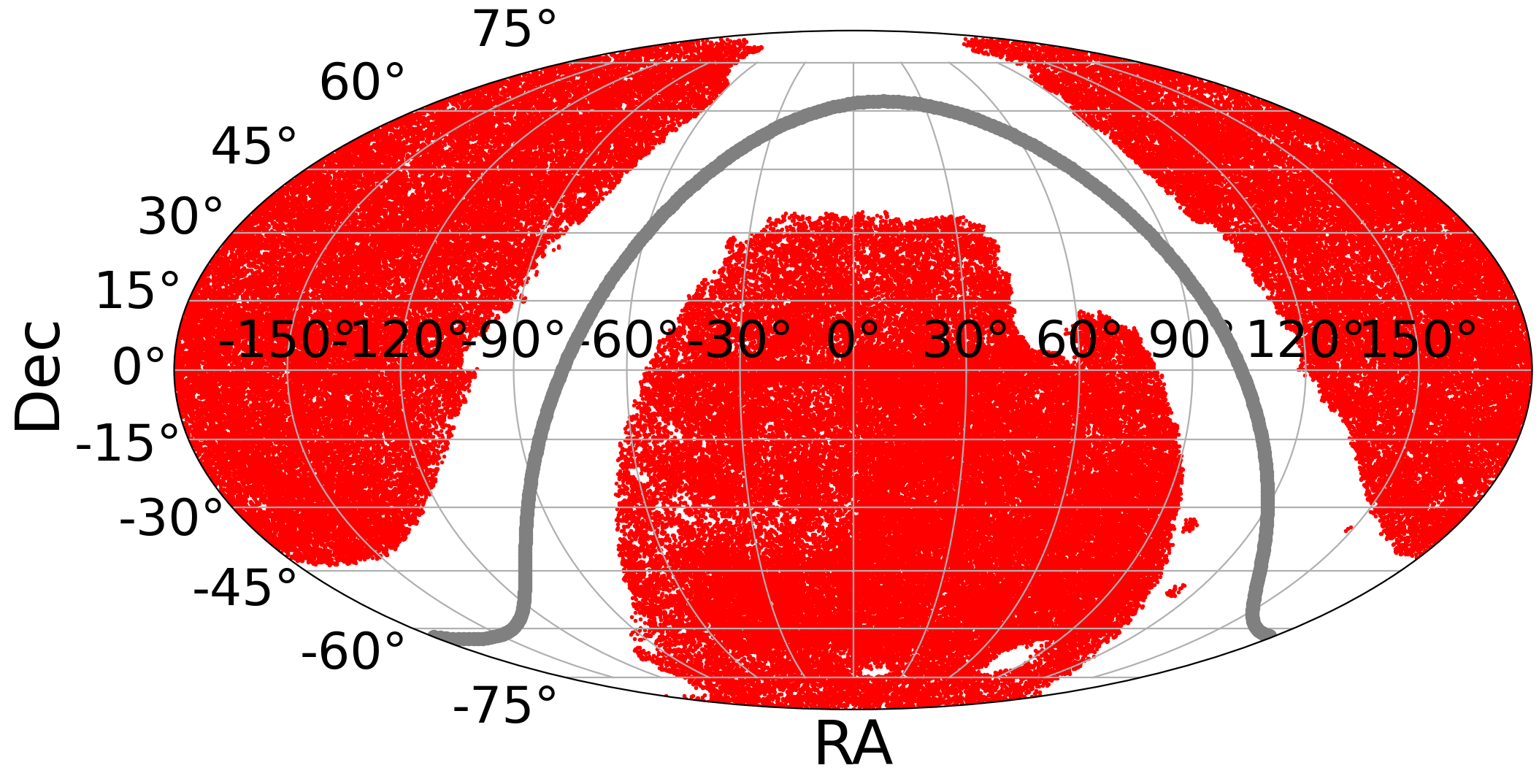}
\caption{
Final distribution of edge-on galaxies from the EGIDE, found in the DESI DR10 footprint, over the sky in the equatorial coordinate system. The grey line symbolically represents the plane of our Galaxy.
}
\label{fig:skymap}
\end{figure}

Throughout this work, we use images from the latest version of DESI Legacy available, which is DR10\footnote{For details about the release, see \url{https://www.legacysurvey.org/dr10/description/}.}. The two most notable changes in this release are, first, the inclusion of $i$-band observations for the first time, and second, the extension of the southern ($\delta \leq 32.375^{\circ}$) part of the survey \citep{2021ApJS..256....2D}. The survey footprint is well represented in Figure~\ref{fig:skymap}. It is easy to see that the full DESI DR10 covers significant fractions of both the northern and southern celestial hemispheres and covers more than 20,000 square degrees of the sky. In addition to its large coverage, DESI Legacy DR10 has another advantage: the significant depth of the images, which reaches on average 28.5-29 mag arcsec$^{-2}$ within a 10 arcsec square box at the $3\sigma$ level in the $g$-band \citep{2026arXiv260102579B}.

%All coadded observations in DESI Legacy are grouped in so-called "bricks". Each brick has approximate size of $0.25^{\circ}\times 0.25^{\circ}$ and defined in terms of RA/DEC coordinates. There are 366898 unique bricks in DR10, which have some small border overlap and this is the reason why we see duplicate detections later on. Each image has identical pixel scale 0.262 arcsec/pix. The median PSF FWHM approximately equals 1.29~arcsec, 1.18~arcsec and 1.11~arcsec for $grz$ accordingly (Dey et al. 2019). Note, that starting from DESI DR9, PSF model includes inner and outer component, where the latter represents large extended wings by a power law (or Moffat profile for $z$).

All coadded observations in DESI Legacy are grouped into so-called "bricks". Each brick has an approximate size of $0.25^{\circ}\times 0.25^{\circ}$ and is defined in terms of RA/DEC coordinates. There are 366,898 unique bricks in DR10, which have some small border overlap, and this is the reason why we see duplicate detections later. Each image has an identical pixel scale of 0.262 arcsec/pix. The median PSF FWHM is approximately 1.29 arcsec, 1.18 arcsec, and 1.11 arcsec for $grz$, respectively \citep{dey2018}. Note that starting from DESI DR9, the PSF model includes an inner and an outer component, where the latter represents large extended wings by a power law for $g$ and $r$ bands or a Moffat profile for $z$-band\footnote{\url{https://www.legacysurvey.org/dr10/psf/}}.

To construct a sample of edge-on galaxies in the DESI survey, we used an approach similar to the one  used in \citep{2024A&C....4600771S} to make the EGIPS database \citep{egips}. We downloaded all available images  (``bricks'') in all available bands, and then performed an initial object search using the \textsc{SExtractor}  package~\citep{1996A&AS..117..393B}. During this search we excluded too small objects by allowing the  \textsc{SExtractor} parameter \texttt{A\_IMAGE} to only be greater then 10 pixels (which roughly corresponds to a lower limit of the Kron radius of 9~arcsec). We also excluded too round  (\texttt{ELLIPTICITY} $<$ 0.4) objects, as they usually cannot represent edge-on galaxies, and too elliptical objects  (\texttt{ELLIPTICITY} $>$ 0.95), because no real galaxies showed such ellipticities, but only image artifacts such as  leaked signal from oversaturated bright stars. This preliminary stage resulted in $1.4\times 10^6$ objects. To find  edge-on galaxies in this catalogue we utilized an artificial neural network classifier.

All necessary technical details about the neural network training are given in Appendix~\ref{ap:nnsearch}. Here, for convenience, we provide only some of them. The main difference from our previous work \citep{2024A&C....4600771S} is that then we trained a neural  network classifier from scratch using solely our training set of edge-on and not edge-on galaxies. In this work, we  use another approach called fine-tuning \citep{2021arXiv210410972R}, where we start with a neural network already  pretrained on a broad generic task and then adapt it to our task of edge-on galaxies search. Starting from a  pre-trained model instead of a random initial condition allows one to achieve a better convergence of the training  process especially for limited training samples \citep{2024arXiv240402973W}. The contemporary and upcoming wide-field  surveys such as Rubin, Roman, and Euclid provide a plethora of data to build heavy foundation models that can be  further fine-tuned to more specific tasks. 

We use \textsc{Zoobot} as the base for our search. \textsc{Zoobot} \citep{2023JOSS....8.5312W} is a Python package that leverages deep learning to characterize the detailed appearance of galaxies. It is trained on around 92 million labels collected from Galaxy Zoo volunteers, who answered a series of questions about each galaxy's morphology (see the example question tree in figure~4 in \citep{2022MNRAS.509.3966W}). \textsc{Zoobot} learns to answer all of these diverse tasks simultaneously, acquiring a rich, transferable representation of galaxy structure that can be efficiently adapted to new related tasks. We trained nine independent models by repeatedly re-running \textsc{Zoobot} fine-tuning process in order to reduce possible missclassifications of individual models. We used edge-on RFGC and EGIPS galaxies from the previous works as positive examples \citep{2024A&C....4600771S}, the training size is several thousands of examples (see Appendix~\ref{ap:nnsearch} for exact numbers). The performance of individual classifiers of the endemble is good, achieving 97\% accuracy (see Figure~\ref{fig:learningcurve}). We select a galaxy as a candidate to be located edge-on if the majority of the ensemble models (at least five out of nine) vote for it.

Automated selection of galaxies works as a "black box" and, in principle, can be susceptible to errors of the first and second kind. Bright bars, stellar spikes due to the PSF, or other elongated structures can easily mimic edge-on galaxies; therefore, additional efforts are always needed to ensure the validity of the sample. The application of our neural ensemble resulted in an considerable number of false detections caused by horizontal lines due to CCD leakage around bright stars. This is a known problem in DESI Legacy, and examples of such stars can be easily found in the DESI Viewer\footnote{\url{https://www.legacysurvey.org/viewer}}. The number of such lines greatly exceeds the number of edge-on galaxies, so even a small fraction of false positive detections of them significantly contaminates the classification results. To solve this issue, we created a sample of 5000 such artifacts and trained another \textsc{Zoobot} classifier to separate these artifacts from edge-on disks. This genuinely fixed the problem.

%To ensure the reliability of our results, as the last step of sample building pipeline we consider to check all pre-selected candidates manually. From the 199... net-selected objects we remove 55... or roughly 25\% of the initial sample. These includes not only clearly false detections, but also situations where there is an edge-on galaxy within the cropped image, which probably activates net weights, or really elongated objects, which are however not edge-on (see examples in Figure~...). We also remove relatively few duplications caused by bricks overlap (see above). We want to emphasize, that the quarter of false detections is a good result consider difficulty of some cases and the small training dataset. It can be clearly lowered using additional training, and clearly analogous applications to the future missions such as wide-field surveys as Rubin/Roman/Euclid will benefit from the EGIDE size.

To ensure the reliability of our results, as the last step of the sample building pipeline we manually check all pre-selected candidates. From the 192593 net-selected objects, we remove 43378 or roughly 25\% of the initial sample. These removals include not only clearly false detections but also cases where there is an edge-on galaxy within the cropped image that probably activates the network weights, as well as genuinely elongated objects that are nevertheless not edge-on (see examples in Figure~\ref{fig:exampleBad}). We also remove relatively few duplications caused by bricks overlap (see above). We want to emphasize that a quarter of false detections constitutes only 3\% of the initial sample pre-selected by \textsc{SExtractor}, and it is a good result considering the difficulty of some cases and the small training dataset. This rate can be clearly lowered with additional training, and future missions such as the wide-field surveys Rubin, Roman, and Euclid will clearly benefit from the EGIDE size, if it will be used for updating weights.

%The final Edge-on Galaxies in the DESI survey (EGIDE) sample consists of ... galaxies. Several examples of edge-on or nearly edge-on galaxies are presented in Figure~\ref{fig:exampleGood} and in Appendix~\ref{ap:gallery}. The sky distribution of all selected candidates is presented in Figure~\ref{fig:skymap}. It is clear, that there is no preferred directions within DESI DR10 footprint except the area around the Milky Way plane, which posses strong extinction and thus is hard to observe. Overall, the distribution of EGIDE galaxies is uniform an covers more than half of the sky. Take in mind however, that not all bands are available for every galaxy, with $i$ band images are most deficient as they cover only 72\% of the whole sample.

The final Edge-on Galaxies in the DESI Survey (EGIDE) sample consists of 149215 galaxies. Several examples of edge-on or nearly edge-on galaxies are presented in Figure~\ref{fig:exampleGood} and in Appendix~\ref{ap:gallery}. 
The sky distribution of all selected candidates is presented in Figure~\ref{fig:skymap}. It is clear that galaxies of our sample uniformly cover the DESI DR10 footprint without significant gaps. Overall, the distribution of EGIDE galaxies covers more than half of the sky. Note, however, that not all bands are available for every galaxy, with $i$-band images being the most deficient, as they cover only 72\% of the entire sample.

%In Table~\ref{tab:inters} we provide intersection of EGIDE sample with various extragalactic databases, including previously constructed catalogues of edge-on systems RFGC, EGIS and EGIPS. We provide this information in case the reader will be interested in edge-on galaxies in the exact survey and may planning future research using this information.

In Table~\ref{tab:inters}, we provide the intersection of the EGIDE sample with various extragalactic databases, including previously constructed catalogs of edge-on systems such as RFGC, EGIS, and EGIPS. We provide this information in case the reader is interested in edge-on galaxies within a specific survey and may be planning future research using this information.

 \begin{figure}
 \centering
 \includegraphics[width=0.23\textwidth]{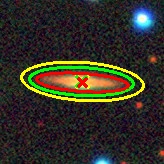}
 \includegraphics[width=0.23\textwidth]{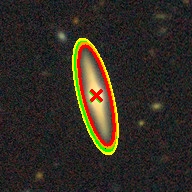}
 \includegraphics[width=0.23\textwidth]{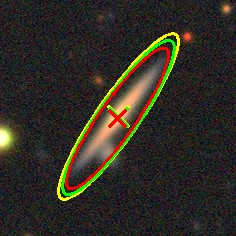}
 \includegraphics[width=0.23\textwidth]{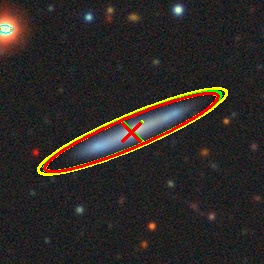}
 \includegraphics[width=0.23\textwidth]{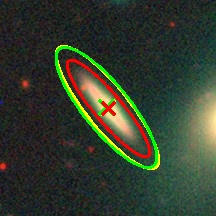}
 \includegraphics[width=0.23\textwidth]{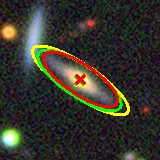}
 \includegraphics[width=0.23\textwidth]{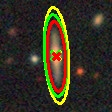}
 \includegraphics[width=0.23\textwidth]{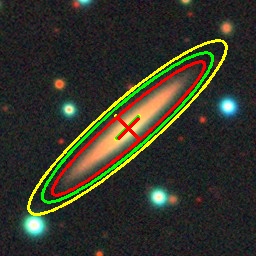}
 \caption{
 Examples of objects classified as truly edge-on galaxies. Ellipses of different colours show segmentation by \textsc{SExtractor} in different bands. More images can be found in the EGIDE project site \url{https://www.sao.ru/edgeon/catalogs.php?cat=EGIDE}.
More images can be found in Appendix~\ref{ap:gallery} and in the EGIDE project site \url{https://www.sao.ru/edgeon/catalogs.php?cat=EGIDE}.
 }
 \label{fig:exampleGood}
 \end{figure}

\section{Photometry}
\label{sec:Photometry}

\subsection{Photometry making}
\label{subsec:Photometry}

To obtain photometric parameters of the galaxies in our sample, we used the \textsc{SExtractor} software package~\citep{1996A&AS..117..393B}. To perform photometry, we downloaded images of each galaxy in all available bands, using preliminary estimates of their coordinates and sizes that were obtained during the initial sample construction. To reduce the contamination of galaxy photometry by foreground stars, we decided to mask them out (even though the deblending algorithm of \textsc{SExtractor} effectively separates objects from each other, in some cases it  fails to do so when a star is projected onto a galactic body). To make a stellar mask, we queried the Gaia database \citep{2023A&A...674A...1G} to obtain coordinates and magnitudes of stars in the image and then covered each star with a circular mask whose radius depends on the star's magnitude in $G$ band. We empirically found that a mask with radius 
%  $$
%  r(G)\mathrm{[arcsec]} = 35.6\exp\left( -0.3(G-12.7)\right) + 8
%  $$
  $$
  r(G)\mathrm{[arcsec]} = 9.33\exp\left( -0.3(G-12.7)\right) + 2.10
  $$
  effectively covers stars over the entire range of magnitudes. For consistency, we used the same mask for all bands.

Once the mask was ready, we automatically ran \textsc{SExtractor} for all bands and obtained a set of
  photometric parameters including coordinates, sizes, fluxes, and position angles. To illustrate the photometry results, we used the estimates of geometric parameters (central coordinates, position angles, and major- and minor-axes sizes) to draw ellipses that contain the Kron fluxes. The resulting ellipses produced by \textsc{SExtractor} were checked\footnote{Images with them are available at the \url{https://www.sao.ru/edgeon/catalogs.php?cat=EGIDE} site.}. There are some problematic situations where a close or foreground object distorts the ellipse boundary, or rare situations where the segmentation process splits objects into pieces. These often result in severe variation of ellipse properties between bands, such as position angle, and most suspicious cases can thus be easily identified; see examples of such cases in Figure~\ref{fig:exampleBad}. We provide an additional test, where we check the difference in position angle of ellipses between bands and find that a difference of more than $2^{\circ}$ occurs for $\sim$9000 cases, which is a small fraction. We validate all results without such problematic cases. Overall, the quality of our \textsc{SExtractor} aperture photometry was decent, and the comparison with other surveys in Appendix~\ref{ap:catalog_comparison} strengthens this point. It is important to mention that as was noted in \citep{2023ApJS..269....3M}, $z$-band  images in the northern portion of the DESI footprint significantly affected by pattern noise subtraction. It is not an issue in this work since $z$-band photometry was not used in the analysis, but presented in EGIDE $z$-band magnitudes should be used with caution.

%In total, for most of the galaxies in EGIDE we measured semi-major $a$ and semi-minor $b$ axes of the ellipse, its position angle and ellipticity, as well as the Petrosian \citep{1976ApJ...209L...1P} and Kron \citep{1980ApJS...43..305K} magnitudes in each image and band available. We cares the comparison of the results with other available measurements in Appendix~\ref{ap:catalog_comparison} and find overall the good match (see Figure~\ref{fig:mag_comparison}). All magnitudes were corrected for the Galactic extinction using SFD map \citep{2011ApJ...737..103S}, accessed via dustmaps \citep{dustmap}. Exact corrections calculated from $E(B-V)$ reddening for $R_V=3.1$ using table~6 coefficients in \citep{2011ApJ...737..103S}. It is important to mention that edge-on galaxies obviously have a great internal extinction, but it is difficult to measure (cf. \citep{2007ApJ...659.1159S}). Finally, corrections for SDSS stellar magnitudes to take into account differences in bands with DESI Legacy were estimated using B4-B6 equations from \citep{dey2018}.

In total, for most of the galaxies in EGIDE, we measured the semi-major $a$ and semi-minor $b$ axes of the ellipse, its position angle and ellipticity, as well as the Petrosian \citep{1976ApJ...209L...1P} and Kron \citep{1980ApJS...43..305K} magnitudes in each available image and band. We compare the results with other available measurements in Appendix~\ref{ap:catalog_comparison} and find overall a good match (see Figure~\ref{fig:mag_comparison}). 
All magnitudes were corrected for Galactic extinction using GMS25 map from \citep{2025RAA....25l5016G}. This map gives similar values as widely-used recalibration of SFD data in \citep{2011ApJ...737..103S}, but constructed 
 with hundred times more stars than one used for SFD recalibration. We additionally check that usage of \citep{2011ApJ...737..103S} extinctions, accessed via \textsc{dustmaps} \citep{dustmap}, will not alter the results. 
The exact corrections were calculated from $E(B-V)$ reddening for $R_V=3.1$ using table~6 coefficients in \citep{2011ApJ...737..103S}. It is important to note that edge-on galaxies obviously have significant internal extinction, but it is difficult to measure (cf. \citep{2007ApJ...659.1159S}). 
We derived K-correction in $griz$ bands using the \textsc{kcorrect} software\footnote{Available at \url{https://github.com/blanton144/kcorrect}, we use version 5.1.9.} \citep{2007AJ....133..734B}. Estimated median K-correction is around 0.2~mag in $g$-band, 0.1~mag in $r$-band and 0.05~mag in $i$-band.
Finally, corrections for SDSS stellar magnitudes to account for differences in bands with DESI Legacy were estimated using the B4-B6 equations from \citep{dey2018}.

\subsection{Oblateness estimation: PSF correction and bias sources}
\label{subsec:psf_smearing}

\begin{figure}
\centering
\includegraphics[width=0.95\textwidth]{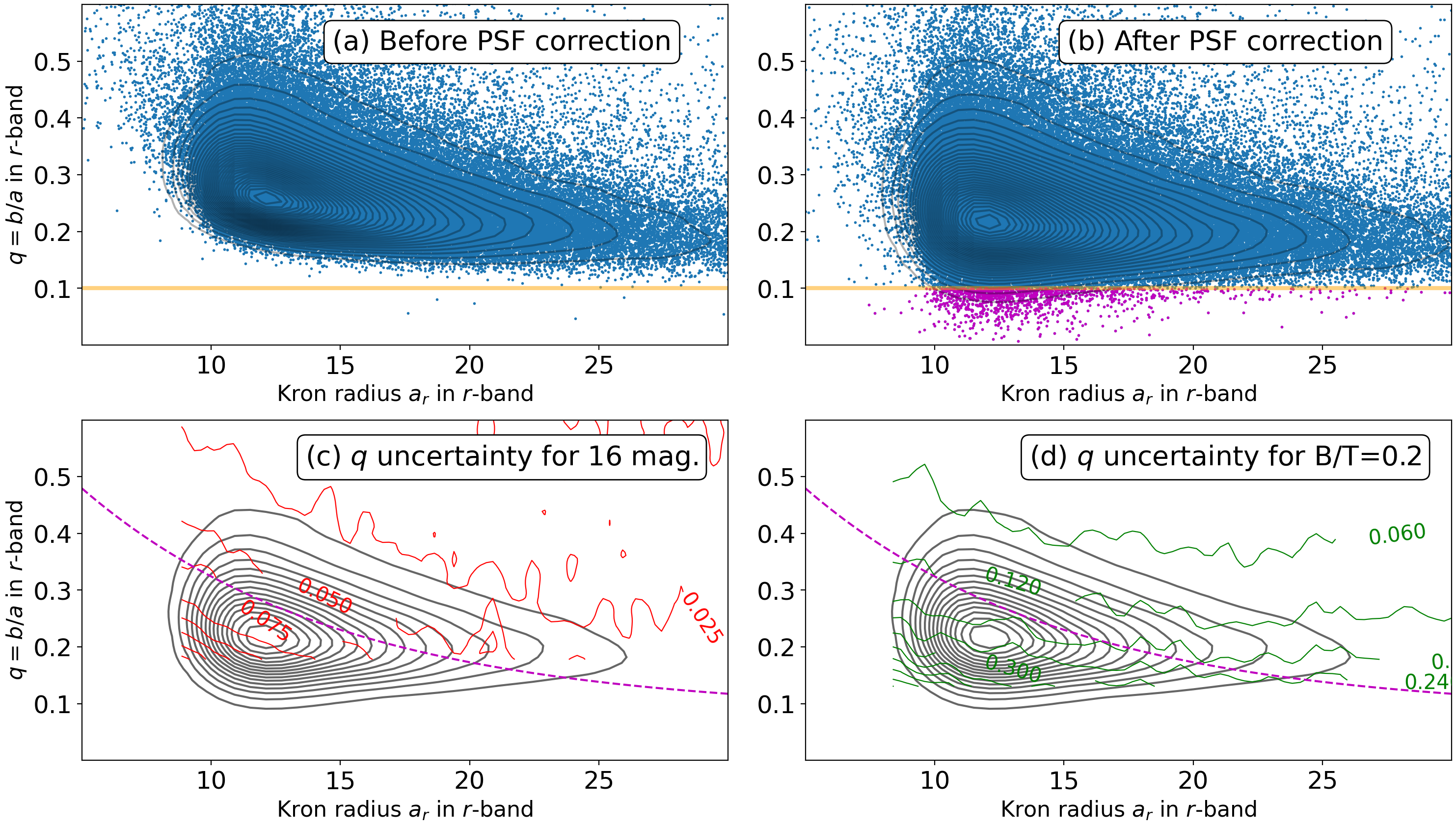}
\caption{
Distribution of $q$ versus Kron radius $a_r$ in the $r$-band. In each plot, black contours show the density of EGIDE points (blue). The data are presented before PSF correction (a) and after correction (b, c, d). The horizontal line shows the boundary at $q=0.1$; points shown in magenta are filtered out. In panels (c) and (d), coloured isolines show contours of constant relative error of the $q=b/a$ after PSF correction estimated without and with a bulge, respectively. The magenta line indicates the adopted error threshold used in the text.
}
\label{fig:qcorrect}
\end{figure}

  % \textcolor{red}{Since oblateness $q=b/a$ estimation using \textsc{SExtractor} is crucial for analysis in several parts of this work, we will discuss in this subsection how it can be affected by various effects. We will touch several of these: image smearing by the PSF, resolution effects and how it makes small galaxies ``rounder'' and also bulge-related incrementation in $b$.}

  % Since oblateness $q=b/a$ estimation using \textsc{SExtractor} is crucial for analysis in several parts of this work, we discuss in this subsection how it can be affected by various effects. We touch on several of these: image smearing by the PSF, resolution effects and how they make small galaxies ``rounder'', and bulge-related incrementation of $b$.
  Oblateness estimates $q = b/a$ made by \textsc{SExtractor} are affected by various effects: image smearing by the PSF, finite resolution, and bulge impact all make small galaxies appear "rounder". Since we use oblateness estimates for analysis throughout this work, in this subsection we discuss the scales of these effects and how they can be reduced in our measurements.

 Image smearing by the PSF alters the observed brightness distribution of astronomical objects, resulting in systematic biases of measured structural parameters, including major and minor axis lengths. As a consequence, the observed axis ratio is shifted toward rounder values, and need to be corrected. The values $a$ and $b$ of major and minor axes presented in this work are based on the \texttt{A\_IMAGE} and \texttt{B\_IMAGE} output parameters of \textsc{SExtractor}, which are derived from the square roots of eigenvalues of the covariance matrix $\hat{C}_{\mathrm{object}}$ of the object brightness distribution. Convolution of a brightness distribution with a kernel results in addition of their covariance matrices:
$$
\hat{C}_{\mathrm{observed}} =
\hat{C}_{\mathrm{object}} +
\hat{C}_{\mathrm{PSF}},
$$
therefore the intrinsic covariance matrix can be estimated as
$$
\hat{C}_{\mathrm{object}} =
\hat{C}_{\mathrm{observed}} -
\hat{C}_{\mathrm{PSF}}.
$$

  Assuming that the PSF can be approximated by a circular Gaussian with dispersion $\sigma$, the intrinsic (deconvolved) object sizes can be estimated as
$$
\mathrm{A\_IMAGE}^2_{\mathrm{object}} =
\mathrm{A\_IMAGE}^2_{\mathrm{observed}} - \sigma^2,
$$
$$
\mathrm{B\_IMAGE}^2_{\mathrm{object}} =
\mathrm{B\_IMAGE}^2_{\mathrm{observed}} - \sigma^2.
$$
And thus:
$$
q = \left(\frac{b}{a}\right)=
\sqrt{
\frac{
\mathrm{B\_IMAGE}^2_{\mathrm{observed}} - \sigma^2
}{
\mathrm{A\_IMAGE}^2_{\mathrm{observed}} - \sigma^2
}
},\,\, \mathrm{where}\,\, \sigma =\frac{\mathrm{FWHM}}{2\sqrt{2\ln 2}}.
$$

  Therefore we correct all $a$ and $b$ values using appropriate PSF $\sigma$ of DESI Legacy survey in each band. We note that this correction is approximate because real galaxy profiles and the PSF are generally non-Gaussian, while \textsc{SExtractor} moments are computed from weighted pixel distributions.

  % \textcolor{red}{The effect of such correction presented in Figure~\ref{fig:qcorrect}, upper row. We can clearly see how after correction $q$ values became smaller, especially for galaxies with small sizes, which appear to be more rounder due to PSF smearing effect. Now both small and large galaxies demonstrate thin objects and the isolines appear more strait, as expected. We additionally filter out galaxies which are smaller than value $q=0.1$, which is interpolated from lower boundary of larger galaxies, to remove those where correction results in unphysically small values. }

  The effect of this correction is presented in Figure~\ref{fig:qcorrect}, where panel (a) shows distibution before correction and panel (b) -- after correction. We can clearly see how after correction the $q$ values become smaller, especially for galaxies with small sizes, which appear rounder due to the PSF smearing effect. Now both small and large galaxies demonstrate thin objects, and the isolines appear straighter, as expected. We additionally filter out galaxies with $q$ smaller than $0.1$, a value interpolated from the lower boundary of larger galaxies, to remove those for which the correction results in unphysically small values.

  % \textcolor{red}{After correction we can have unrealistically small $b$ values due to PSF variation and resolution effects, which are limited by signal-to-noise level and other effects. In order to estimate the number of affected objects we performed a numerical experiment. In this experiment we created images of artificial galaxies with given values of $a$ and $b$ and then pefromed {\small SEXTRACTOR} protometry the same way we did it for real galaxies of our sample. Then we compared true (input) values of galactic parameters and measured ones. Since the effect of rounding should strongly depend on both true values of galaxy size and its axes ratio (small and thin galaxies should be more affected by rounding), we performed this experiment for a range of combinations of $a$ and $q=b/a$.}

  After correction, we can obtain unrealistically small $b$ values due to PSF variation and resolution effects, which are limited by the signal-to-noise level and other factors. To estimate the number of affected objects, we performed a numerical experiment. In this experiment, we created images of artificial galaxies with given values of $a$ and $b$ and then performed \textsc{SExtractor} photometry in the same way we did for real galaxies in our sample. We then compared the true (input) values of galactic parameters with the measured ones. Since the rounding effect should depend strongly on both the true values of galaxy size and its axis ratio (small and thin galaxies should be more affected by rounding), we performed this experiment for a range of combinations of $a$ and $q=b/a$.

  % \textcolor{red}{We used {\small IMFIT MAKEIMAGE} package to create a model image of a galaxy with a pure exponential brightness distribution, normalized it's flux to a 16-th magnitude and 17-th magnitude in $r$-band (which are typical for galaxies of our sample, see next Section), convolved it with $r$-band PSF image of a Legacy survey and then coadd it into a random position of a real DESI Legacy field in order to simulate the impact of the noise of real images. Then we used {\small SEXTRACTOR} to perform photometry and obtain observed values of $a$ and $b$ parameters.}

  We used the \textsc{Imfit} \textsc{MakeImage} \citep{2015ApJ...799..226E} package to create a 5000 model images of a galaxy with a pure exponential brightness distribution, normalized its flux to 16th and 17th magnitude in the $r$-band (which are typical for galaxies in our sample; see the next Section), convolved it with the $r$-band PSF image of the Legacy Survey, and then co-added it to a random position in a real DESI Legacy field in order to simulate the impact of the noise from real images. We then used \textsc{SExtractor} to perform photometry and obtain observed values of the $a$ and $b$ parameters.

%   \textcolor{red}{The results of such experiment are presented in Figure~\ref{fig:qcorrect}, lower row, where for each combination of observed values of $a$ and $b/a$
% we show a relative error of $q$ measurement:
% $$
% \frac{q_{\mathrm{measured}} - q_{\mathrm{true}}}{q_{\mathrm{true}}}.
% $$
% As expected, the error in measured $q$ value is positive, i.e. galaxies are observed to be rounder, then
% they are in reality, and the smaller and thinner the galaxy, the larger the error. The errors for 16-th magnitude and 17-th magnitude models are similar and can be used interchangeably. Using such isolines as presented, we can arguably select only galaxies with relatively small error of $q$ estimation. It is also clear from this Figure, that majority of simulated galaxies in this experiment have $\delta q < 50\%$. Note also the good coincide between density distribution of estimated $b/a$ after correction and presented islines.}

The results of this experiment are presented in Figure~\ref{fig:qcorrect}, panel (c), where for each combination of observed values of $a$ and $b/a$ we show the relative error of the $q$ measurement:
$$
\frac{q_{\mathrm{measured}} - q_{\mathrm{true}}}{q_{\mathrm{true}}}.
$$
As expected, the error in the measured $q$ value is positive, i.e., galaxies are observed to be rounder than they are in reality, and the smaller and thinner the galaxy, the larger the error. The errors for the 16th magnitude and 17th magnitude models are similar and can be used interchangeably. Using the isolines as presented, we can arguably select only galaxies with a relatively small error of $q$ estimation. It is also clear from this figure that the majority of simulated galaxies in this experiment have $\delta q < 15\%$ after PSF correction. Note also the good agreement between the density distribution of estimated $b/a$ after correction and the presented isolines.

  % \textcolor{red}{The last but not least important to discuss here is the bulge influence, which can arguably increase measured thickness if not taken into account. The bulge component light can be properly measured only after photometric decomposition, which is a tedious task if performed well. We will try to make a glimpse on its effect indirectly using several approaches below.}

Last but not least, it is important to discuss here the bulge influence, which can increase the measured thickness if not taken into account. The bulge component light can be properly measured only after photometric decomposition, which is a tedious task if performed well. We will try to glimpse its effect indirectly using several approaches below.

  % \textcolor{red}{In principle, the knowledge of morphological Hubble types can provide the  needed $B/T$ (see, for example, figure~8 in \citep{bizyaev_etal2014}). For some edge-on galaxies in the intersection with HyperLEDA database \citep{leda}, see Table~\ref{tab:inters}, we have estimated Hubble type $t$. There are only several thousands object in total with available $t$, which are not enough for full analysis, but in principle this information can be used for bulge influence on $q$ estimation. According to HyperLEDA database, in intersection we have 176 Sa, 289 Sb, and 1863 Sc galaxies. For them we can estimate oblateness $q$ and its standard deviation equal $0.26\pm0.10$, $0.24\pm0.08$ and $0.24\pm0.07$ accordingly. If we assume $B/T$ values similar to those found from photometric decomposition in EGIS \citep{bizyaev_etal2014} (see figure 8), then bulge-to-total ratio will be approximately 0.6/0.3/0.2 for Sa/Sb/Sc bins accordingly. Given that there is no drastic difference in linear sizes between galactic types under consideration, we can assume only moderate influence of bulge on $q$.}

 In principle, the knowledge of morphological Hubble types can provide the  needed $B/T$ (see, for example, figure~8 in \citep{bizyaev_etal2014}). For some edge-on galaxies in the intersection with HyperLEDA database \citep{leda}, see Table~\ref{tab:inters}, we have estimated Hubble type $t$. There are only several thousands object in total with available $t$, which are not enough for full analysis, but in principle this information can be used for estimation of bulge influence on $q$. According to HyperLEDA database, in intersection we have 176 Sa, 289 Sb, and 1863 Sc galaxies. For them we can estimate oblateness $q$ and its standard deviation equal $0.26\pm0.10$, $0.24\pm0.08$ and $0.24\pm0.07$ accordingly. If we assume $B/T$ values similar to those found from photometric decomposition in EGIS \citep{bizyaev_etal2014} (see figure 8), then bulge-to-total ratio will be approximately 0.6/0.3/0.2 for Sa/Sb/Sc bins accordingly. Given that there is no drastic difference in linear sizes between galactic types under consideration, we can assume only moderate influence of bulge on $q$.

% (176, 289, 1863)
% np.median(data[0]), np.median(data[1]), np.median(data[2])
% (0.26490043454365153, 0.24301397282880685, 0.24355317835423465)
% np.std(data[0]), np.std(data[1]), np.std(data[2])
% (0.09840603857469465, 0.08346195332247712, 0.07421492487345369)

    % \textcolor{red}{Bulge influence can potentially be approximated by metrics of galactic bulge strength, such as the concentration index $C$, the Sérsic index $n$, or the so-called Gini-$M_{20}$ bulge parameter (GMB), which is less sensitive to dust and mergers \citep{2019MNRAS.483.4140R}. All of these, however, require calculating additional nontrivial parameters, which estimation may also be unstable. We utilize the advantage of the already calculated CAS \citep{2003ApJS..147....1C} statistics for EGIPS data and estimate the GMB parameter as $GMB = -0.693\times M_{20} + 4.95\times \text{Gini} - 3.96$, see \citep{2019MNRAS.483.4140R}. A GMB value greater than zero indicates greater bulge domination, while a lower value indicates greater disk domination. We see a statistically significant anticorrelation between GMB and $a/b$ in EGIPS galaxies, with a Spearman's coefficient $\rho = -0.47$, and $\rho = 0.43$ for GMB versus $(g-i)$ colour. This could provide a clue about bulge influence, which may reasonably be expected to be of the same order in the EGIDE sample.}

   Bulge influence can potentially be approximated by metrics of galactic bulge strength, such as the concentration index $C$, the Sérsic index $n$, or the so-called Gini-$M_{20}$ bulge parameter (GMB), which is less sensitive to dust and mergers \citep{2019MNRAS.483.4140R}. All of these, however, require calculating additional nontrivial parameters, whose estimation may also be unstable. We utilize the advantage of the already calculated CAS \citep{2003ApJS..147....1C} statistics for EGIPS data and estimate the GMB parameter as $GMB = -0.693\times M_{20} + 4.95\times \text{Gini} - 3.96$ (see \citep{2019MNRAS.483.4140R}). A GMB value greater than zero indicates greater bulge domination, while a lower value indicates greater disk domination. We see a statistically significant anticorrelation between GMB and $a/b$ in EGIPS galaxies, with Spearman's coefficient $\rho = -0.47$, and $\rho = 0.43$ for GMB versus $(g-i)$ colour. This could provide a clue about bulge influence, which may reasonably be expected to be of the same order in the EGIDE sample.

    % \textcolor{red}{Finally, we carried out the same experiment as above with a model image of a galaxy, but with additional bulge component to estimate relative error of $q$ measurement. All parameters were similar and we add bulge with $B/T=0.2$, which is expected to be a typical value in the sample (see figure~6 in \citep{bizyaev_etal2014}). The results are presented in Figure~\ref{fig:qcorrect}, lower right. As one can expect, for smaller galaxies effect is significant, combined with PSF it reaches relative error of 60\%--100\% and more. Same limit as we impose for reliable $q$ estimation will give 30\% error in $q$, which is smaller than variance within individual Hubble types, as we show previously.}

   Finally, we carried out the same experiment as above with a model image of a galaxy, but with an additional bulge component to estimate the relative error of the $q$ measurement. All parameters were similar, and we added a Sersic bulge with $B/T=0.2$, which is expected to be a typical value in the sample (see Figure~6 in \citep{bizyaev_etal2014}). The results are presented in Figure~\ref{fig:qcorrect}, lower right. As one might expect, for smaller galaxies the effect is significant; after PSF correction, it reaches a relative error of 30\%--50\%. The same limit that we impose for reliable $q$ estimation gives a conservative 25\% error in $q$, which is smaller than the variance within individual Hubble types, as we showed previously.

  % \textcolor{red}{To ensure the validity of obtained results, in all places where it matters (Sections~\ref{sec:color_flatness}--\ref{sec:countdrop}) we use only subsample with $b/a$ estimated above the threshold line, presented in Figure~\ref{fig:qcorrect} and consisted of 61775 galaxies.}

To ensure the validity of the obtained results, in all places where it matters (Sections~\ref{sec:color_flatness}--\ref{sec:countdrop}) we use only the subsample with $b/a$ estimated above the threshold line presented in Figure~\ref{fig:qcorrect}, consisting of 61,775 galaxies.

\subsection{Photometry results}
\label{subsec:photoresults}

\begin{figure}
\centering
\includegraphics[width=0.95\textwidth]{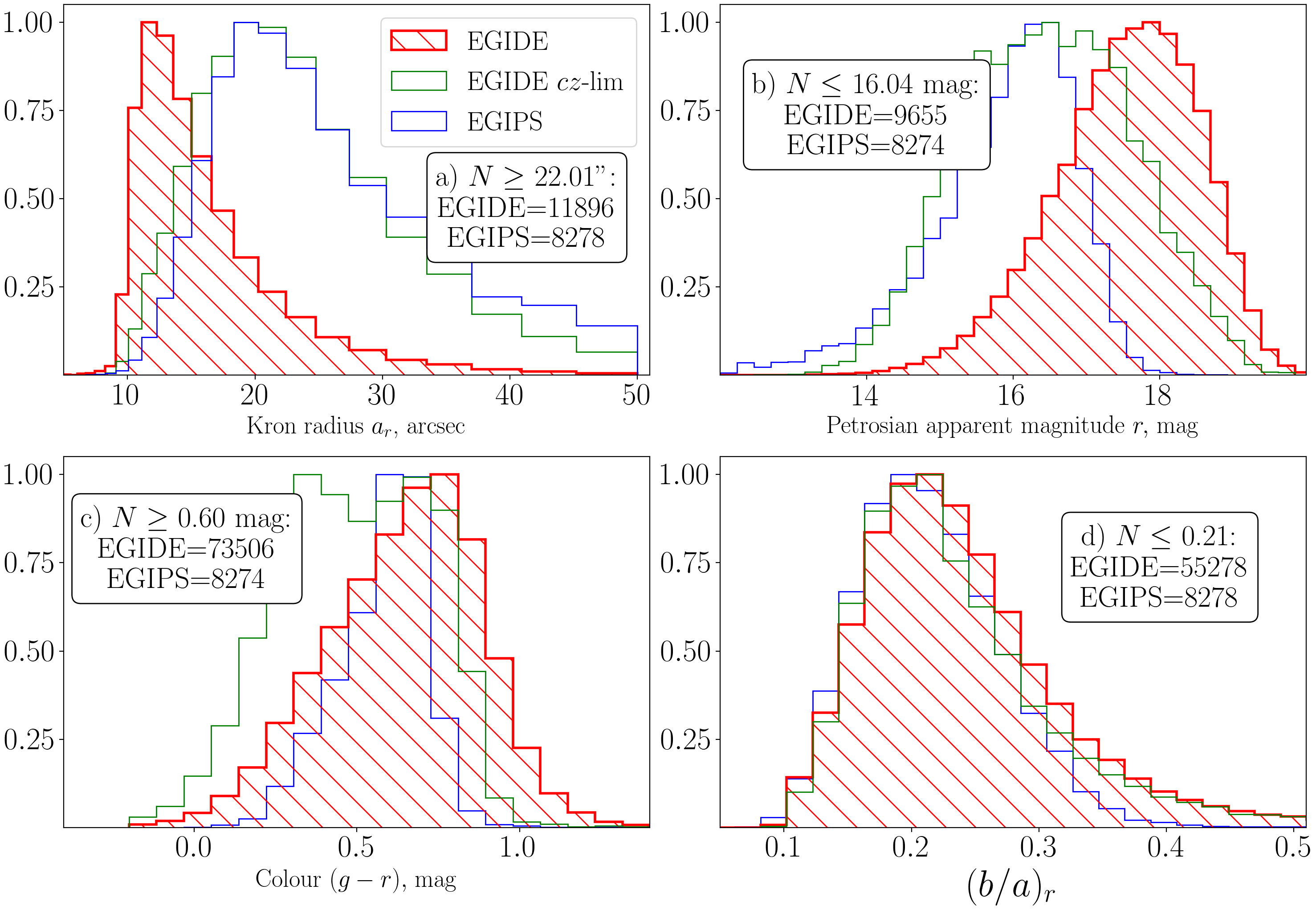}
\caption{
Distributions of the photometric properties of galaxies in EGIDE (red hatched), EGIPS (blue), and the volume-limited subsample of EGIDE (green, for 5000~km/s~$< cz <$~15000~km/s, see Section~\ref{sec:completeness}). Presented values are the Kron radius in the $r$ band (upper left, panel \textbf{a}), apparent Petrosian magnitude in the $r$~band (upper right, panel \textbf{b}), colour $(g-r)$ (lower left, panel \textbf{c}), and $b/a$ ($q$-value) in the $r$ band (lower right, \textbf{d}). In each panel, we also show the number $N$ of galaxies in EGIPS and EGIDE that are greater than (or lower than) the displayed value, which is chosen as the median of the EGIPS distribution.
}
\label{fig:phot_properties}
\end{figure}

A summary of some of the obtained photometry results and a comparison with the analogous EGIPS measurements are presented in Figure~\ref{fig:phot_properties}. With the green line in Figure~\ref{fig:phot_properties} we also show properties for a $cz$-limited subsample of EGIDE galaxies between 5000~km/s and 15000~km/s (see Section~\ref{fig:completeness} for justification). In the top left panel, we show the angular Kron radius in arcseconds. The EGIDE distribution is shifted toward smaller angular sizes than EGIPS, with the average size almost twice as small (13.5~arcsec versus 22~arcsec). The reason is that we detect more distant galaxies in DESI Legacy (see Section~\ref{sec:redshifts}). Despite this, there are still many extended galaxies in our data, including 12,599 objects larger than the EGIPS median of $a>22$ arcsec in the $r$-band. Note that the size parameter presented here differs from figure~6 in \citep{egips}, where the semi-major axis $a_r$ was shown before multiplication by the Kron coefficient.

%Second panel in Figure~\ref{fig:phot_properties} shows distribution for the apparent Petrosian magnitude in $r$. The EGIDE and the EGIPS distributions are drastically different: there is almost no galaxies with $m_r > 16$~mag in new sample, and the distribution has rather abrupt side with sharp $m_r = 19$~mag limit. The second fact is due to parameters used in \textsc{SExtractor} detection stage. The absence of galaxies with bright apparent $m_r$ does not mean that EGIDE contains dimmer objects than EGIPS. Quite contrary, we in fact have more massive and more luminous galaxies in EGIDE, as Figure~\ref{fig:completeness} shows, they are just much more distant (see Figure~\ref{fig:cz}).

The top right panel in Figure~\ref{fig:phot_properties} shows the distribution of the apparent Petrosian magnitude in $r$-band. The EGIDE and EGIPS distributions are drastically different: there are almost no galaxies with $m_r \leq 16$~mag in the new sample, and the distribution has a limit at $m_r \approx 20$~mag. 
% The latter is due to the parameters used in the \textsc{SExtractor} detection stage. 
The absence of galaxies with bright apparent $m_r$ does not mean that EGIDE contains fainter objects than EGIPS. Quite the contrary, we actually have more massive and more luminous galaxies in EGIDE\footnote{Compare also with Figure~\ref{fig:completeness} in the Section~\ref{sec:completeness}.}, they are simply much more distant (see Figure~\ref{fig:cz}).

In the bottom left panel of Figure~\ref{fig:phot_properties}, we show the $(g-r)$ colour distribution from EGIPS and EGIDE in the same photometric system. Here it is evident that the majority of galaxies in our sample exhibit rather red colours. These optical colours are similar to those from EGIPS: it's median value $(g-r)=0.60$~mag divides the EGIDE sample in half, i.e. close to its median. At the same time, for the volume-limited sample from EGIDE (green line) we see two clearly separated peaks, which are absent in the other distributions. Thus it is clear that from the colours alone, red and blue galaxies are well separated only in a small $cz$ range.

%The last panel in Figure~\ref{fig:phot_properties} shows $q$ flattening in $r$-band. We could see, that indeed all galaxies in EGIDE are highly flattened objects with median value $q \approx 0.25$. We want to emphasize again that, as Section~\ref{sec:CandidateSelection} describes, we do not use $q$ information directly during selection. Since many obtained $b/a$ values are small, there are indeed edge-on galaxies.  Comparing to previous EGIPS sample \citep{egips} new distribution is shifted by $0.05-0.1$ in terms of quartiles. Another difference is a larger tail with rounder galaxies in EGIDE, in total .... of them have measured $b/a>0.5$ ([outlier in this band?]). These differences may be caused by different reasons,  which may relate to deeper $\mu$-limit in DESI images compared to Pan-STARRS, or that few candidates have different orientation rather than edge-on. In any case, the amount of thin galaxies is still enormous in EGIDE: we have 8278 objects with $q<0.21$ in EGIPS, while in EGIDE same limitation gives almost three times bigger sample consisted of 22958 objects. For `superthin' galaxies with $q<0.1$ the proportion is opposite: 63 and 32 in EGIPS and EGIDE accordingly. 

In the bottom right panel in Figure~\ref{fig:phot_properties}, we show the $b/a$ distribution in the $r$-band. We can see that indeed all galaxies in EGIDE are highly flattened objects, with a median value of $q \approx 0.25$. We want to emphasize again that, as Section~\ref{sec:CandidateSelection} describes, we do not use $q$ information directly during selection. Since many of the obtained $b/a$ values are small, these are indeed edge-on galaxies. Compared to the previous EGIPS sample \citep{egips}, the new distribution is shifted by $+0.05$-$+0.10$ in terms of quartiles. Another difference is a larger tail of galaxies with rather round isophotes of the ellipses produced by \textsc{SExtractor} in EGIDE. As we discuss in Section~\ref{subsec:psf_smearing}, these differences may be caused by various factors, which may relate to the deeper surface brightness limit in DESI images compared to Pan-STARRS, PSF and bulge influence or to the fact that some candidates have orientations other than edge-on. Nevertheless, we still have many thin galaxies in EGIDE. For instance, in \citep{2024RAA....24g5019H}, the condition $a/b>5$ is used as the boundary for "superthin" galaxies. We prefer to traditionally define superthin galaxies as those with major-to-minor axial ratios $a/b > 10$ \citep{2021ApJ...914..104B,1979BAAS...11..668G,1981ApJ...250...79G}, and refer to objects with $a/b>5$ as "thin." We have 8,278 thin galaxies in EGIPS, while the same criterion in the newly formed EGIDE yields an almost three times larger sample consisting of 22,958 objects. For "superthin" galaxies with $q<0.1$, the proportion is reversed: 63 and 32 in EGIPS and EGIDE, respectively.

For all bands under consideration, we show the measured $q$ and the standard deviations in Table~\ref{tab:qbands}. Besides the already mentioned difference between EGIPS and EGIDE, we also note that the flattening in the $g$-band is on average smaller than in the $r$-band in all cases, and that the dispersion is largely similar between bands. Galaxies with a semi-major axis larger than 20 arcsec show smaller relative thickness probably due to lower impact of the PSF roundening, as Table~\ref{tab:qbands} suggests, and have lower uncertainties in the EGIPS case.

%Last but not least important is a comparison with independent oblateness measurements, available in other surveys. We provide such comparison in Appendix~\ref{ap:flatness}. In total, we compare with ellipticity components from DESI, $R2/R1$ or minor-to-major axis ratio measurements in REGALADE, same ratio from HyperLEDA data and $q$ modeling from \citep{2025arXiv251211035B}. As Figure~\ref{fig:q_appendix} demonstrates, all of these measurements have distributions very similar to what measured here and thus validate small $q$ found here. One visible peculiarity is significantly smaller values obtained from DESI DR, which may probably relate to methodology they used [???]. Another interesting feature of Figure~\ref{fig:q_appendix} comes from the comparison with \citep{2025arXiv251211035B} model for galaxies in $\log M_{\star}$ from 9.0 to 9.5 mass bin. We can see that $b/a$ for subsamples, obtained according to different methods and data employed, is actually fit left and right wings of the reconstructed model in this mass bin relatively well. This may have a meaning that oblate model in \citep{2025arXiv251211035B} fit actual thickness of galaxies we obtain directly for edge-on galaxies very well. On the other hand, this may be a coincidence, which needs further testing. 

Last but not least important is a comparison with independent oblateness $q$ measurements available in other surveys. We provide such a comparison in Appendix~\ref{ap:flatness}. In total, we compare with ellipticity components from Dark Energy Spectroscopic Instrument Data Release 1 (DESI DR1, \citep{2025arXiv250314745D}), $R2/R1$ (minor-to-major axis ratio) measurements in REGALADE \citep{regalade}, the same axis ratio  \textsc{logr25} from HyperLEDA data \citep{leda}, and also statistical $q$ modeling from \citep{2025arXiv251211035B}. As Figure~\ref{fig:q_appendix} demonstrates, all of these measurements have distributions very similar to what we measure here, thus validating the small $q$ values found in this work. One visible peculiarity is the significantly smaller values obtained from DESI DR1, which may likely relate to the methodology they used: in all examined cases probabilistic \textsc{TRACTOR} models results in much bigger $a$ than observed. Another interesting feature of Figure~\ref{fig:q_appendix} comes from the comparison with the \citep{2025arXiv251211035B} model for galaxies in the $\log M_{\star}$ mass bin from 9.0~$M_{\odot}$ to 9.5~$M_{\odot}$. We can see that $b/a$ for subsamples obtained according to different methods and data actually fit the left and right wings of the reconstructed model in this mass bin relatively well. This may imply that the oblate model in \citep{2025arXiv251211035B} fits the actual thickness of galaxies we obtain directly for edge-on galaxies very well. On the other hand, this may be a coincidence and that requires further testing (see also discussion in Section~\ref{sec:mass}).

 \begin{table}
 \centering
 \caption{
Number of galaxies and median $q=b/a$ value for EGIPS and EGIDE samples. Values are presented for each band under consideration for the whole samples, and additionally for the large ($a_{band}>20$~arcsec) galaxies.
 }
 \begin{tabular}{c|cc|cc|cc|cc}
 \hline\hline
 &  \multicolumn{2}{c}{EGIPS} & \multicolumn{2}{c}{EGIDE} &  \multicolumn{2}{c}{EGIPS: $a>20"$} & \multicolumn{2}{c}{EGIDE: $a>20"$}\\
 band & N & $q$ & N & $q$ & N & $q$ & N & $q$\\
 \hline
$g$ & 16550 & 0.205 $\pm$ 0.070 & 148064 & 0.239 $\pm$ 0.079 & 10745 & 0.180 $\pm$ 0.050 & 25351 & 0.203 $\pm$ 0.076\\
$r$ & 16557 & 0.210 $\pm$ 0.057 & 142076 & 0.249 $\pm$ 0.079 & 10226 & 0.186 $\pm$ 0.046 & 17897 & 0.212 $\pm$ 0.080\\
$i$ & 16558 & 0.215 $\pm$ 0.055 & 103210 & 0.241 $\pm$ 0.075 & 9690 & 0.191 $\pm$ 0.046 & 9834 & 0.206 $\pm$ 0.075\\
$z$ & 16556 & 0.214 $\pm$ 0.058 & 145070 & 0.240 $\pm$ 0.073 & 8524 & 0.187 $\pm$ 0.045 & 11625 & 0.199 $\pm$ 0.069\\
 \hline\hline
 \end{tabular}
 \label{tab:qbands}
 \end{table}

\section{Redshift distribution}
\label{sec:redshifts}

%For better understanding of the objects we works with, i.e. derive stellar mass, we need information of the distance to them. To find this we use recently compiled database called REGALADE (the Revised Galaxy List for the Advanced Detector Era, \citealt{REGALADE}). This new all-sky catalog combines data from various surveys (DESI, Pan-STARRS, Cosmicflows, SDSS, etc) for more than 80 million galaxies out to the distance 2000~Mpc ($z\leq 0.37$, $cz \leq 111000$~km/s). 

For a better understanding of the objects we work with, for example, to derive stellar masses, we need information about their distances. To obtain this, we use a recently compiled database called REGALADE (the Revised Galaxy List for the Advanced Detector Era, \citep{regalade}). This new all-sky catalog combines data from various surveys (DESI, Pan-STARRS, Cosmicflows, SDSS, etc.) for more than 80 million galaxies out to a distance of 2000 Mpc ($z\leq 0.37$ or $cz \leq 111000$ km/s).

%For 98\% of galaxies in EGIDE (146203 objects) we find $z$ information within 5~arcsec distance of REGALADE sources, but for majority of our data (137678 galaxies) the euclidian distance between coordinates is less than 1~arcsec. Since in \citep{REGALADE} authors mix all spectroscopic, photometric, and redshift-independent distances in the same table, this can potentially bias the results. To ensure the reliability we additionally collect $z$ measurements from three sources. First is the HyperLEDA database \citep{leda}, where we take data in the CMB frame of reference. The second source is the Dark Energy Spectroscopic Instrument survey Data Release 1 (DESI DR1) \citep{2025arXiv250314745D}, which contains spectroscopic redshift measurements for $z<4$. However, since DESI DR1 is included in REGALADE, we use the specific Stellar Mass and Emission Line Catalog (EMLines) Value Added Catalogue (VAC), which contains emission line information for galaxies with reliable redshift measurements. In both cases we match the galaxies within 1~arcsec distance. The last one we consult with is RCSEDv2\footnote{\url{https://rcsed2.voxastro.org/}} \citep{2017ApJS..228...14C}. We find good agreement between redshift $z$ from different databases, see the comparison in Appendix~\ref{ap:catalog_comparison}. Throughout the paper we check, that all dependent results stay the same when using different $z$.

For 98\% of the galaxies in EGIDE (145,876 objects), we find $z$ information within 5~arcsec of REGALADE sources, and for the majority of our data (137,678 galaxies or 92\%), the Euclidean distance between coordinates is less than 1~arcsec. Since the authors of \citep{regalade} mix all spectroscopic, photometric, and redshift-independent distances in the same table, this could potentially bias the results. To ensure reliability, we additionally collect $z$ measurements from three other sources. The first is the HyperLEDA database \citep{leda}, from which we take data in the CMB frame of reference. The second source is the DESI DR1 \citep{2025arXiv250314745D}, which contains spectroscopic redshift measurements for $z<4$. However, since DESI DR1 is included in REGALADE, we use the specific Stellar Mass and Emission Line Catalog Value Added Catalogue (EMLines VAC), which contains information for galaxies with reliable redshift measurements. In both cases, we match galaxies within 1~arcsec distance. The third source we consult is second version of Reference Catalog of Spectral Energy Distributions of galaxies (RCSEDv2\footnote{\url{https://rcsed2.voxastro.org/}}, \citep{2017ApJS..228...14C}). We find good agreement between the redshifts $z$ from the different databases; see the comparison in Appendix~\ref{ap:catalog_comparison}. Throughout the paper, we verify that all dependent results remain the same when using different $z$ sources.

%The resulted redshift distribution presented in Figure~\ref{fig:cz}. It is clear, that effective depth of the EGIDE catalogue is significantly increased comparing with EGIPS: the median velocity is three times bigger and equal $\sim$33750~km/s, which corresponds to luminosity distance about 520~Mpc for given cosmology. Note also that $cz$ distribution of EGIDE galaxies have a large tail, which lasts until REGALADE boundary $D=2000$~Mpc. We could also see slightly less large galaxies in the $cz<5000$~km/s covered local volume, which is due to selection criteria (cf. with Table~\ref{tab:inters}). These galaxies are too big to be detected by neural net, and were manually added in EGIPS. Thus we want to emphasize, that EGIPS and RFGC are still can be used as a good supplement to complete EGIDE in $cz<5000$~km/s velocity range.

\begin{figure}
\centering
\includegraphics[width=0.88\textwidth]{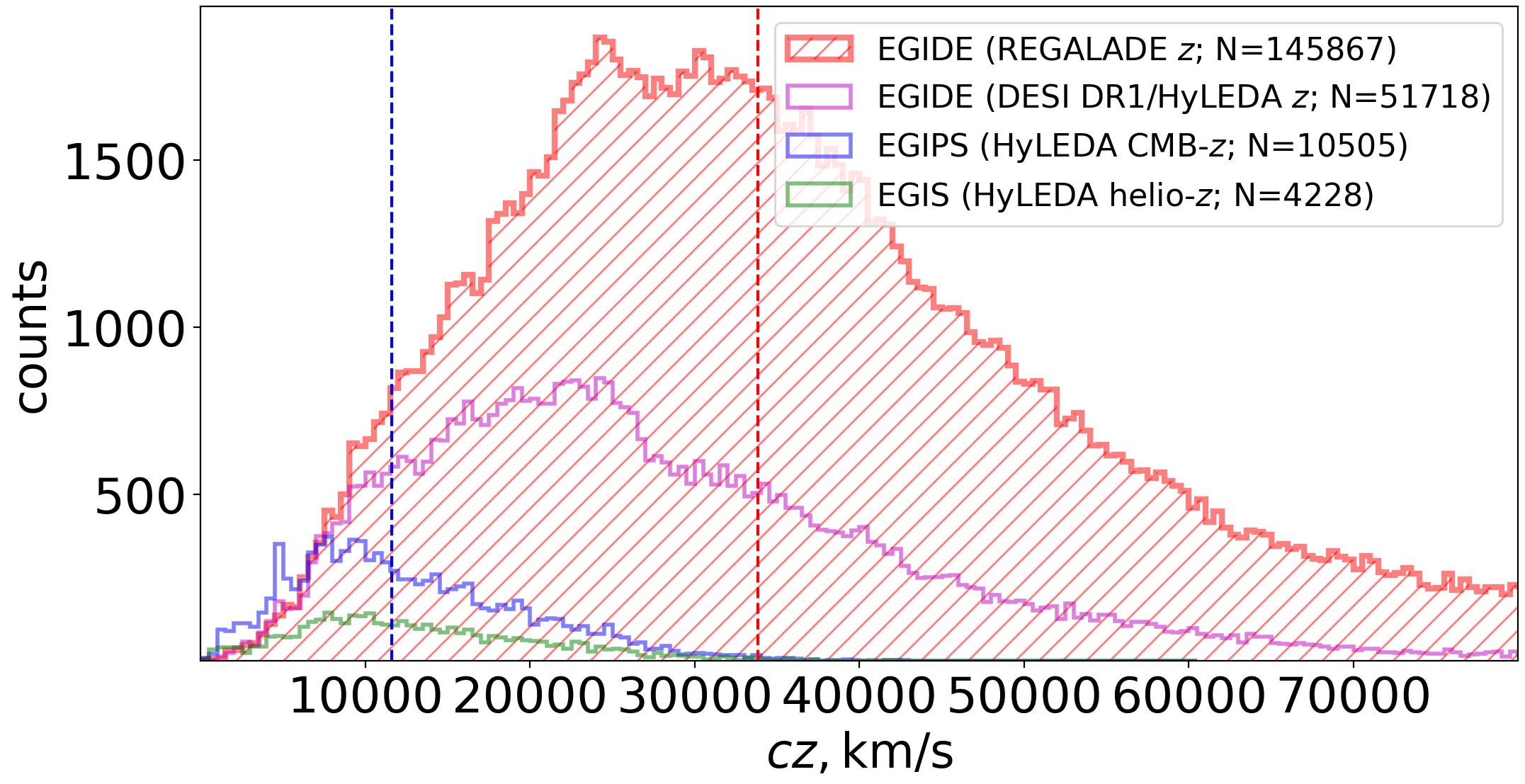}
\caption{
The redshift distribution of edge-on galaxies for the EGIS (green), EGIPS (blue), and EGIDE (red hatched for the whole dataset and magenta for the smaller subsample with information from other surveys) samples. The median $cz$ values for EGIPS and EGIDE are indicated by the vertical dashed lines. Each bin size is 500~km/s. For illustration purposes, we limit the figure to an 80000~km/s range, which contains 94\% of the EGIDE sample.
}
\label{fig:cz}
\end{figure}

The resulting redshift distribution is presented in Figure~\ref{fig:cz}. It is clear that the effective depth of the EGIDE catalog is significantly increased compared to EGIPS: the median velocity is three times larger, equal to $\sim$33,750 km/s, which corresponds to a luminosity distance of about 520 Mpc for the adopted cosmology. Note also that the $cz$ distribution of EGIDE galaxies has a long tail that extends to the REGALADE boundary of $D=2000$ Mpc. Compared to EGIPS, we also see slightly fewer large galaxies in the $cz<5000$ km/s covered local volume, which is due to the selection criteria (see also Table~\ref{tab:inters}). These galaxies are too large to be detected by the neural network and were manually added in EGIPS. Thus, we want to emphasize that EGIPS and RFGC can be used as good supplements to complete EGIDE in the $cz<5000$ km/s velocity range.

\begin{table*}
    \caption{The size of the intersection of the EGIDE sample with the surveys we are comparing with. The first column shows the acronym for the survey; the next columns show the reference, the intersection size, and the median Kron radius $a_r$ of the galaxies in the intersection. The lower part of the table, below the horizontal line, is dedicated to the surveys included in the RCSEDv2 project.}
    \begin{tabular}{llcc|llcc}
    \hline
         Survey & Ref. & Size, $N$   & $a_r$ [arcsec] & Survey & Ref. & Size, $N$ & $a_r$ [arcsec] \\
         \hline
         RFGC & \citep{1999BSAO...47....5K} & 135 & 32.1 & EGIS & \citep{bizyaev_etal2014} & 2726 & 22.9 \\  
         EGIPS & \citep{egips} & 8294 & 22.6 & HyperLEDA & \citep{hyperleda} & 60,838 & 16.0 \\ 
		 DESI DR1 & \citep{hyperleda} & 30,833 & 13.6 & REGALADE & \citep{sdss2011} & 145,876  & 13.6 \\	
		 \, & \, & \, & \, & \, & \, & \, & \, \\
		 RCSEDv2 & \citep{hyperleda} & 30,163 & 16.0 & \, & \, & \, & \, \\ 
         \hline
SDSS & \citep{2020ApJS..249....3A} & 17,027 & 16.0 & FAST & \citep{fast} & 406 & 26.3 \\
2dFGRS & \citep{2001MNRAS.328.1039C} & 3869 & 14.3 & 2dFLenS & \citep{2016MNRAS.462.4240B} & 381 & 13.8 \\
6dFGS & \citep{2004MNRAS.355..747J} & 3067 & 22.1 & Hectospec & \citep{2005PASP..117.1411F} & 279 & 13.3 \\
eBOSS & \citep{2020ApJS..249....3A} & 2392 & 14.8 & CfA & \citep{1992BICDS..41...31H} & 125 & 19.2 \\
LAMOST & \citep{2019ApJS..240....6Y} & 1621 & 16.8 & UZC & \citep{1999PASP..111..438F} & 55 & 49.3 \\
         \hline
         \end{tabular}
    \label{tab:inters}
\end{table*}

\section{Completeness}
\label{sec:completeness}

%To find out the completeness of the constructed EGIDE sample we use several approaches. The first one is to run the classical $V/V_m$ test, also known as the luminosity-volume test, which shows whether a sample of objects is uniformly distributed in space \citep{1979ApJ...231..680T} according to expectations from their fluxes. For 40\% of most luminous galaxies in EGIDE (58302 objects) we get $V/V_m = 0.4636 \pm 0.0012$ in $g$-band, which means that it is mostly complete for these galaxies. Inclusion of galaxies with less flux results in more incomplete subsample ($V/V_m = 0.4398 \pm 0.0010$ for 50\% quartile). From the comparison with EGIPS we also know that size-selected sample also incomplete: $V/V_m = 0.5781 \pm 0.0016$ for galaxies with Kron radius bigger than 15~arcsec.

To assess the completeness of the constructed EGIDE sample, we use several approaches. The first is to perform the classical $V/V_m$ test, also known as the luminosity-volume test, which indicates whether a sample of objects is uniformly distributed in space \citep{1979ApJ...231..680T} according to expectations from their fluxes. For the 40\% most luminous galaxies in EGIDE (58,302 objects), we obtain $V/V_m = 0.4636 \pm 0.0012$ in the $g$-band, which means that the sample is mostly complete for these galaxies. Inclusion of galaxies with lower flux results in a more incomplete subsample ($V/V_m = 0.4398 \pm 0.0010$ for the 50\% quantile). From the comparison with EGIPS in Section~\ref{sec:Photometry}, we also know that the size-selected sample is also incomplete: $V/V_m = 0.5781 \pm 0.0016$ for galaxies with a Kron $r$-band radius larger than 15~arcsec.

%Another two tests are shown in Figure~\ref{fig:logNlogA}. Here we plot $\log N$--$\log a_r$ cumulative distribution (completely similar to figure~3 in \citep{egips}), which should have a slope equals -3 for uniform distribution of galaxies. We can see that as indeed average slope is close to -3, but there is a knee in the distribution around $a_r \approx 17$~arcsec. Before the knee the slope is closer to -4, and after it is -2.2. For the population of the blue $(g-r)<0.95$ and red $(g-r)>0.95$ galaxies the situation  is the same. On the right side of Figure~\ref{fig:logNlogA} we plot $\log N$--apparent Petrosian magnitude cumulative distribution for all $griz$ bands. For uniform distribution the slope for such distribution should be 0.6. For a part between 12~mag and 17~mag, where increase is steady, we see slopes slightly bigger, around 0.65-0.7, which may be a sign of excess of faint galaxies in EGIDE. As in Figure~\ref{fig:phot_properties}, we see the asymptote magnitude to be $m=19$~mag.

\begin{figure}
\centering
\includegraphics[width=0.49\textwidth]{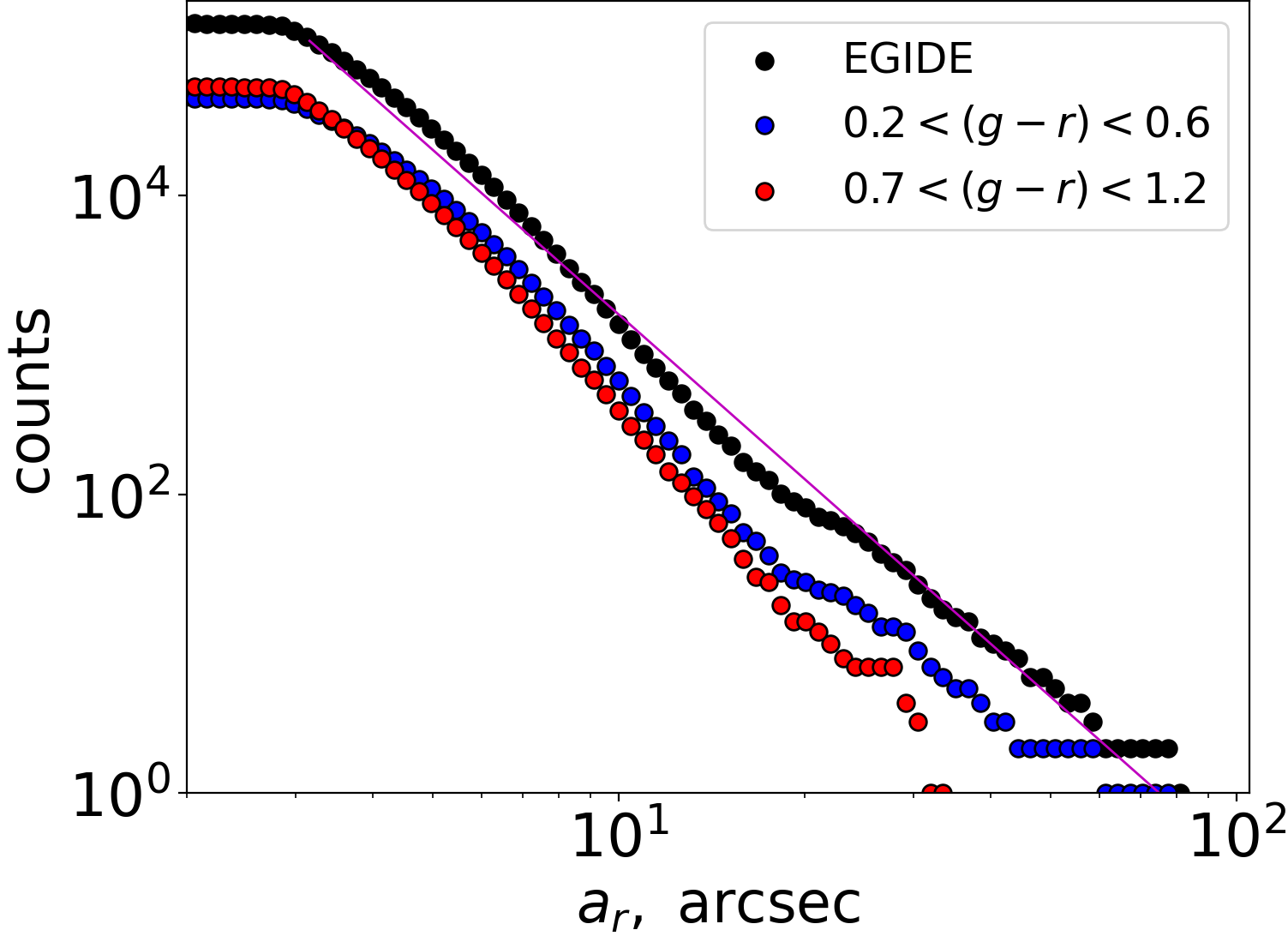}
\includegraphics[width=0.49\textwidth]{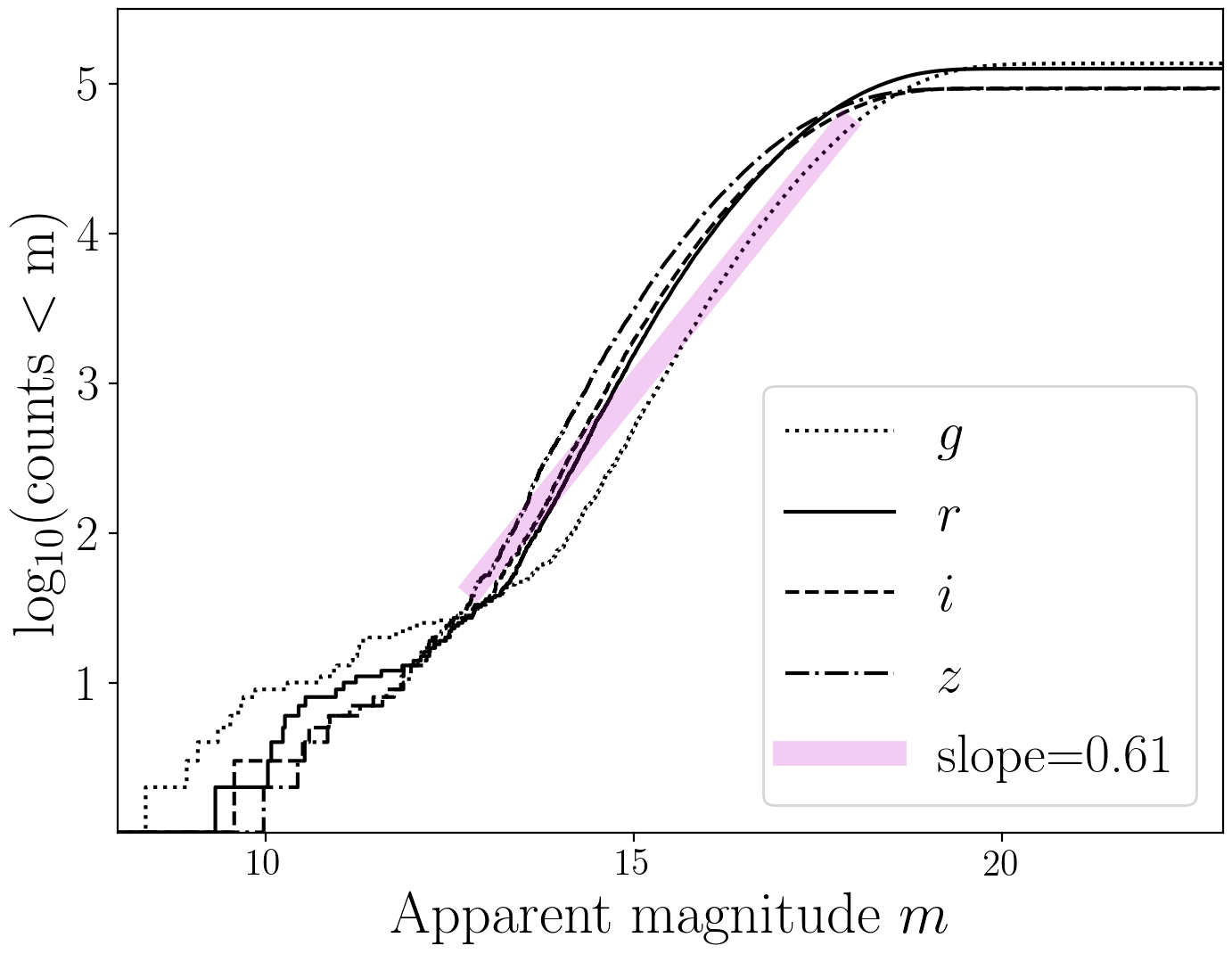}
\caption{
\textbf{Left:} Completeness function in the form of $\log N$ versus $\log a_r$ for EGIDE galaxies. Different colours show the whole sample (black), galaxies with colour $0.7<(g-r)<1.2$ (red), and galaxies with colour $0.2<(g-r)<0.6$ (blue). The line marks the linear fit to the whole distribution. \textbf{Right:} Cumulative distribution of apparent magnitude. Individual black lines correspond to the $g$, $r$, $i$, and $z$ bands, and the line shows the linear fit in the given magnitude range for the $r$-band.
}
\label{fig:logNlogA}
\end{figure}

Two additional tests are implemented and shown in Figure~\ref{fig:logNlogA}. Here we plot the $\log N$-$\log a_r$ cumulative distribution (completely analogous to figure~3 in \citep{egips}), which should have a slope of $-3$ for a uniform distribution of galaxies if we neglect cosmological effects. We can see that indeed the average slope is close to $-3$, but there is a knee in the distribution around $a_r \approx 17$~arcsec. Before the knee, the slope is closer to $-4$, and after it, it is $-2.2$. For the populations of blue $0.2 < (g-r)<0.6$~mag and red $0.7<(g-r)<1.2$~mag galaxies (see motivation in Section~\ref{sec:GalaxyCMD}), the situation is the same. On the right side of Figure~\ref{fig:logNlogA}, we plot the $\log N$-apparent Petrosian magnitude cumulative distribution for all $griz$ bands. For a uniform distribution, the slope of such a distribution should be $0.6$. For the part between 12~mag and 17~mag, where the increase is steady, we see slopes slightly larger, around $0.65$-$0.7$, which may indicate an excess of faint galaxies in EGIDE. As in Figure~\ref{fig:phot_properties}, we see the limiting magnitude to be $m = 19$~mag.

%As a final test we plot stellar mass versus redshift in Figure~\ref{fig:completeness}. Total stellar mass   $M_{\star}$ estimation were maid using \citep{2025A&A...704A.232E} calibrations using DESI $g$ and $r$ magnitudes. These estimations are well fitted with SED-estimated masses with CIGALE (Boquien et al. 2019; Yang et al. 2020) from DESI DR1 \citep{2025arXiv250314745D} and with $M_{\star}$ estimation in REGALADE, see Appendix~\ref{ap:catalog_comparison}. The distribution demonstrate traditional shape with visible selection bias, where only more massive (i.e. more luminous) galaxies are visible on larger distances. As comparing with EGIPS sample, we can clearly see that EGIDE lacks some galaxies in the nearby volume, but have significantly improved the depth in terms of $cz$ and also have more massive galaxies (median value $\log M_{\star} = 10.68$ in EGIDE versus $\log M_{\star} = 10.34$ in EGIPS). We mark the observational limitations for both surveys schematically with lines in Figure~\ref{fig:completeness}. It is reasonable to analyze two distinct subsamples: 1) with galaxies $5000<cz<15000$, which includes both faint and luminous objects and correctly represents flux dynamic range; 2) with galaxies in $\log M_{\star} = 10.7-11.4$ mass bin, which correctly represents the distances range. These subsamples can be seen as volume- or mass-complete, thus we test all aforementioned results with both of them in case to see if there is any difference in the conclusions.

\begin{figure}
\centering
\includegraphics[width=0.80\textwidth]{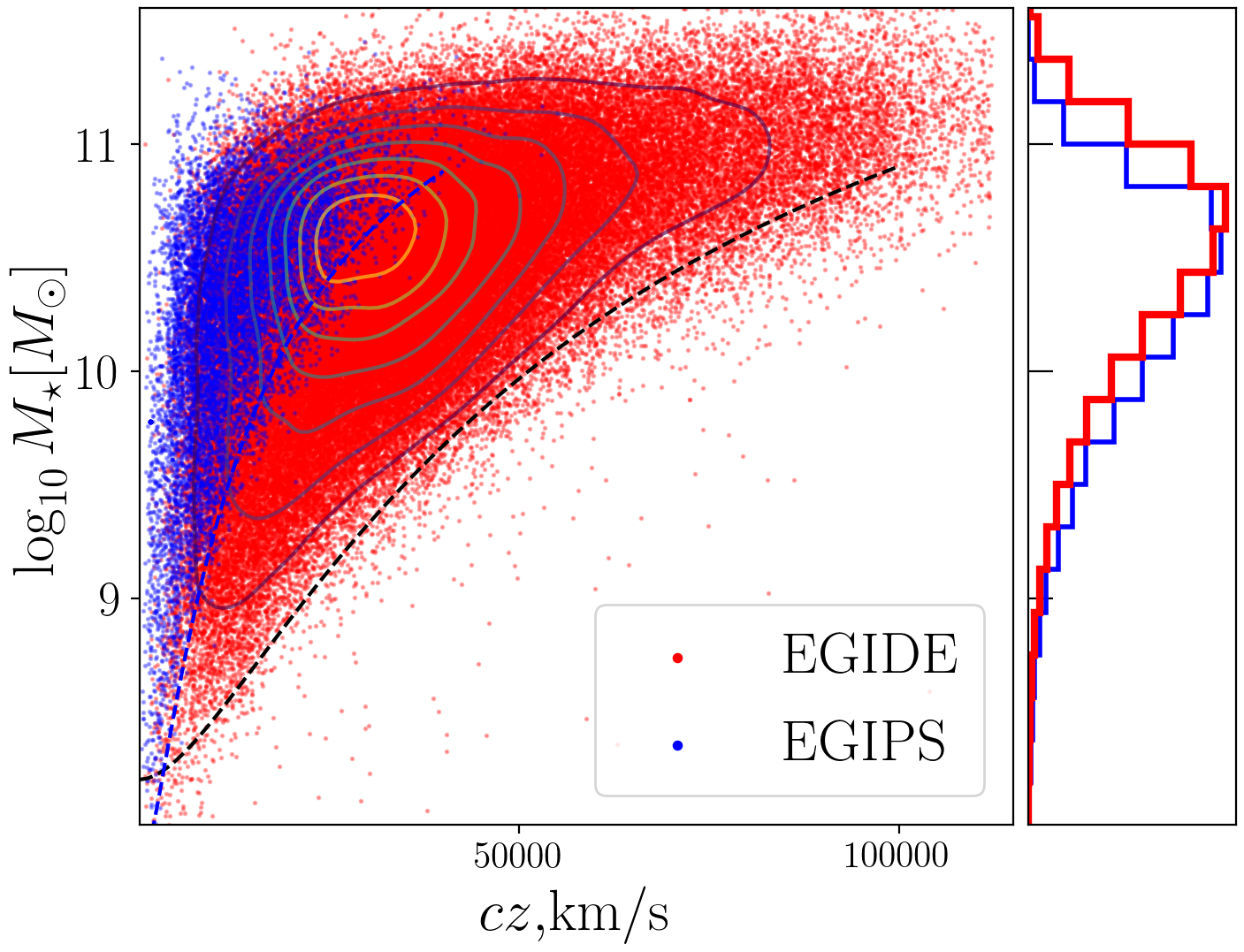}
\includegraphics[width=0.80\textwidth]{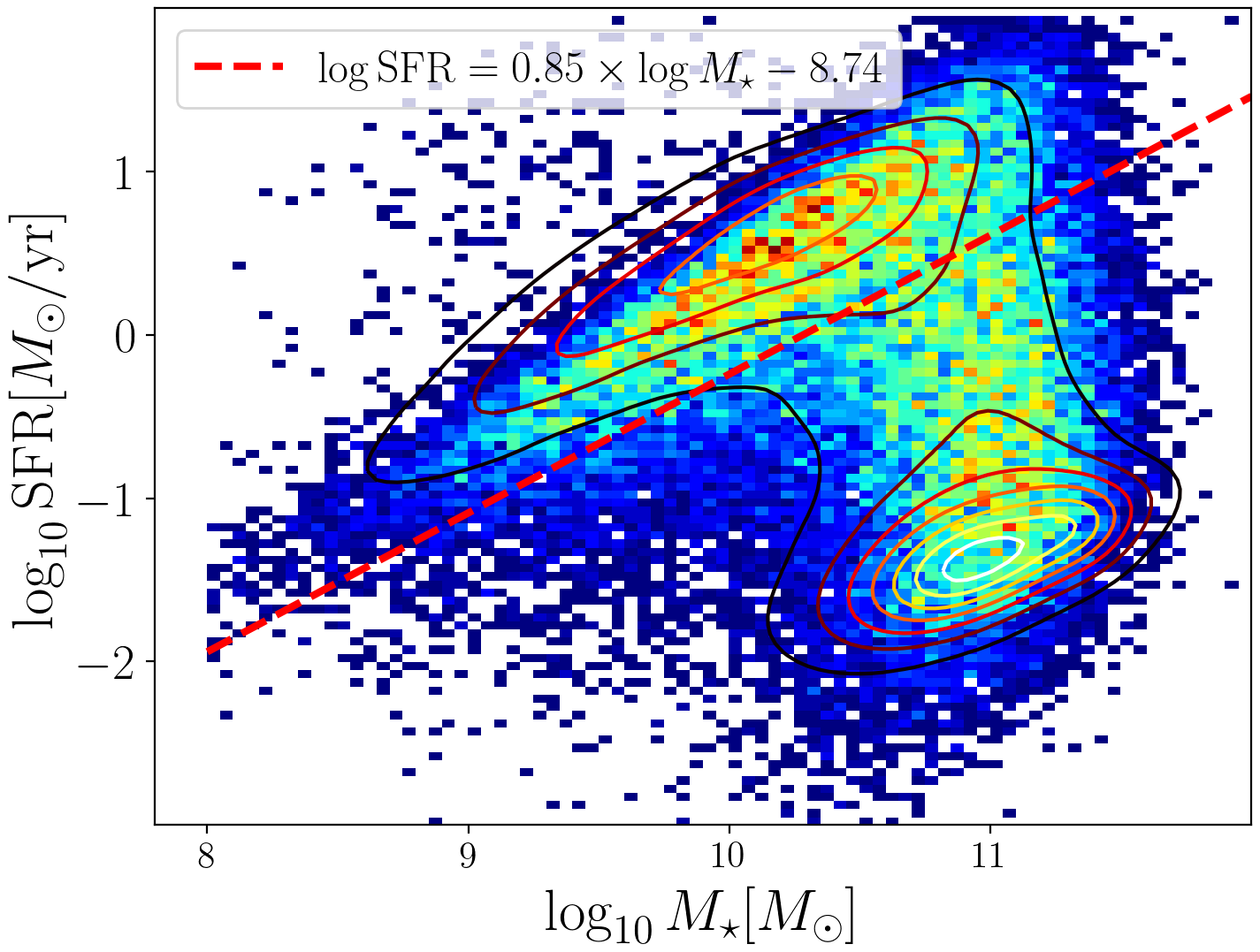}
\caption{
\textbf{Upper:} Total stellar mass versus redshift for EGIPS (blue) and EGIDE (red). Isocontours show the density distribution for the EGIDE sample. The blue and black dashed lines schematically show the observational limitations for each mass. In the right subplot, we show histograms of $M_{\star}/M_{\odot}$ for both surveys. \textbf{Lower:} Total SFR versus stellar mass relation for galaxies in the intersection of EGIDE with DESI DR1. Isocontours show the general DESI EDR sample (see text for details). The red dashed line separates SFMS galaxies from early-type galaxies.
% , and the dotted line is from \citep{2015ApJ...801L..29R}.
}
\label{fig:completeness}
\end{figure}

As a final test, we plot stellar mass versus redshift in Figure~\ref{fig:completeness}. Total stellar mass $M_{\star}$ estimates were made using the calibrations from \citep{2025A&A...704A.232E} with DESI $g$ and $r$ absolute stellar magnitudes. These estimates are in good agreement with SED-estimated masses from CIGALE \citep{2019A&A...622A.103B} from DESI DR1 \citep{2025arXiv250314745D} and with $M_{\star}$ estimates in REGALADE, see details in Appendix~\ref{ap:catalog_comparison}. The distribution shows a traditional shape with a visible selection bias, where only more massive (i.e., more luminous) galaxies are visible at larger distances. Comparing with the EGIPS sample, we can clearly see that EGIDE lacks some galaxies in the nearby volume but has significantly improved depth in terms of $cz$ and also contains more massive galaxies (median value $\log M_{\star}/M_{\odot} = 10.56$ in EGIDE versus $\log M_{\star}/M_{\odot} = 10.34$ in EGIPS). We schematically mark the observational limitations for both surveys with lines in Figure~\ref{fig:completeness}. It is reasonable to analyze two distinct subsamples: (1) galaxies with 5000~km/s~$< cz <$~15000~km/s, which includes both faint and luminous objects and correctly represents the flux dynamic range; and (2) galaxies in the $\log M_{\star}/M_{\odot} = 10.7\textrm{--}11.4$ mass bin, which correctly represents the distance range. These subsamples can be considered volume- or mass-complete, and we test all the aforementioned results with both to see if there is any difference in the conclusions.

\section{Discussion}
\label{sec:results}

In this Section, we present some derived results about the EGIDE sample. Specifically, we discuss the galaxy colour–magnitude diagram, the colour–flattening relation, and the disk thickness dependence on stellar mass.

\subsection{Galaxy colour-magnitude diagram}
\label{sec:GalaxyCMD}

\begin{figure}
\centering
\includegraphics[width=0.95\textwidth]{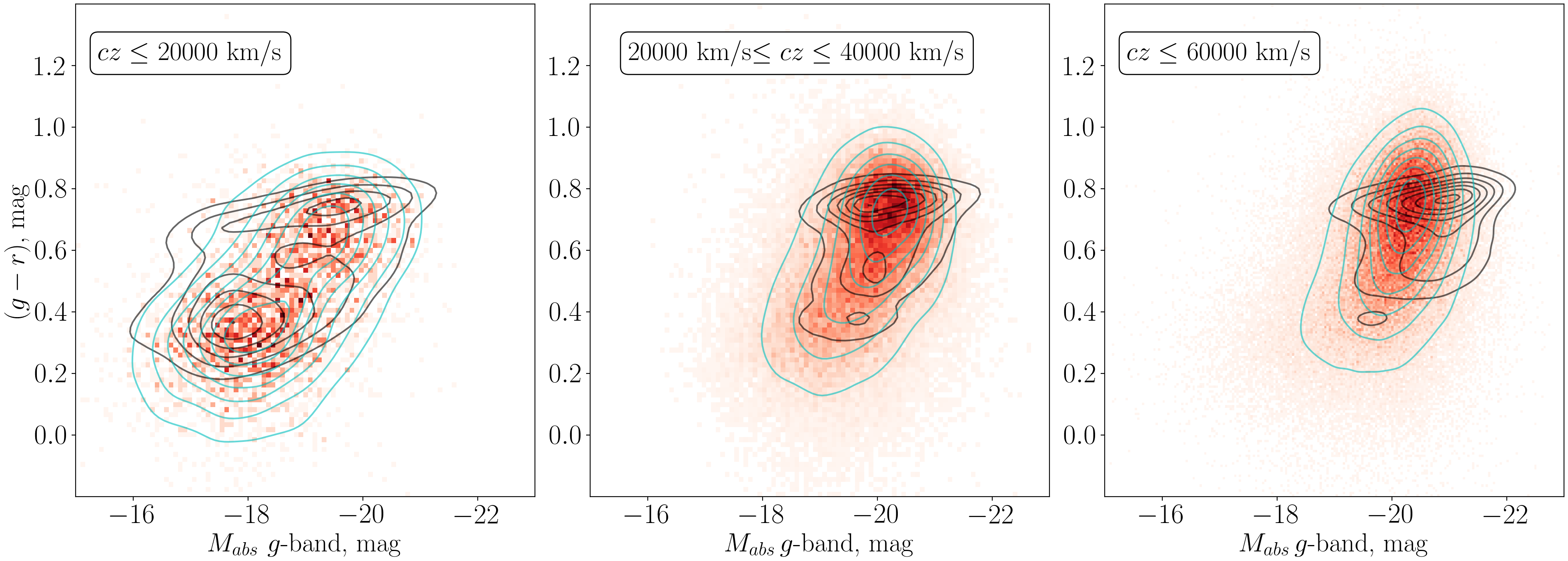}
\caption{
The galaxy colour $(g-r)$ versus absolute magnitude in the $g$-band diagram. The density distribution of the EGIDE galaxies is shown in reddish colours and cyan contours. We show subsamples in three $cz$-limited bins, marked at the top of each panel. The black solid isocontours illustrate the distribution of a general sample of galaxies from the SDSS DR7 survey, converted into the same photometric system. 
}
\label{fig:galaxyCMD}
\end{figure}

In Figure~\ref{fig:galaxyCMD}, we present the $(g-r)$ colour versus absolute magnitude $M_g$ diagram (CMD) for galaxies in EGIDE. We use the $r$-band instead of $i$ because the latter is available for only 70\% of the sample. For easier comparison with previous results, we show three separate redshift bins: the nearby bin $cz<10000$~km/s, the intermediate bin $10000 < cz < 30000$~km/s, and the bin for 85\% of all galaxies in the sample with $cz<60000$~km/s. For comparison, we also overlay SDSS DR12 \citep{2015ApJS..219...12A} data as contours for the same $cz$ bins, corrected to fit DESI magnitudes (see Section~\ref{sec:Photometry}). In all three subpanels of the SDSS data, we can clearly see the red cloud, preferentially populated by early-type disk galaxies, which are also brighter, and the dimmer blue cloud, formed by spiral galaxies \citep{2003ApJ...594..186B}. For $cz<10000$~km/s, we see good agreement between EGIDE and SDSS DR12 data, with edge-on galaxies equally populating the red and blue clouds. For galaxies with larger redshifts (second and third panels), as well as for the entire EGIDE sample, we see, first, that galaxies tend primarily to occupy the red sequence with no clear presence of the so-called green valley, and second, that the colour $(g-r)$ is shifted to redder and, to a lesser extent, to bluer values compared to SDSS photometry. This fact---that the SDSS contours are shifted toward bluer galaxies---may be explained by higher internal extinction in edge-on galaxies relative to galaxies with arbitrary inclinations, as well as by the deeper limit of the DESI Legacy observations. Note, however, that this effect should be more important for star-forming galaxies with high dust influence. The high prominence of red-sequence galaxies in Figure~\ref{fig:galaxyCMD} may be due to a significant fraction of lenticular galaxies in the constructed EGIDE sample. Unlike late-type galaxies, edge-on lenticular galaxies may be harder to detect due to the absence of spiral arms or prominent dust lanes. Potentially, there can also be some elongated elliptical galaxies of type E7 with $q\geq0.3$, but this type is rare, and the detection of S0 galaxies is much more probable.

Optical colours alone are not optimal to separate clearly red and blue clouds in CMD, with $NUV -r$ often implemented as better choice \citep{2014SerAJ.189....1S}. However, when available, it is even better to use for this purpose physically motivated star formation rate (SFR)-$M_{\star}$ plane.
% A similar situation can be found in the star formation rate (SFR)-$M_{\star}$ plane. 
In Figure~\ref{fig:completeness}, we show these values for 30,986 EGIDE galaxies with available SED models with estimated SFR from DESI DR1. Overlaid on the density distribution in Figure~\ref{fig:completeness} are contours for the general sample from the DESI Early Data Release\footnote{\url{https://data.desi.lbl.gov/doc/releases/edr/}} (EDR). We choose to use EDR over DR1 data for practical reasons because it produces essentially the same contours but faster due to being 16 times smaller. To plot the contours, we select only morphological EDR data with exponential, Sérsic, and de Vaucouleurs models (\textsc{EXP}, \textsc{SER}, and \textsc{DEV} codes) and with $z<0.37$ to match the EGIDE distance volume. We can clearly see the star-forming main sequence (SFMS), which is essentially the equivalent of the blue cloud in the galactic CMD. A slightly less populated "red and dead" area with more massive early-type galaxies (ETGs) is also evident in Figure~\ref{fig:completeness}. We see that the EGIDE galaxies coincide well with the general population contours from DESI EDR. For later use, we mark the separation line between ETG and SFMS galaxies as $\log \text{SFR} = 0.85\log M_{\star} - 8.74$, which is similar to H$\alpha$-related local SFMS in \citep{2019MNRAS.482.1557S}. 
% We also show in Figure~\ref{fig:completeness} a ridge line approximation for SDSS DR7 data from the \citep{2015ApJ...801L..29R} work, which is roughly consistent with the presented values.

\subsection{Colour-flatness relation}
\label{sec:color_flatness}

\begin{figure}
\centering
\includegraphics[width=0.95\textwidth]{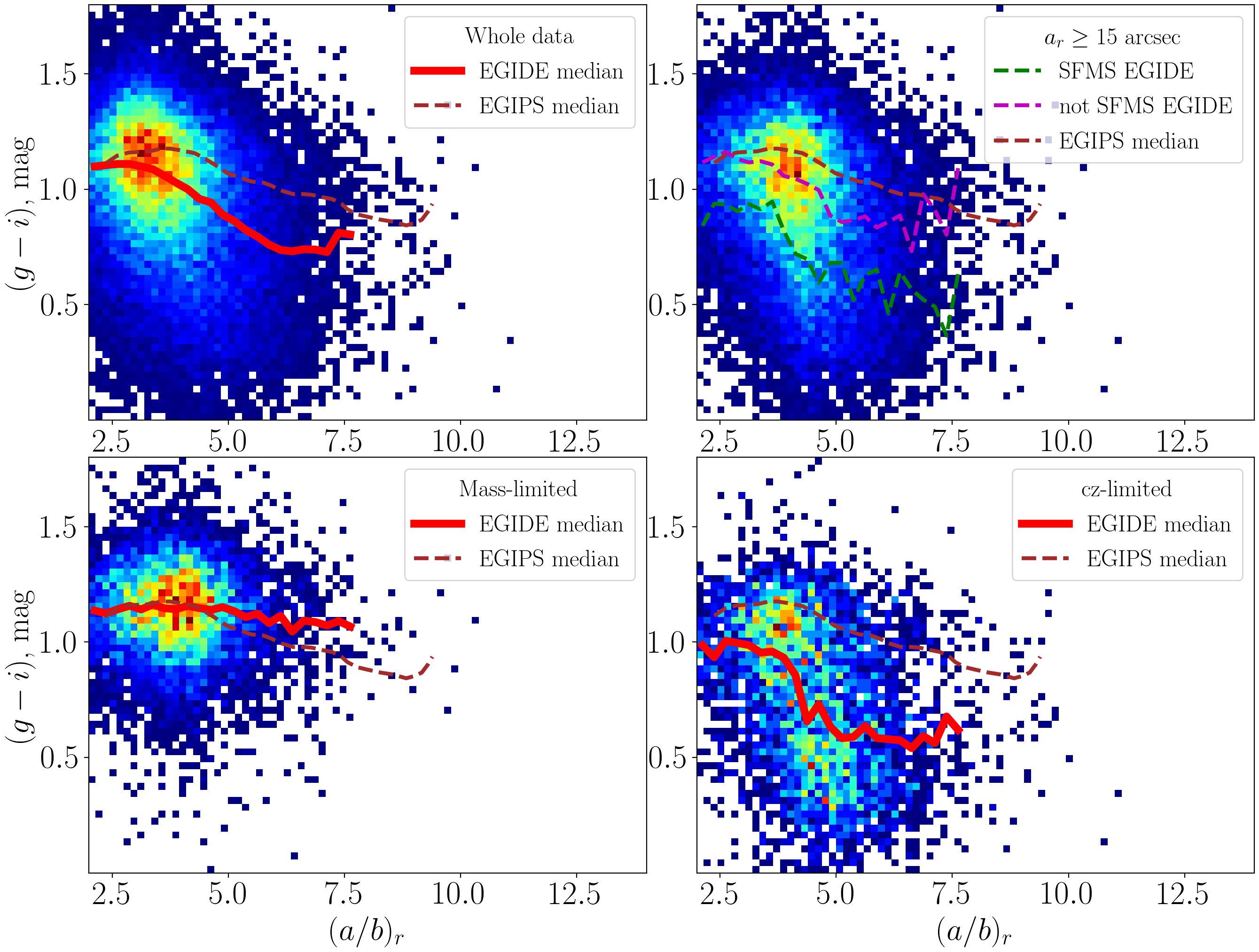}
\caption{
The colour $(g-i)$ versus inverse thickness $(a/b)$ in the $r$-band for the EGIDE galaxies. In each panel, for comparison, the dashed brown line shows the running median over vertical bins for the EGIPS galaxies. Colours show the density distribution for the whole sample (upper left), for galaxies with a Kron radius larger than 15~arcsec (upper right), and for the mass-limited (lower left) and volume-limited (lower right) subsamples. The red line shows the running median for the corresponding subsample. Additionally, in the upper right subplot, we show the running median for SFMS and non-SFMS galaxies in EGIDE (see the separation line in Figure~\ref{fig:completeness}, right).
}
\label{fig:color_axisRatio}
\end{figure}

%In Figure~\ref{fig:color_axisRatio} we show $(g-i)$ color versus $q^{-1}$ for the EGIDE galaxies. Despite the ten times bigger size and other principal differences between samples, we observe exactly the same picture as it was found for EGIPS \citep{egips}. The obvious cluster/overdensity of galaxies represent thick ($a/b\sim3-5$) and red ($(g-i)\sim1.2-1.3$~mag) objects. The thinner galaxies with $a/b>5$ demonstrate on average bluer color by $0.2-0.4$~mag. Overall the distribution is visually shifted towards  smaller $a/b$ than analogous figure~10 for EGIPS \citep{egips}, with heavy visible "tail" around $a/b\sim5$. However, despite this fact the running medians in both samples are very close to each other, meaning the same trend is confirmed. Result in Figure~\ref{fig:color_axisRatio} stands if we use subsamples with only large galaxies ($a_r>15$~arcsec), [...smth else?]. 

In Figure~\ref{fig:color_axisRatio}, we show $(g-i)$ colour versus $a/b$ for the EGIDE galaxies. Despite the ten times larger sample size and other principal differences between the samples, we observe similar picture to what was found for EGIPS \citep{egips}. The obvious cluster or overdensity of galaxies represents thick, $a/b\sim3$-$5$, and red, $(g-i)\sim1.2$-$1.3$~mag, objects. The thinner galaxies with $a/b>5$ are on average bluer by $0.2$-$0.4$~mag. Overall, the distribution is visually shifted toward smaller $a/b$ compared to the analogous figure~10 for EGIPS \citep{egips}, with a heavily visible "tail" around $a/b\sim5$. The running medians in both samples are very close to each other, confirming the same trend. The result in Figure~\ref{fig:color_axisRatio} holds if we use subsamples with only large galaxies ($a_r>15$ arcsec), as we show in the upper right panel.

%Similar conclusions were already made previously for smaller samples of galaxies too in $g-r$ color (e.g. figure~4 in \citep{2009PASP..121.1297K} or figure~7 in \citep{bizyaev_etal2014}). The presence of internal extinction can't an explanation, because existing models \citep{2007ApJ...659.1159S} predicts that $(g-i)$ increase or galaxy become redder, when $a/b$ increase. The size of the effect is similar to what we observe, but for reversed trend. Thus, there must be another explanation, which may the effect if the bulge. Indeed, as Figure~\ref{fig:msfr} (and SDSS?) clearly shows that we have many galaxies with significant spherical component. While the matter if the disc thickness depends on morphology is a debated question \citep{1972MmRAS..75...85H,1998MNRAS.299..595D,2015MNRAS.451.2376M}, such dependency can potentially have a similar effect on the color (see figure~8 for EGIS in \citep{bizyaev_etal2014}). Note, that in SAMI survey \citep{2021ApJ...906..100B} authors measure that $g-i$ color of bulges is just slightly redder by $0.12\pm 0.02$~mag than color of discs, which they attribute to higher metallicities rather than older age of stellar population. Such difference is rather small to be important for measured effect if $B/T\sim 0.5$ or less.

Similar conclusions were already made previously for smaller samples of galaxies in $(g-r)$ colour as well (e.g., figure~4 in \citep{2009PASP..121.1297K} or figure~7 in \citep{bizyaev_etal2014}). The internal extinction variation due to a small inclination difference from the edge-on orientation cannot be the explanation, because when $a/b$ increases we expect to find galaxies closer to edge-on, and thus they must be redder. Existing statistical models from \citep{2007ApJ...659.1159S} that connect extinction and inclination predict an effect that is similar in magnitude to what we observe, but reversed.
% The presence of internal extinction cannot be the explanation because existing models \citep{2007ApJ...659.1159S} predict that $(g-i)$ increases, i.e., galaxies become redder, when $a/b$ increases. The magnitude of the effect is similar to what we observe, but for the reversed trend.
Thus, there must be another explanation, which may potentially be the effect of the bulge, because light from a bulge would geometrically lead to a greater axis ratio (see Section~\ref{subsec:psf_smearing}). Indeed, as Figure~\ref{fig:completeness} and Figure~\ref{fig:galaxyCMD} clearly show, we have many galaxies with a significant spherical component. While whether disk thickness depends on morphology is a debated question \citep{1972MmRAS..75...85H,1998MNRAS.299..595D,2015MNRAS.451.2376M}, such a dependency could potentially have a similar effect on colour (see figure~8 for EGIS in \citep{bizyaev_etal2014}). Note that in the SAMI survey \citep{2021ApJ...906..100B}, the authors measure that the $(g-i)$ colour of bulges is only slightly redder by $0.12\pm 0.02$ mag than the colour of disks, which they also attribute to higher metallicities rather than older stellar populations. Such a difference is rather small to be important for the measured effect if $B/T \sim 0.5$ or less.

%The bulge component light can be properly measured only after photometric decomposition, which is a tedious task. Nevertheless, it can be potentially approximated by metrics of galactic bulge strength, such as the concentration index $C$, S{\'e}rsic index $n$, or so-called Gini-$M_{20}$ bulge parameter (GMB), which is less sensitive to dust and mergers \citep{2019MNRAS.483.4140R}. All of these, however, need to calculate additional nontrivial parameters, which also may be unstable. We instead use star-forming galaxies subsample, which are selected as in Figure~\ref{fig:msfr}, to check if such separation affects $(g-i)$ versus $a/b$ relation. This results in almost constant shift between two subpopulations in about $0.15-0.2$~mag, even for thin galaxies with $a/b>7$, where galaxies around main sequence are bluer. This experiment shows that there is indeed a difference between these subpopulations, but the trend is almost the same between them. As another approach, we utilize advantage of already calculated CAS statistics for EGIPS data and estimate GMB parameter as $GMB = -0.693\times M_{20} + 4.95\times Gini - 3.96$ \citep{2019MNRAS.483.4140R}. The greater/lower GMB is above zero means greater bulge/disk domination. We could see anticorrelation between GMB and $a/b$ in EGIPS galaxies  with spearman's coefficient $\rho = -0.47$ (and $\rho = 0.43$ for GMB versus $(g-i)$ color). This could give a clue about bulge influence, which way be reasonable to expect to be the same order in EGIDE sample. 

We also plot in Figure~\ref{fig:color_axisRatio} the star-forming galaxy subsample, selected as in Figure~\ref{fig:completeness}, to check whether such separation affects the $(g-i)$ versus $a/b$ relation. This results in an almost constant shift between the two subpopulations of about $0.15$-$0.2$ mag, even for thin galaxies with $a/b>7$, where galaxies around the SFMS are bluer (see lines in the upper right panel of Figure~\ref{fig:color_axisRatio}). This experiment shows that there is indeed a difference between these subpopulations, but the trend is almost the same for both. Note that even for star-forming galaxies, those with larger bulges should show redder colours, and therefore selecting a subsample of SFMS galaxies cannot fully remove the bulge effect.

On the bottom row of Figure~\ref{fig:color_axisRatio} we show the same $q$ versus $(g-i)$ dependence for the mass-limited ($\log M_{\star}/M_{\odot} = 10.7\textrm{--}11.4$, left) and $cz$-limited (5000~km/s~$< cz <$~15000~km/s, right) subsamples. These subplots demonstrate interesting results. The most massive galaxies are indeed redder and lie above the median line for the whole EGIDE sample, and demonstrate an almost constant colour of $(g-i)\approx 1.15\pm 0.1$~mag with flattening. On the other side, the volume-limited subsample shows clear bimodality with only a few galaxies in the transition zone between two distinct clouds. These two clouds are equally well populated; the redder one demonstrates rounder galaxies with $a/b\sim 3.5\pm 1.0$ and $(g-i)\approx 1.0\pm 0.14$~mag, while the blue galaxies are thinner with $a/b\sim 5.0\pm 0.9$ and a colour of $(g-i)\approx 0.5\pm 0.2$~mag. Further investigation of this dependence is beyond the scope of this paper, but such a clear separation, combined with the fact that massive $\log M_{\star}/M_{\odot} > 10.7$ galaxies have almost the same colour regardless of axis ratio, clearly needs to be explained. One of the possible directions to explain such dependence is a well-known relation of dispersion versus stellar age population 
\citep{2024arXiv241212304T}.

\subsection{Mass-thickness relation}
\label{sec:mass}

%As another scientific application of the newly-builded EGIDE sample we compare how flatness $q$ depend on total stellar mass $M_{\star}$. To ensure the consistency of the methodology, i.e. that the same photometry used for mass derivation, we use \citep{2025A&A...704A.232E} calibrations (see also Appendix~\ref{ap:catalog_comparison} for the information). In Figure~\ref{fig:q_vs_mass} we show results in $r$-band across 5 orders of magnitude mass bins. Despite the small $q\sim 0.2-0.3$ in the whole $M_{\star}$, we could see in that Figure that it is not constant: $q$ increase towards the high-mass end of the distribution. We also see the increase of the $b/a$ in dwarf galaxies, but it is far less significant.

\begin{figure}
\centering
\includegraphics[width=0.95\textwidth]{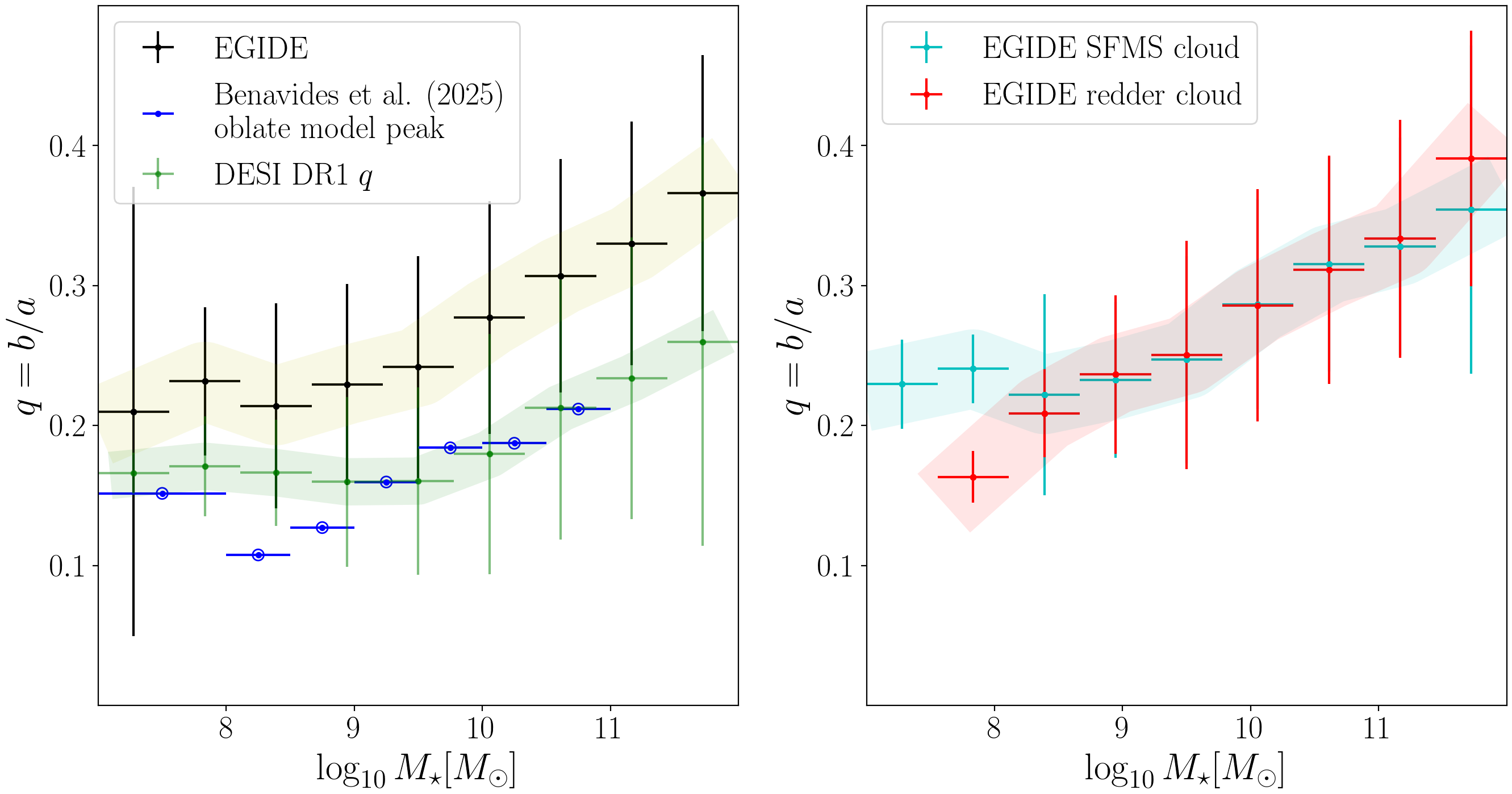}
\caption{
\textbf{Left:} Black points and the latte-colored line of arbitrary width show the flattening $b/a$ of galaxies measured for EGIDE in different stellar mass bins. Blue points represent the $b/a$ peak value for statistical oblate models in \citep{2025arXiv251211035B} (obtained via private communication for most of the bins). The green line and points show $b/a$ measured in DESI DR1 for $\sim20\%$ of the EGIDE galaxies. \textbf{Right:} Same values as in the left panel for EGIDE, but separated into star-forming main sequence (SFMS; see Figure~\ref{fig:completeness}, right) galaxies (blue line and points) and galaxies from the so-called green valley and red cloud (red line and points).
}
\label{fig:q_vs_mass}
\end{figure}

As another scientific application of the newly built EGIDE sample, we compare how the flattening $q$ depends on the total stellar mass $M_{\star}$. To ensure consistency of the methodology, i.e., that the same photometry is used for mass derivation, we use the \citep{2025A&A...704A.232E} calibrations for mass derivation (see also Appendix~\ref{ap:catalog_comparison} for details and comparison with other mass measurements). In Figure~\ref{fig:q_vs_mass}, we show results in the $r$-band across five orders of magnitude in mass bins. Despite the small $q \sim 0.2$-$0.3$ across the whole $M_{\star}$ range, we can see in the Figure that $q$ is not constant: it increases toward the high-mass end of the distribution. We also see a slight increase in $b/a$ for dwarf galaxies, but it is far less significant.

%In order to validate this result we compare it with other studies. Recently, in \citep{2025arXiv251211035B} the broad range of surveys (GAMA, DESI, ALFALFA) with hundreds of galaxies under arbitrary inclinations were used to infer the abundance of intrinsically flat galaxies needed to reproduce the observed abundance of highly elongated systems in projection. To do so, authors in \citep{2025arXiv251211035B} use two statistical models for $r$-band images: oblate model, where $a=b>c$ and triaxial model, where $a>b>c$ (in this notation $q=c/a$). We show in Figure~\ref{fig:q_vs_mass} peaks of $q$ distribution  for oblate model from their work (for $\log M_{\star} = 9..9.5 M_{\odot}$ see their figure~6; images for other mass bins obtained in private communication with Jos{\'e} Benavides). It is obvious, that while derived statistically $q$ from \citep{2025arXiv251211035B} are smaller, which is by definition, we can see perfect coincidence between both dataset. Their sample is limited by $10^11\, M_{\odot}$, but the steepness of the trend, minimum location and slightly bigger $q$ values for dwarfs are the same (correlation $\rho=...$). This similarity between two explicitly different samples, which represent direct measurements from observations and statistical models approach, makes result more reliable and cross-validates two works.

To validate this result, we compare it with other studies. Recently, in \citep{2025arXiv251211035B}, a broad range of surveys (GAMA, DESI, ALFALFA) with hundreds of thousands of galaxies at arbitrary inclinations was used to infer the abundance of intrinsically flat galaxies needed to reproduce the observed abundance of highly elongated systems in projection. To do so, the authors in \citep{2025arXiv251211035B} used two statistical models for $r$-band images: an oblate model, where $a = b > c$, and a triaxial model, where $a > b > c$ (in this notation, $q = c/a$). In Figure~\ref{fig:q_vs_mass}, we show the peaks of the $q$ distribution for the oblate model from their work (for mass bin $\log M_{\star}/M_{\odot} = 9\textrm{--}9.5$, see their figure 6; data for other mass bins were obtained via private communication with Jos{\'e} Benavides). It is evident that while the statistically derived $q$ values from \citep{2025arXiv251211035B} are smaller, which is expected by definition, we can see agreement between the trends in two datasets. Their sample is limited to $10^{11} M_{\odot}$, but the steepness of the trend, the location of the minimum, and the slightly larger $q$ values for dwarfs are all consistent (Spearman correlation $\rho = 0.61$). This similarity between two explicitly different samples, when one representing direct measurements from observations and the other a statistical modeling approach cross-validates both studies. Note, however, that measurements of the axis ratio $q$ in both samples may suffer from PSF smearing, resolution, and bulge effects, as discussed in Section~\ref{subsec:psf_smearing}.

%Several other findings support this result. Thus, for example lower envelope curve of $b/a$ versus $M_{\star}$ for galaxies in SDSS DR7 from figure~1 in \citep{2010MNRAS.406L..65S} clearly should describe only edge-on systems, and it is consistent with Figure~\ref{fig:q_vs_mass}. In \citep{2015MNRAS.451.2376M} authors do photometric decomposition for the subsample of EGIS galaxies and find that $h_z/h_R$ increase for brighter galaxies in SDSS $i$ (see their figure~12, left). In \citep{2013MNRAS.436L.104R} authors notice that dwarf galaxies, which are located in the left side of the mass distribution, are systematically thicker than the general population. Finally, for a subsample of our galaxies from DESI DR1 we plot $q_{DESI}$, which is derived using two ellipticity components $\epsilon_1$ and $\epsilon_2$ from TRACTOR models, versus same masses as we already use. Despite these are automatic measurements and can be subscriptable to errors, $q_{DESI}$ broadly shows the same picture. However, the $q_{DESI}$ values are lower ([which band it is??]) and the minimum is shifted towards larger galaxies. 

Several other findings support this result. For example, the lower envelope curve of $b/a$ versus $M_{\star}$ for galaxies in SDSS DR7 from figure~1 in \citep{2010MNRAS.406L..65S} should clearly describe only edge-on systems, and it is consistent with Figure~\ref{fig:q_vs_mass}. In \citep{2015MNRAS.451.2376M}, the authors perform photometric decomposition for a subsample of EGIS galaxies and find that $h_z/h_R$ increases for brighter galaxies in SDSS $i$ band (see their figure~12, left). In \citep{2013MNRAS.436L.104R}, the authors note that dwarf galaxies, which are located on the left side of the mass distribution, are systematically thicker than the general population. Finally, for a subsample of our galaxies from DESI DR1, we plot $q_{\text{DESI}}$, derived using the two ellipticity components $\epsilon_1$ and $\epsilon_2$ from the \textsc{TRACTOR} models, versus the same masses we use. Although these are automatic measurements and can be susceptible to errors, $q_{\text{DESI}}$ broadly shows the same trend. However, the $q_{\text{DESI}}$ values are lower and the minimum is shifted toward larger galaxies.

The observed trend in Figure~\ref{fig:q_vs_mass} can be justified at least qualitatively. If disks are in a submarginal stability regime with Toomre $Q$ assumed constant and slightly above unity, then we can show as in \citep{2002AstL...28..527Z} that the vertical-to-radial disk scale is proportional to the ratio of disk mass to total mass: $z_0/h \propto M_d/M_{tot} \propto M_{\star}/M_{h},$ where $M_h$ is the mass of the dark halo, for example $M_{200}$. If we take into account any stellar mass--halo mass (SMHM) relationship, for example as presented in figure~14 in \citep{2013ApJ...770...57B}, we can broadly assume $M_{\star} \propto M_h^{\alpha}$, where $\alpha \approx 1.4$ for $\log M_{h}/M_{\odot} \leq 12$. Substituting into the previous relationship, we can derive that 
$$z_0/h \propto M_{\star}/M_{h} \propto M_{\star}^{1 - 1/\alpha}.$$ 
This can in principle justify the observed trend in Figure~\ref{fig:q_vs_mass}, if we assume that $b/a$ and $z_0/h$ are close and well-correlated values. Note, however, that the derived simple power law predicts an increase in flattening larger than observed for realistic $\alpha$, thus it is probably also governed by additional processes.

% \textcolor{red}{Interestingly, that in \citep{2004ApJ...608..189D} it was shown for 49 edge-on bulgeless disk galaxies that $z_0/h$ is higher when rotation speed $V_c$ is lower, with a clear transition region around $V_c = 120$~km/s. This is a reverse trend to the picture presented here and in \citep{2025arXiv251211035B}, if we assume bulge influence to $b/a$ to be small. Note, however, that sample in \citep{2004ApJ...608..189D} is rather small and that the decision to use bulgeless galaxies can in principle bias results, since we can expect that processes that form or grow a massive bulge can also heat the disk \citep{2009ApJ...707L...1B}. From the other side, if we roughly estimate $M_{200}$ using approximation $M_{200}/M_{\odot} = 1.074\times10^5\times V_c^{3.115}$ from \citep{2016MNRAS.460.3610O} and transition value $V_c = 120$~km/s in \citep{2004ApJ...608..189D}, we get $M_{200}/M_{\odot}\approx 3\times 10^{11}$. This value is close to the beginning of the SMHM peak in figure~14 in \citep{2013ApJ...770...57B}, showing that indeed such rotation speed $V_c$ can be in transition region and form $z_0/h$ due halo different contribution as descibed in previous paragraph.}

Interestingly, it was shown in \citep{2004ApJ...608..189D} for 49 edge-on bulgeless disk galaxies that $z_0/h$ is higher when rotation speed $V_c$ is lower, with a clear transition region around $V_c = 120$~km/s. This is a reverse trend to the picture presented here and in \citep{2025arXiv251211035B}, if we assume the influence of the bulge on $b/a$ to be small. Note, however, that the sample in \citep{2004ApJ...608..189D} is rather small and that the decision to use bulgeless galaxies can in principle bias the results, since we can expect that processes that form or grow a massive bulge can also heat the disk \citep{2009ApJ...707L...1B}. On the other hand, if we roughly estimate $M_{200}$ using the approximation $M_{200}/M_{\odot} = 1.074\times10^5\times V_c^{3.115}$ from \citep{2016MNRAS.460.3610O} and the transition value $V_c = 120$~km/s from \citep{2004ApJ...608..189D}, we get $M_{200}/M_{\odot}\approx 3\times 10^{11}$. This value is close to the beginning of the SMHM peak in figure~14 of \citep{2013ApJ...770...57B}, showing that indeed such rotation speed $V_c$ can be in the transition region and affect $z_0/h$ due to different halo contributions as described in the previous paragraph.

Of course, the trend presented in Figure~\ref{fig:q_vs_mass} could be related to factors other than the disk, such as the spheroidal component. Thus, for galaxies at the high-mass end, we can expect to find lenticular galaxies, which may have a large bulge-to-total ratio $B/T$, which in turn will increase $b$ value. In \citep{2024ApJ...974...88X}, the authors found the fraction of stellar mass in the spheroidal component in TNG50 simulated galaxies and evaluated concentration versus stellar mass for galaxies from the HSC-SSP survey. They indeed found (figures~5 and 6) that the spherical component fraction shows a similar U-shaped trend with $M_{\star}$. In \citep{2025arXiv250215581Q}, the authors show for early Euclid data how $b/a$ changes during decomposition if a bulge with different $B/T$ is taken into account. Their figure~8 gives a sense that the effect size can be significant for red points with disk $q\approx 0.2$, which are likely edge-on systems. Note also that the axis ratio can systematically increase when the S{\'e}rsic index increases \citep{2025arXiv250315309E}.. Another piece of evidence is \citep{2015MNRAS.451.2376M}, where figure~13 exactly shows that flattening $h_z/h_R$ increases as $B/T$ increases.

%As another independent test of spheroidal component influence we can do the follows. Based on SFR-stellar mass relation presented in Figure~\ref{fig:msfr} we separate star-forming MS galaxies from red cloud using $...$ condition. For each of these two classes we present separately mass-$q$ relation in the right of Figure~\ref{fig:q_vs_mass}. Note, that these are not exactly galaxies as on the left subplot, because we know SFR only for a subsample of 30... objects, which is roughly 20\% of EGIDE size. We see in Figure~\ref{fig:q_vs_mass} that galaxies from the so-called red cloud with greater spherical component follows the previously demonstrated trend, while galaxies from SFMS demonstrate more constant line around $q\sim0.2-0.25$ with just a little variation. This test additionally strengthen point above that $B/T$ may influence the found relation significantly.

As another independent test of the spheroidal component influence, we can proceed as follows. Based on the SFR-stellar mass relation presented in Figure~\ref{fig:completeness}, we separate the star-forming main-sequence galaxies from the red cloud using the separation line shown in that Figure. For each of these two classes, we present the mass-$q$ relation separately on the right side of Figure~\ref{fig:q_vs_mass}. Note that these are not exactly the same galaxies as in the left subplot, because we have SFR information only for a subsample of 30833 objects, which is roughly 20\% of the EGIDE sample size. In Figure~\ref{fig:q_vs_mass}, we see that galaxies from the so-called red cloud, which potentially have a larger spherical component, follow exactly the same previously demonstrated trend, as do galaxies from the main sequence.
% In Figure~\ref{fig:q_vs_mass}, we see that galaxies from the so-called red cloud, which have a larger spherical component, follow the previously demonstrated trend, while galaxies from the main sequence show a more constant line around $q \sim 0.2$-$0.25$ with smaller variation. This test further strengthens the point above that $B/T$ may significantly influence the found relation. 
Similarly to that, if we instead split galaxies into blue and red according to the CMD presented in Figure~\ref{fig:galaxyCMD} with colours $0.2 \le (g-r) \le 0.6$ and $0.6 \le (g-r) \le 1.2$, then for both types of galaxies we will see an increase similar to that presented in Figure~\ref{fig:q_vs_mass}.

% \citep{2025arXiv251211035B} If anyone wanted to see the three NFW authors again in one article, here it is! (In a different order, though.) And what an article, I really find it interesting. They use a huge dataset from GAMA + DESI + ALFALFA to look at the q oblateness over a wide mass range. And in all the data and mass bins, they claim to find 40\% or more very flat galaxies with a true oblateness of 0.2 or less (they do this in the current statistically accepted sense, rotating various figures and measuring the resulting projection). They insist that this results in an extremely large number of "flat" dwarfs, which is not even close to what you'll find in space simulations, but overall, their oblateness ratio of 1in5 or more for full-size galaxies really surprised me. We don't see that many in our samples, and we need to understand why (I don't think TNG has that many either).

\subsection{Count drop with colour}
\label{sec:countdrop}

%Previously for galaxies in EGIPS sample we detect the difference in number change of galaxies with $a/b$ for galaxies with different colour (see figure~11 in \citep{egips}). Thus, the axis ratio functions of red, $1.0\le(g-i)_0\le1.4$, and blue, $0.4\le(g-i)_0\le0.8$, galaxies are noticeably different, i.e. number of the redder galaxies drops with increasing $a/b$ faster than for the bluer galaxies. Overall, slope coefficient $k$ in relation $\log N\propto\-k \times a/b$ was found to form U-shape with color (see figure~2 in \citep{egips} and also \citep{1994AstL...20....8K}). 

Previously for the EGIPS sample, we detected a difference in the number distribution of galaxies with $a/b$ for galaxies of different colours (see figure~11 in \citep{egips}). Thus, the axis ratio functions of red ($1.0 \le (g-i) \le 1.4$) and blue ($0.4 \le (g-i) \le 0.8$) galaxies are noticeably different; i.e., the number of redder galaxies drops with increasing $a/b$ faster than for the bluer galaxies. Overall, the slope coefficient $k$ in the relation $\log N \propto -k \times a/b$ was found to form a U-shape with colour (see figure~2 in \citep{egips} and also \citep{1994AstL...20....8K}).

%Here in Figure~\ref{fig:logN_AxisRatio} we present similar analysis but for the $(g-r)$ color. Motivated by Figure~\ref{fig:galaxyCMD} distributions, we present fits for $\log N$ versus $a/b$ in $g$-band for blue and red galaxies with conditions $0.2<(g-r)<0.6$ and $0.7 < (g-r) < 1.2$ respectively. We see indeed the same picture as before: decline of red galaxies is faster, than for the bluer ones. The effect is smaller than presented in \citep{egips}, probably because of smaller and more distant galaxies, but it is still noticeable. The slope values are bigger, because we use different color and number of galaxies in $a/b < 5$ bins is much bigger than in EGIPS, while the right-side end of the distribution is the same as before (see also Figure~\ref{color_axisRatio} in Section~\ref{sec:color_flatness}). On the right panel of Figure~\ref{fig:logN_AxisRatio} we present slope $k$ in various color bins. The rate of decline becomes gradually flatter for bluer colors and stay near the same for galaxies with $(g-r) < 0.7$ and $(g-r)>0.9$. This can be explained by the fact that younger and bluer stellar populations form thinner discs, as well-known vertical-dispersion versus age relation dictates [cite], thus such galaxies have more counts  in $a/b>7-8$ bins. Note, that the same conclusions about $k$ in $ N\propto \exp (-k \times a/b)$ can be in principle derived from the Figure~\ref{fig:color_axisRatio}. 

\begin{figure}
\centering
\includegraphics[width=0.48\textwidth]{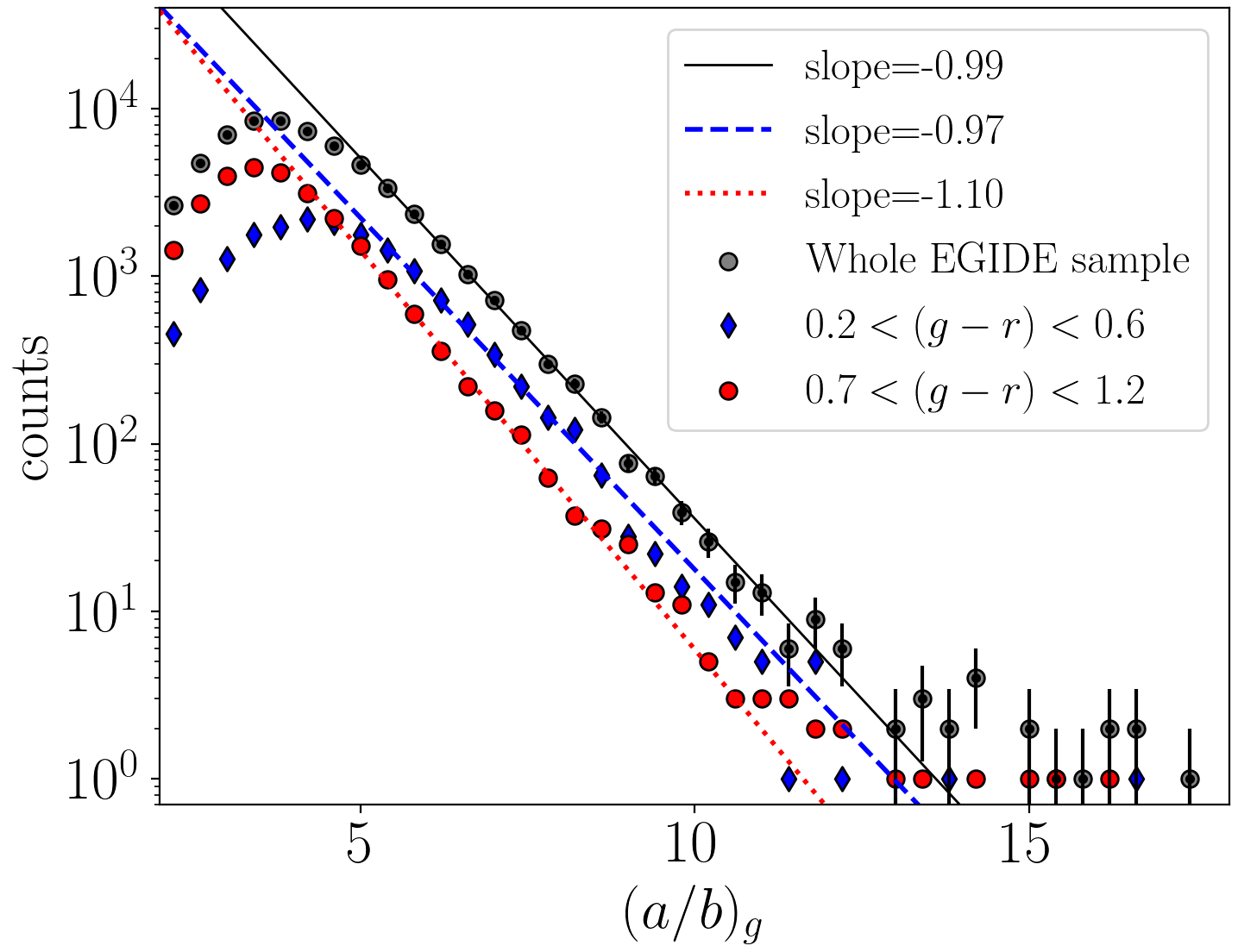}
\includegraphics[width=0.48\textwidth]{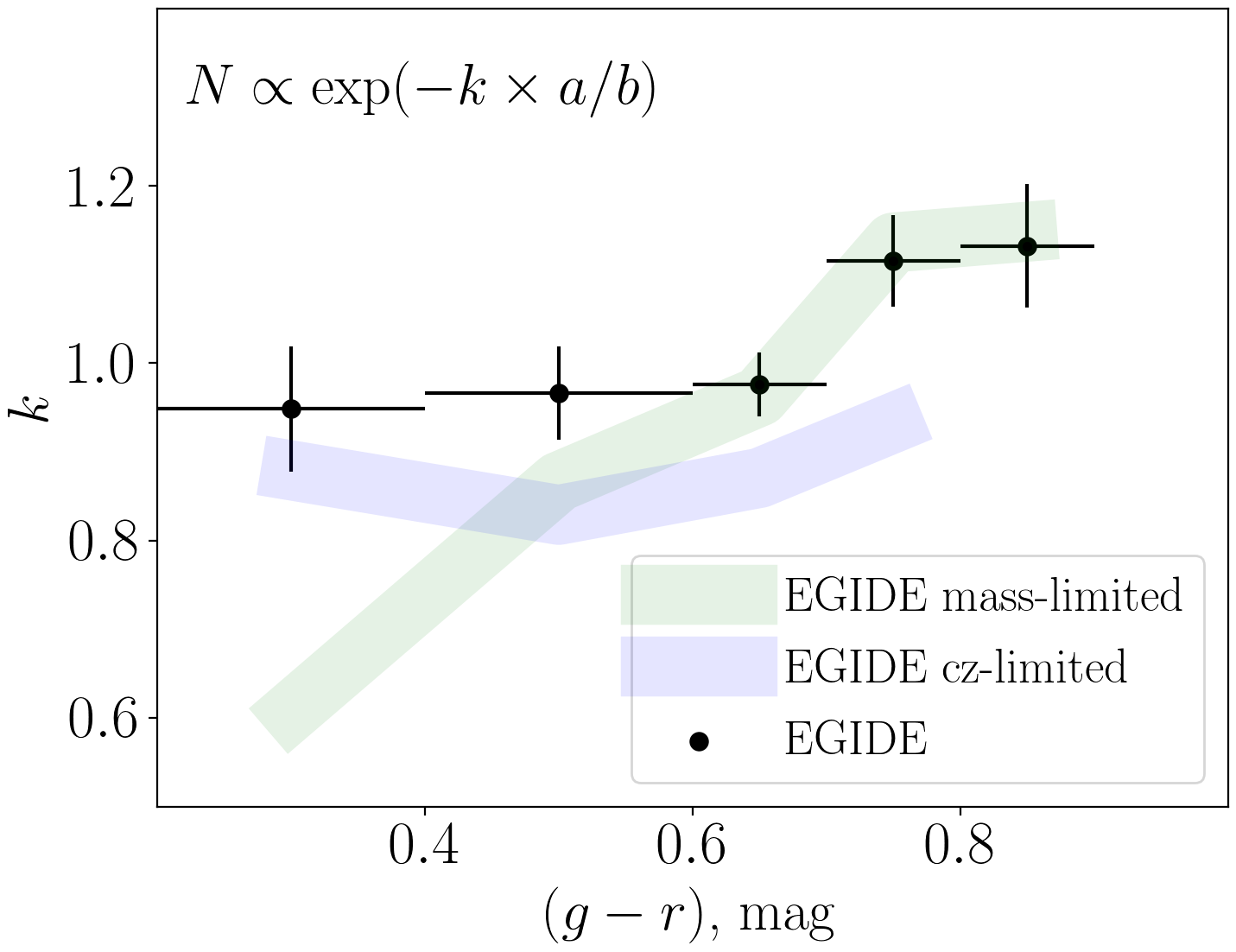}
\caption{
\textbf{Left:} distribution of the EGIDE galaxies by the $a/b$ axis ratio in the $g$ band. Distributions for different $(g-r)$-selected subsamples are also shown. Lines represent the linear fits in the given coordinates. \textbf{Right:} points show the dependence of the decline rate of the distribution function of the axis ratios for a given colour bin. Vertical error bars represent 1-$\sigma$ uncertainties of the fit. Lines with transparent colours show the same dependence for the mass-limited (green) and volume-limited (blue) subsamples (see Section~\ref{sec:completeness} for details).
}
\label{fig:logN_AxisRatio}
\end{figure}

Here in Figure~\ref{fig:logN_AxisRatio}, we present a similar analysis but for the $(g-r)$ colour. Motivated by the CMD distributions in Figure~\ref{fig:galaxyCMD}, we present fits for $\log N$ versus $a/b$ in the $g$-band for blue and red galaxies with conditions 0.2~mag~$< (g-r) < 0.6$~mag and 0.7~mag~$< (g-r) < 1.2$~mag, respectively. We indeed see the same picture as before: the decline of red galaxies is faster than that of the bluer ones. The effect is smaller than that presented in \citep{egips}, probably because of the smaller and more distant galaxies, but it is still noticeable. The slope values $k$ are larger because the number of galaxies in the $a/b < 5$ bins is much larger than in EGIPS, while the right-side end of the distribution is similar to that in EGIPS (see also Figure~\ref{fig:color_axisRatio} in Section~\ref{sec:color_flatness}). On the right panel of Figure~\ref{fig:logN_AxisRatio}, we present the slope $k$ in various colour bins. The rate of decline becomes gradually flatter for bluer colours and remains nearly the same for galaxies with $(g-r) < 0.6$~mag and $(g-r) > 0.7$~mag. This can be explained by the fact that younger and bluer stellar populations form thinner disks, as the well-known vertical dispersion versus age relation dictates \citep{2025ApJ...979..103S,1977A&A....60..263W}; thus, such galaxies have higher counts in the $a/b > 7\textrm{--}8$ bins. Note that the same conclusions about $k$ in $ N \propto \exp (-k \times a/b)$ can in principle be derived from Figure~\ref{fig:color_axisRatio}.

%We also plot in Figure~\ref{fig:logN_AxisRatio} the same dependency for mass- and $cz$-volume limited subsamples, defined in Section~\ref{sec:completeness}. For the latter there is not enough galaxies with $(g-r)>0.9$ to fit linear regression properly, thus we limit $k$ estimation for it with this value. We can clearly see that for volume-limited sample galaxies count $N$ drop is almost the same: $k \approx 1$. For mass-limited data decline coefficient $k$ increase in the whole presented colour range. From the comparison with the full EGIDE it is clear, that its peak value $k\approx 1.7$ as well as dependence for red galaxies are indeed defined by massive galaxies.

We also plot in Figure~\ref{fig:logN_AxisRatio} the same dependency for the mass-limited and $cz$-volume-limited subsamples, defined in Section~\ref{sec:completeness}. For the latter, there are not enough galaxies with $(g-r)>0.9$ to fit the linear regression properly, so we limit the $k$ estimation for it to this value. We can clearly see that for the volume-limited sample, the galaxy count $N$ drop is almost the same: $k \approx 0.9$. For the mass-limited data, the decline coefficient $k$ increases across the entire presented colour range. From the comparison with the full EGIDE, it is clear that its peak value of $k\approx 1.1$, as well as the whole dependence for redder galaxies, is indeed defined more by massive galaxies.

%\begin{figure}
%\centering
%\includegraphics[width=0.45\textwidth]{paper_imgs/ab_slope_gi.png}
%\caption{
%Dependence of the rate of the exponential decline for the distribution function of the axis ratios on colour.
%Each value is calculated for a colour range indicated by the horizontal lines.
%}
%\label{fig:ColorAxisRatio}
%\end{figure}

\section{Conclusions}
\label{sec:conclusions}

We have presented the EGIDE (Edge-on Galaxies in the DESI Survey) catalogue, a new large-scale sample of 149,215 edge-on galaxy candidates selected from the DESI Legacy Imaging Surveys DR10. The catalogue is publicly available through the Edge-on Galaxy Database\footnote{\url{https://www.sao.ru/edgeon/catalogs.php?cat=EGIDE}}. EGIDE is approximately ten times larger than its predecessor Edge-on Galaxies in the Pan-STARRS survey (EGIPS) and covers more than half of the sky, providing a homogeneous dataset for studying the vertical structure of disk galaxies and statistical properties of edge-on systems. The edge-on selection methodology combines a neural network search using a fine-tuned \textsc{Zoobot} model with manual visual inspection, resulting in a reliable sample.

For each galaxy in EGIDE, the catalogue provides homogeneous \textsc{SExtractor} photometry in the $griz$ bands (Kron and Petrosian magnitudes, ellipse parameters), total stellar masses estimated using colour-based calibrations, and redshifts for 98\% of the sample, which were collected primarily from REGALADE. Cross-identifications with HyperLEDA, RCSEDv2, DESI DR1 and other surveys (see Table~\ref{tab:inters}) are also included.

The EGIDE dataset has the following key properties:

\begin{enumerate}

\item Galaxies in EGIDE have a median angular half-size of the major axis equal to 13.5~arcsec and are highly flattened objects with $q=b/a\approx 0.24$ (Figure~\ref{fig:phot_properties}, Table~\ref{tab:qbands} and Figure~\ref{fig:q_appendix}). Measurements of the axis ratio $q$ may suffer from PSF smearing, resolution, and bulge effects, especially for smaller galaxies. We correct for PSF smearing, impose a constraint on the relative error of $q$, and discuss the bulge influence in Section~\ref{subsec:psf_smearing}.

\item Due to methodological reasons the sample lacks very close and extended edge-on galaxies; thus previous surveys of edge-on galaxies such as RFGC, EGIS and EGIPS should be seen as complementary to the EGIDE dataset.

\item The $V/V_m$ test indicates that the sample is mostly complete for the 40\% most luminous galaxies ($V/V_m = 0.4636 \pm 0.0012$), with a depth limit of $m \approx 19$~mag in the $r$-band (Figure~\ref{fig:logNlogA}).

\item The median redshift of EGIDE is $cz \approx 33,750$~km~s$^{-1}$ and is three times larger than that of EGIPS (see Figure~\ref{fig:cz}). The galaxies are distributed up to a distance of 2000~Mpc and are slightly more massive than those from EGIPS (median value $\log M_{\star}/M_{\odot} = 10.56$ in EGIDE versus $\log M_{\star}/M_{\odot} = 10.34$ in EGIPS, see Figure~\ref{fig:completeness}, left).

\item Edge-on galaxies in EGIDE populate both the red and blue sequences (see Figure~\ref{fig:galaxyCMD} and Figure~\ref{fig:completeness}, right). Compared to the SDSS DR7 reference sample, EGIDE galaxies are shifted to fainter absolute magnitudes and exhibit a broader range of colours, which is potentially explained by the greater depth of the DESI Legacy images and by internal extinction.

\end{enumerate}

The analysis presented in this paper leads to the following results:

\begin{enumerate}

\item The colour–inverse flattening diagram confirms the previously found bimodality: a dense cloud of thicker ($a/b \sim 3$--$5$) and redder ($(g-i) \sim 1.1$--$1.3$~mag) galaxies is followed by a population of thinner ($a/b > 5$) galaxies that are bluer by $0.2$--$0.4$~mag (Figure~\ref{fig:color_axisRatio}). This bimodality appears to be a fundamental property of disc galaxies, and is better visible for the volume-limited subsample, while massive galaxies form a single cluster with near-constant colour. Despite the tenfold increase in sample size, the running medians of EGIDE and EGIPS are nearly identical on these plots, but there is also a noticeable difference between galaxies from the star-forming main sequence (SFMS) and non-SFMS areas.

\item The flattening $q=b/a$ increases with total stellar mass $M_{\star}$ at the high-mass end (Figure~\ref{fig:q_vs_mass}, left). The same trend is independently recovered from the statistical oblate models of \citep{2025arXiv251211035B}, and from DESI DR1 ellipticity measurements, cross-validating our approach. We found that the increase in $b/a$ is driven by both red cloud early-type galaxies as well as SFMS galaxies (Figure~\ref{fig:q_vs_mass}, right).
% We found that the increase in $b/a$ is driven primarily by red cloud early-type galaxies, while SFMS galaxies show a nearly constant flattening of $q \approx 0.25$ (Figure~\ref{fig:q_vs_mass}, right).

\item The number of red galaxies drops with increasing $a/b$ faster than that of blue galaxies (Figure~\ref{fig:logN_AxisRatio}). The decline coefficient $k$ in $ N \propto \exp (-k \times a/b)$ is $k \approx 1.10$ for red galaxies and $k \approx 0.97$ for blue galaxies in the full sample. For the volume-limited subsample, $k \approx 0.9$ for all $(g-r)$ colours, while for the mass-limited subsample, $k$ increases across the entire colour range, indicating that the change in $k$ is primarily driven by massive systems.

\item Comparisons with HyperLEDA, DESI DR1, RCSEDv2 and REGALADE show good agreement in redshift (NMAD $\approx$ 20--170~km/s, Figure~\ref{fig:cz_comparison}), total stellar mass $M_{\star}$ (NMAD $\approx$ 0.15~dex, Figure~\ref{fig:masses_comparison}), and apparent magnitude (NMAD $\approx$ 0.10~mag, Figure~\ref{fig:mag_comparison}). Flatness measurements from these independent sources also confirm the small $b/a$ values found in EGIDE (see Figure~\ref{fig:q_appendix}).

\end{enumerate}

In summary, EGIDE provides a large, reliable, and publicly accessible catalogue of edge-on galaxies. The first scientific results strengthen, confirm and extend earlier findings on colour-flattening and mass-flattening relations, and highlight the hypothetical importance of the spheroidal component in shaping the observed flattening of massive galaxies. The methodology developed here can be directly applied to upcoming wide-field surveys such as Euclid, Roman, and LSST, where samples of similarly large size will become routinely available. The EGIDE catalogue is well suited for the search of specific objects (e.g., superthin galaxies, polar rings, boxy bulges, galactic fountains, etc.), as well as for future studies of disk vertical structure and comparisons with cosmological simulations with robust statistical analyses across mass, colour, and environment.

%\section*{Acknowledgements}
%% We thank the anonymous referee for her/his kind and helpful comments.
%
%This research was supported by the Russian Science Foundation grant 24--72--10084. Thanks Jos{\'e} Benavides for the data.
%
%
%\section*{Data availability}
%
% The public access to the EGIDE catalogue is supported by the Edge-on Galaxy Database\footnote{\url{https://www.sao.ru/edgeon/}} \citep{2021AstBu..76..218M}.
%

%%%%%%%%%%%%%%%%%%%%%%%%%%%%%%%%%%%%%%%%%%
\vspace{6pt} 

\authorcontributions{Conceptualization, A.A., S.S., D.I.; Methodology, A.A., S.S., D.I.; Validation, A.A., I.V., S.S., D.I.; Software, A.A., S.S., D.I.; Formal analysis, A.A.; Investigation, A.A., S.S., I.V.; Data curation, A.A., D.I., E.V.; Resources, D.I.; Writing -- Original Draft Preparation, A.A.; Writing -- Review and Editing, A.A., S.S., D.I., V.P., I.V., M.D., A.V., A.M., E.V., D.V.; Visualization, A.A., S.S., I.V.; Supervision, A.A. All authors have read and agreed to the published version of the manuscript.}

% \Author{Alexander A. Marchuk $^{1,2}$*\orcidA{}, 
% Ilia V. Chugunov $^{1}$\orcidC{},
% Fr{\'e}d{\'e}ric Galliano$^{6}$\orcidG{},
% Aleksandr V. Mosenkov$^{3}$\orcidB{},
% Polina V. Strekalova$^{1}$,
% Valeria S. Kostiuk$^{5}$\orcidF{},
% George A. Gontcharov$^{1}$\orcidD{},
% Vladimir B. Il’in$^{1,2,4}$\orcidE{},
% Sergey S. Savchenko$^{1,2}$,
% Anton A. Smirnov$^{1}$  and
% Denis M. Poliakov$^{1}$}

\funding{D.V.B. is partly supported by Russian Science Foundation grant no. 22--12--00080, which helps to collect and process needed data from RCSEDv2 database. For all other work, we acknowledge financial support from the Russian Science Foundation, grant no. 24--72--10084. }

\dataavailability{The public access to the EGIDE catalogue is supported by the Edge-on Galaxy Database\footnote{\url{https://www.sao.ru/edgeon/}} \citep{2021AstBu..76..218M}.} 

% Only for journal Nursing Reports
%\publicinvolvement{Please describe how the public (patients, consumers, carers) were involved in the research. Consider reporting against the GRIPP2 (Guidance for Reporting Involvement of Patients and the Public) checklist. If the public were not involved in any aspect of the research add: ``No public involvement in any aspect of this research''.}

% Only for journal Nursing Reports
%\guidelinesstandards{Please add a statement indicating which reporting guideline was used when drafting the report. For example, ``This manuscript was drafted against the XXX (the full name of reporting guidelines and citation) for XXX (type of research) research''. A complete list of reporting guidelines can be accessed via the equator network: \url{https://www.equator-network.org/}.}

% Only for journal Nursing Reports
%\useofartificialintelligence{Please describe in detail any and all uses of artificial intelligence (AI) or AI-assisted tools used in the preparation of the manuscript. This may include, but is not limited to, language translation, language editing and grammar, or generating text. Alternatively, please state that “AI or AI-assisted tools were not used in drafting any aspect of this manuscript”.}

%% We thank the anonymous referee for her/his kind and helpful comments.
\acknowledgments{We thank two anonymous referees for their work and identified issues with the original draft of the paper, and acknowledge that they help substantially improve the text.  We thank Jos{\'e} Benavides for kindly providing the data. We thank Vasilisa I. Sorokina for help with extinction data. The Legacy Surveys consist of three individual and complementary projects: the Dark Energy Camera Legacy Survey (DECaLS; Proposal ID \#2014B-0404; PIs: David Schlegel and Arjun Dey), the Beijing-Arizona Sky Survey (BASS; NOAO Prop. ID \#2015A-0801; PIs: Zhou Xu and Xiaohui Fan), and the Mayall z-band Legacy Survey (MzLS; Prop. ID \#2016A-0453; PI: Arjun Dey). DECaLS, BASS and MzLS together include data obtained, respectively, at the Blanco telescope, Cerro Tololo Inter-American Observatory, NSF’s NOIRLab; the Bok telescope, Steward Observatory, University of Arizona; and the Mayall telescope, Kitt Peak National Observatory, NOIRLab. Pipeline processing and analyses of the data were supported by NOIRLab and the Lawrence Berkeley National Laboratory (LBNL). The Legacy Surveys project is honored to be permitted to conduct astronomical research on Iolkam Du’ag (Kitt Peak), a mountain with particular significance to the Tohono O’odham Nation.

NOIRLab is operated by the Association of Universities for Research in Astronomy (AURA) under a cooperative agreement with the National Science Foundation. LBNL is managed by the Regents of the University of California under contract to the U.S. Department of Energy.

This project used data obtained with the Dark Energy Camera (DECam), which was constructed by the Dark Energy Survey (DES) collaboration. Funding for the DES Projects has been provided by the U.S. Department of Energy, the U.S. National Science Foundation, the Ministry of Science and Education of Spain, the Science and Technology Facilities Council of the United Kingdom, the Higher Education Funding Council for England, the National Center for Supercomputing Applications at the University of Illinois at Urbana-Champaign, the Kavli Institute of Cosmological Physics at the University of Chicago, Center for Cosmology and Astro-Particle Physics at the Ohio State University, the Mitchell Institute for Fundamental Physics and Astronomy at Texas A\&M University, Financiadora de Estudos e Projetos, Fundacao Carlos Chagas Filho de Amparo, Financiadora de Estudos e Projetos, Fundacao Carlos Chagas Filho de Amparo a Pesquisa do Estado do Rio de Janeiro, Conselho Nacional de Desenvolvimento Cientifico e Tecnologico and the Ministerio da Ciencia, Tecnologia e Inovacao, the Deutsche Forschungsgemeinschaft and the Collaborating Institutions in the Dark Energy Survey. The Collaborating Institutions are Argonne National Laboratory, the University of California at Santa Cruz, the University of Cambridge, Centro de Investigaciones Energeticas, Medioambientales y Tecnologicas-Madrid, the University of Chicago, University College London, the DES-Brazil Consortium, the University of Edinburgh, the Eidgenossische Technische Hochschule (ETH) Zurich, Fermi National Accelerator Laboratory, the University of Illinois at Urbana-Champaign, the Institut de Ciencies de l’Espai (IEEC/CSIC), the Institut de Fisica d’Altes Energies, Lawrence Berkeley National Laboratory, the Ludwig Maximilians Universitat Munchen and the associated Excellence Cluster Universe, the University of Michigan, NSF’s NOIRLab, the University of Nottingham, the Ohio State University, the University of Pennsylvania, the University of Portsmouth, SLAC National Accelerator Laboratory, Stanford University, the University of Sussex, and Texas A\&M University.

BASS is a key project of the Telescope Access Program (TAP), which has been funded by the National Astronomical Observatories of China, the Chinese Academy of Sciences (the Strategic Priority Research Program ''The Emergence of Cosmological Structures'' Grant \# XDB09000000), and the Special Fund for Astronomy from the Ministry of Finance. The BASS is also supported by the External Cooperation Program of Chinese Academy of Sciences (Grant \# 114A11KYSB20160057), and Chinese National Natural Science Foundation (Grant \# 12120101003, \# 11433005).

The Legacy Survey team makes use of data products from the Near-Earth Object Wide-field Infrared Survey Explorer (NEOWISE), which is a project of the Jet Propulsion Laboratory/California Institute of Technology. NEOWISE is funded by the National Aeronautics and Space Administration.

The Legacy Surveys imaging of the DESI footprint is supported by the Director, Office of Science, Office of High Energy Physics of the U.S. Department of Energy under Contract No. DE-AC02-05CH1123, by the National Energy Research Scientific Computing Center, a DOE Office of Science User Facility under the same contract; and by the U.S. National Science Foundation, Division of Astronomical Sciences under Contract No. AST-0950945 to NOAO.}

\conflictsofinterest{The authors declare no conflicts of interest.} 

%%%%%%%%%%%%%%%%%%%%%%%%%%%%%%%%%%%%%%%%%%
%% Optional

%% Only for journal Encyclopedia
%\entrylink{The Link to this entry published on the encyclopedia platform.}

%\abbreviations{Abbreviations}{
%The following abbreviations are used in this manuscript:\\
%\noindent 
%\begin{tabular}{@{}ll}
%SED & Spectral Energy Distribution\\
%BIC & Bayesian Information Criterion\\
%CR & Corotation radius\\
%PSF & Point spread function\\
%FWHM & Full width at half maximum\\
%SPIRE & Spectral and Photometric Imaging REceiver\\
%PACS & Photodetector Array Camera and Spectrometer\\
%IRAC & InfraRed Array Camera\\
%NIR & Near-infrared\\
%MIR & Mid-infrared\\
%FIR & Far-infrared
%\end{tabular}
%}

\appendixtitles{yes} % Leave argument "no" if all appendix headings stay EMPTY (then no dot is printed after "Appendix
%A"). If the appendix sections contain a heading then change the argument to "yes".
\appendixstart
\appendix

\section[\appendixname~\thesection]{Neural net training and pipeline details}
\label{ap:nnsearch}

 \begin{figure}
\centering
\includegraphics[width=0.89\textwidth]{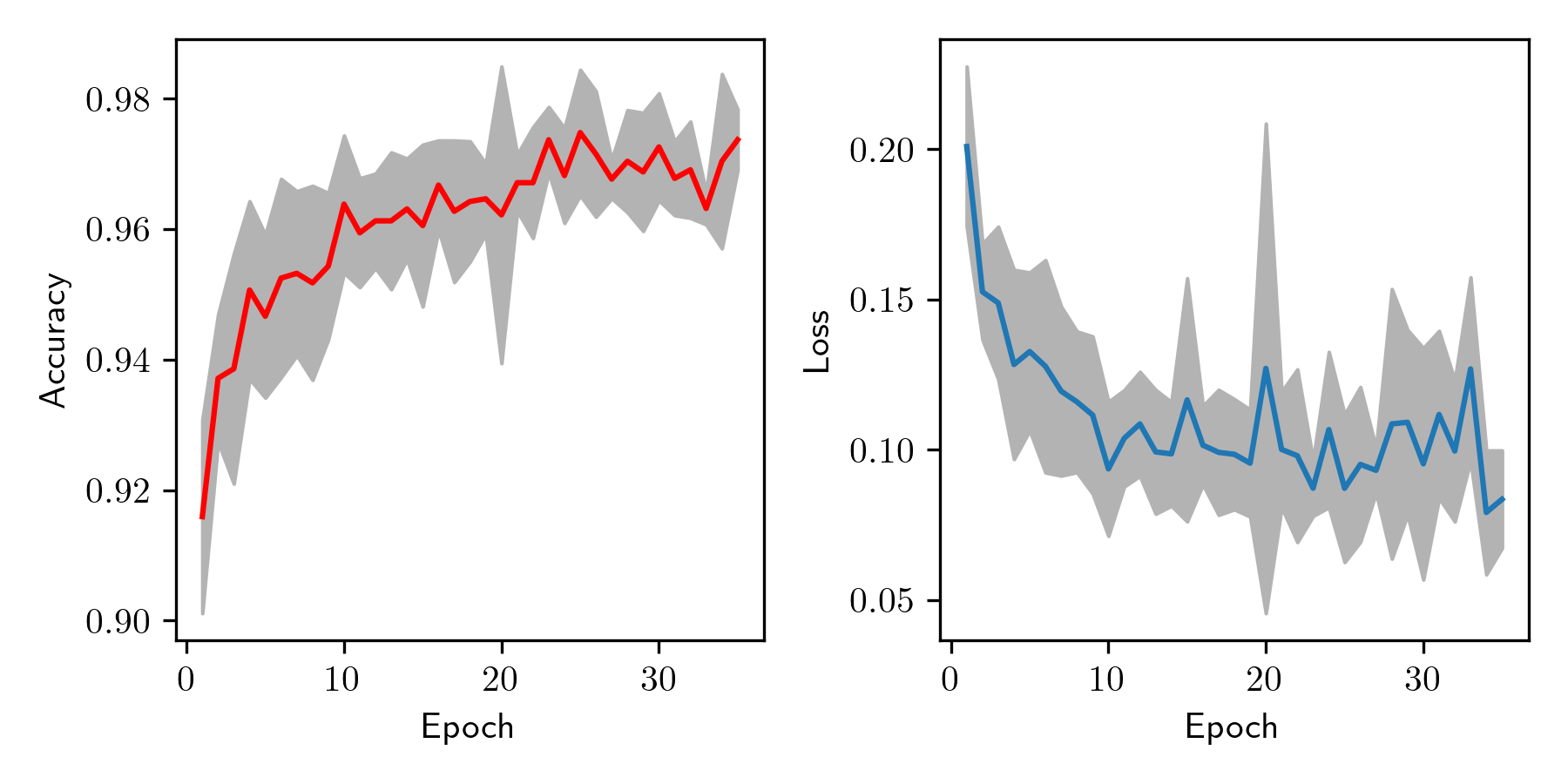}
\caption{ Learning curve of the edge-on/not edge-on classifier: test sample accuracy (left) and loss function as a
  function of epoch number. Solid lines -- mean value among all classifiers of the ensemble, shaded regions show $\pm$
  standard deviation.}
\label{fig:learningcurve}
\end{figure}

\begin{figure}
 \centering
 \includegraphics[width=0.23\textwidth]{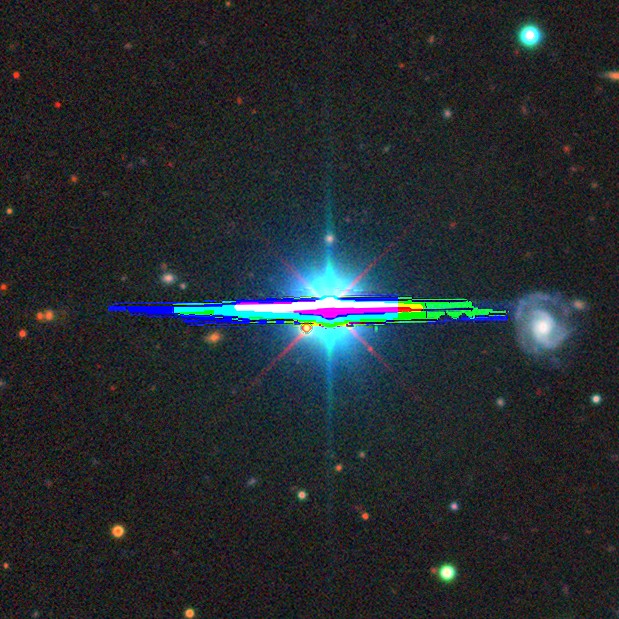}
 \includegraphics[width=0.23\textwidth]{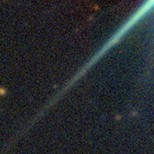}
 \includegraphics[width=0.23\textwidth]{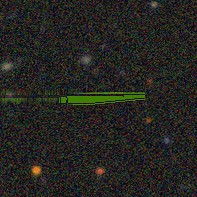}
 \includegraphics[width=0.23\textwidth]{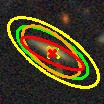}
 \includegraphics[width=0.23\textwidth]{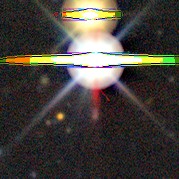}
 \includegraphics[width=0.23\textwidth]{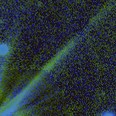}
 \includegraphics[width=0.23\textwidth]{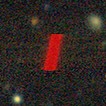}
 \includegraphics[width=0.23\textwidth]{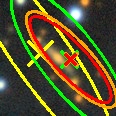}
 \caption{
 Examples of various problems related to edge-on detection and processing. Columns from left to right: CCD leakage detected as edge-on, star rays, artifacts in DESI Legacy. Last column in the example of bad photometry measurements using \textsc{SExtractor}.
 }
 \label{fig:exampleBad}
 \end{figure}
 
To build our classifier we used \texttt{convnext\_nano} model from the  \textsc{Zoobot} project \citep{2023JOSS....8.5312W}. \textsc{Zoobot} models predict morphological classification of  galaxies based on their images. To do so they were trained on a huge dataset of about 92 million labels collected from  Galaxy Zoo volunteers.

The DESI survey does not have a full coverage by all $griz$ bands, some regions are covered only be some  subset of bands like $gri$ or $gi$. Since the input dimension of a neural classifier is fixed (so it should have a  fixed number of bands) it is not possible to make a single classifier to work with all the regions of  survey. Therefore we decided to train three independent classifiers for band subsets $gri$, $grz$, and $gi$. Together  these three classifiers cover the vast majority of the survey. Like in our previous work instead of training a single  classifier, we decided to train an ensemble of nine models for each band combination.

To make a training sample we used a combinations of EGIPS and RFGC samples as a source of positive  examples of edge-on galaxies. To achieve better training results we performed an additional visual inspection and  excluded some galaxies from this sample what have considerable deviations from edge-on orientations. As negative  examples we took random sample of galaxies of various types and orientations from HyperLEDA database and excluded  edge-on disks from it. The galaxies that were excluded from the positive sample after the visual inspection were  included in the negative sample. Our experiments showed that this significantly improved the performance of resulting  trained models. Since edge-on/not edge-on classification does not have a strict condition (like, for example  spirals/ellipticals) it is necessary to provide enough borderline examples during the training process. Without this  model underperforms for such borderline cases. The number of positive examples in our training samples are:  $gri$-classifier -- 1226, $grz$-classifier -- 2543, $gi$-classifier -- 1229. The difference in number reflect the  difference in overlapping between the parent sample (EGIPS+RFGC) and the DESI footprint in different passbands. The  number of negative examples is the same as of positive ones.  The learning curve for the $gri$-classifier is  demonstrated in Fig.~\ref{fig:learningcurve}, two other classifiers show similar statistics.

\section[\appendixname~\thesection]{Surveys data cross-validation}
\label{ap:catalog_comparison}

%In this Appendix we provide comparison and cross-validation of the same parameters between different surveys. We doing this to find the independent evidence that galaxies in EGIDE are correctly cross-identified and that methods we use for parameters estimation are robust. We compare redshifts, total stellar mass estimation and apparent magnitude. Notice also that flatness $b/a=q$ distributions across various used datasets are compared in the next Appendix~\ref{ap:flatness}.

In this Appendix, we provide comparison and cross-validation of the same parameters between different surveys. We do this to find independent evidence that galaxies in EGIDE are correctly cross-identified and that the methods we use for parameter estimation are robust. We compare redshifts, total stellar mass estimates, and apparent magnitudes. Also the flattening $b/a = q$ distributions across the various datasets used are compared in the next Appendix~\ref{ap:flatness}.

%In Figure~\ref{fig:cz_comparison} we compare $cz$ estimations. For surveys used consult with Section~\ref{sec:redshifts}. On the left panel we show $cz$ from HyperLEDA database \citep{leda}, which is a compilation of redshift sources in CMB frame versus Dark Energy Spectroscopic Instrument survey Data Release 1 (DESI DR1) \citep{2025arXiv250314745D}, which collects spectroscopic redshift measurements up to $z\sim 4$. Specifically, we use Stellar Mass and Emission Line Catalog (EMLines) Value Added Catalogue (VAC), which contains information for galaxies with reliable redshift measurements. In the center of Figure~\ref{fig:cz_comparison} we compare REGALADE redshifts, which we use as reference for EGIDE, with combined data from DESE DR1 and HyperLEDA. On both panels we see the good coincidence between various $cz$ estimates from different sources, with normalized median absolute deviation (NMAD, a robust statistical measure of dispersion) about 170~km/s in both cases. In the majority of cases, as bottom subplots show, at all distances difference of $|\Delta cz|$ lie within 500~km/s. On the right panel in Figure~\ref{fig:cz_comparison} we compare REGALADE data with 30410 galaxies from RCSEDv2 database \citep{2017ApJS..228...14C}, which include DESI among other surveys (see Table~\ref{tab:inters}). In this case NMAD is drastically smaller around 20~km/s and the overall scatter lie within 100~km/s, and also some systematic difference is much more obvious.

\begin{figure}
\centering
\includegraphics[width=0.32\textwidth]{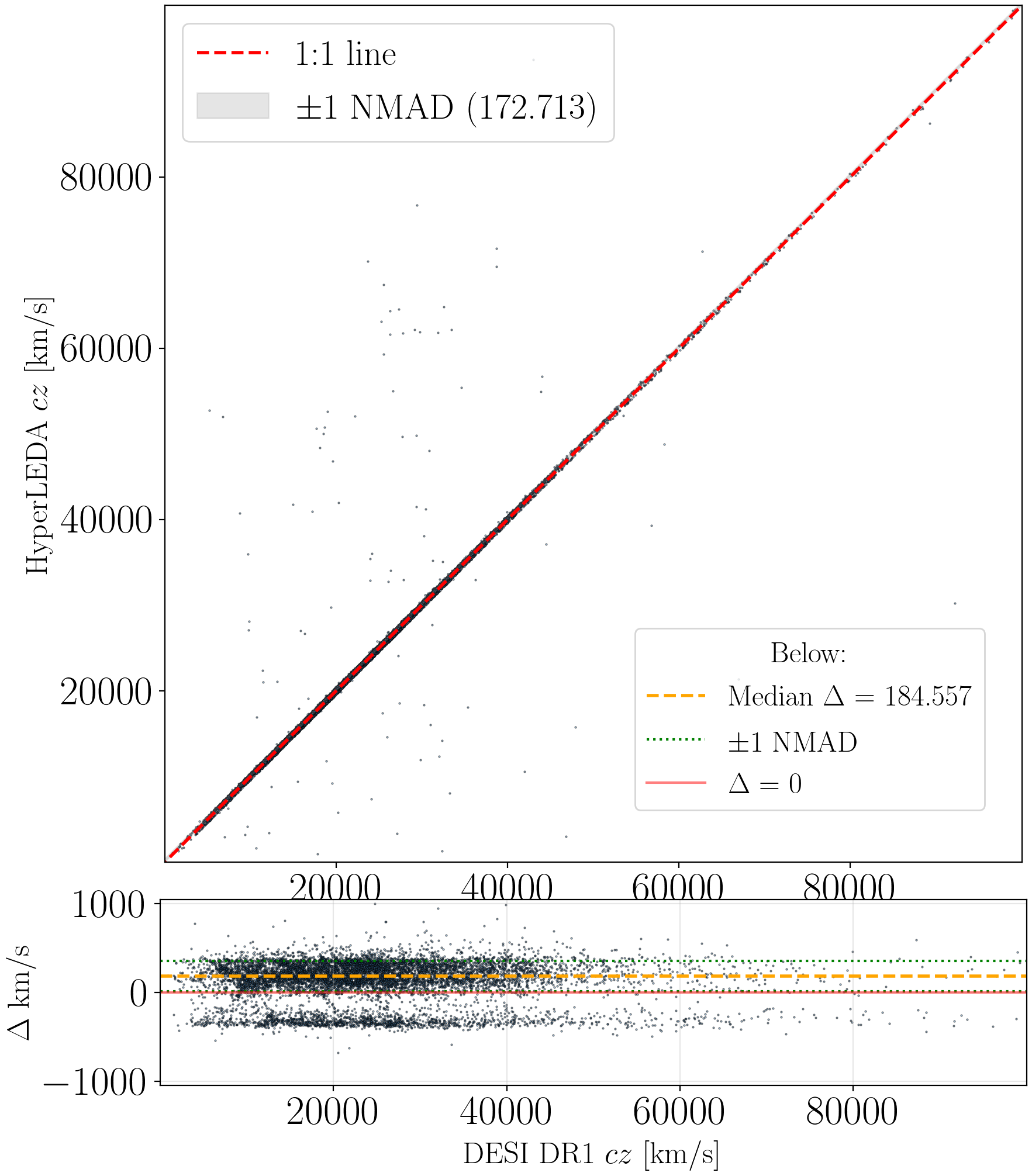}
\includegraphics[width=0.32\textwidth]{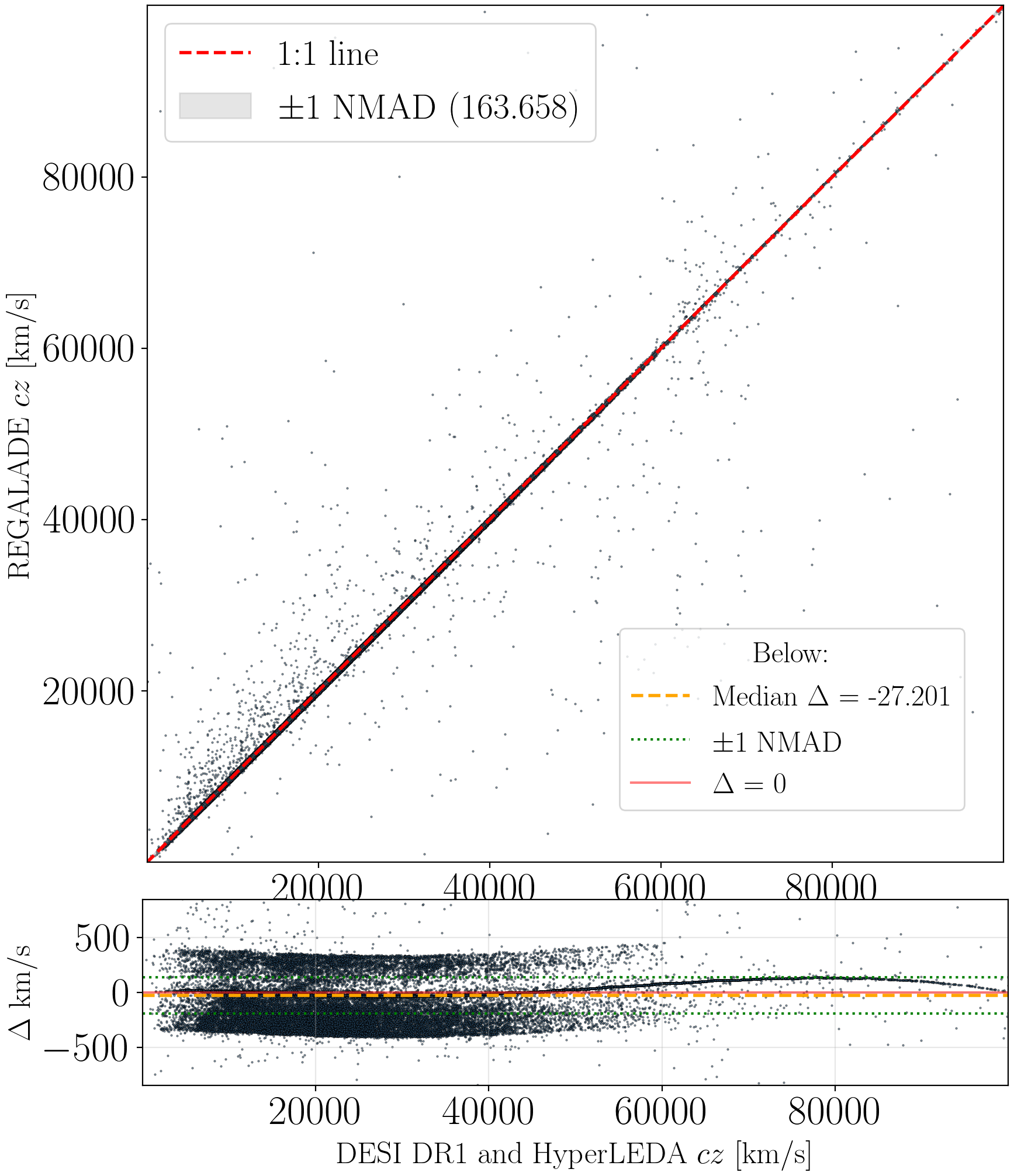}
\includegraphics[width=0.32\textwidth]{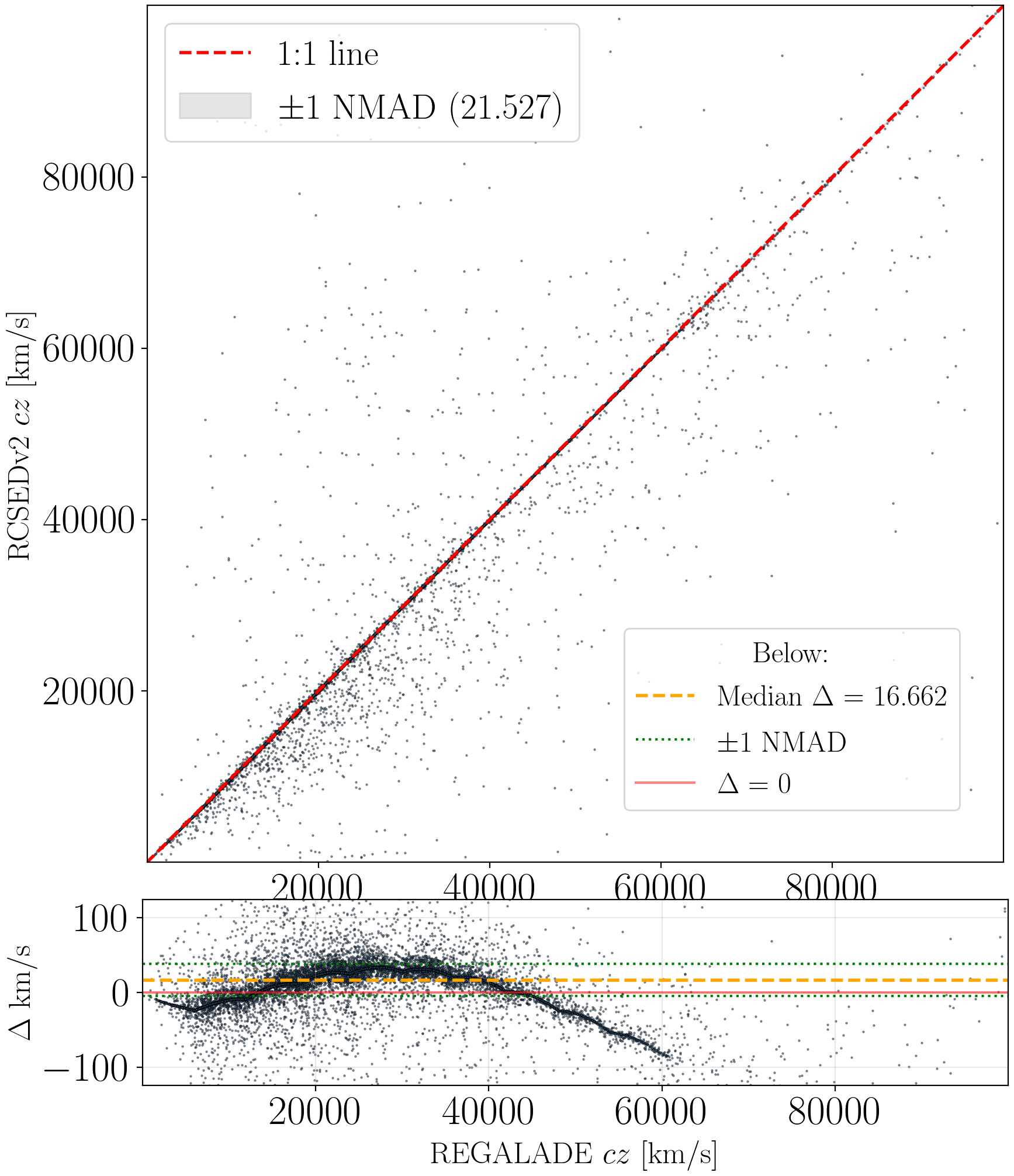}
\caption{Comparison of $cz$ from DESI DR1, HyperLEDA, RCSEDv2, and REGALADE, where the latter is used for the EGIDE sample. In the bottom panel below each subplot, we show the difference between values, and the lines indicate the median, zero, and $\pm$NMAD levels.
}
\label{fig:cz_comparison}
\end{figure}

In Figure~\ref{fig:cz_comparison}, we compare $cz$ estimates. For the list of surveys used consult with  Section~\ref{sec:redshifts}. On the left panel, we show $cz$ from the HyperLEDA database \citep{leda}, which is a compilation of redshift sources in the CMB frame, versus the Dark Energy Spectroscopic Instrument Data Release 1 (DESI DR1) \citep{2025arXiv250314745D}, which collects spectroscopic redshift measurements up to $z \sim 4$. Specifically, we use the Stellar Mass and Emission Line Catalog Value Added Catalogue (EMLines VAC), which contains information for galaxies with reliable redshift measurements. In the center of Figure~\ref{fig:cz_comparison}, we compare REGALADE redshifts, which we use as the reference for EGIDE, with combined data from DESI DR1 and HyperLEDA. In both panels, we see good agreement between the various $cz$ estimates from different sources, with a normalized median absolute deviation (NMAD, a robust statistical measure of dispersion) of about 170 km/s in both cases. In the majority of cases, as the bottom subplots show, the difference $|\Delta cz|$ lies within 500 km/s at all distances. On the right panel of Figure~\ref{fig:cz_comparison}, we compare REGALADE data with 30,410 galaxies from the RCSEDv2 database \citep{2017ApJS..228...14C}, which includes DESI among other surveys (see Table~\ref{tab:inters}). In this case, the NMAD is drastically smaller, around 20 km/s, the overall scatter lies within 100 km/s, and some systematic difference is much more evident.

%In Figure~\ref{fig:masses_comparison} we present comparison of total stellar masses $M_{\star}$ for galaxies in EGIDE obtained in three different ways. In DESI DR1 EMLines VAC stellar masses were derived using Code Investigating GALaxy Emission (CIGALE; Boquien et al. 2019; Yang et al. 2020), which combines spectra with broadband photometry from the DESI Legacy Surveys ($g$, $r$, $z$, $W1$, and $W2$ bands)\footnote{Available in column MASS\_CG of Stellar Mass and Emission Line Catalog.}. In REGALADE survey \citep{REGALADE} authors measure stellar masses using Kron magnitudes from Pan-STARRS in optical $griz$ bands as well as distance $D$ and photometric redshift $z_{ph}$ estimation and additional information from two catalogues, based on DESI Legacy DR9 and DR10, which includes WISE photometry in $W1$ and $W2$. In total REGALADE-related masses were derived in page 6 \citep{REGALADE} using next calibrations: $$log M_{\star} = log D^2 + a \times z_{mag} + b\times r_{mag} + c \times W1_{mag} + d\times z_{ph} + e.$$ From the other hand, \citep{2025A&A...704A.232E} compare optical and IR-counterparts from $S4G$ ([cite]) and find a simple formula for estimating galaxy masses from DESI using absolute stellar magnitudes in the $g$ and $r$ bands. We use available distance measurements and these calibrations to find stellar mass as (see details in \citealt{2025A&A...704A.232E})
%$$log(M[M_{\odot}]) = 0.673M_g - 1.108M_r + 0.996.$$

\begin{figure}
\centering
\includegraphics[width=0.49\textwidth]{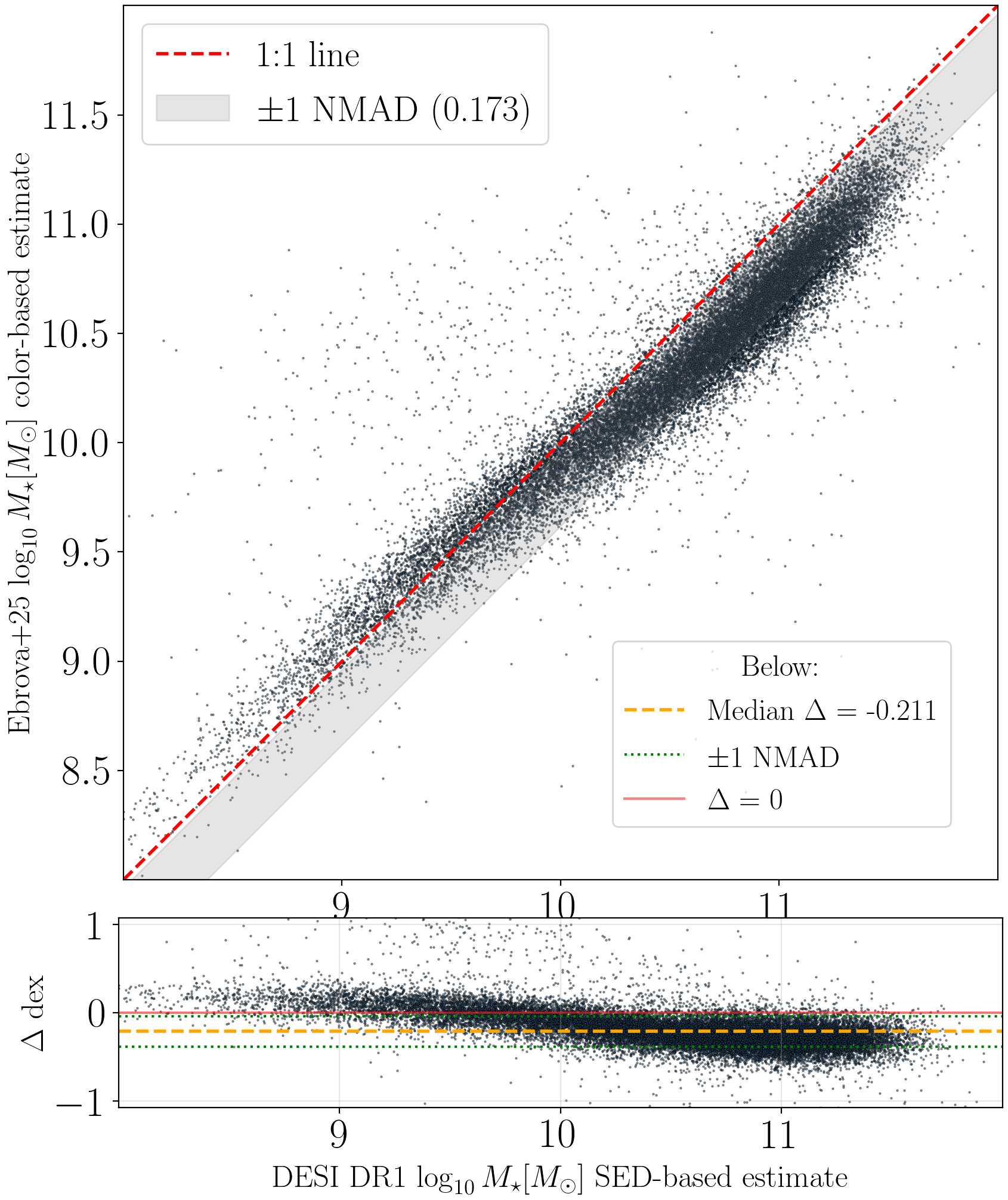}
\includegraphics[width=0.49\textwidth]{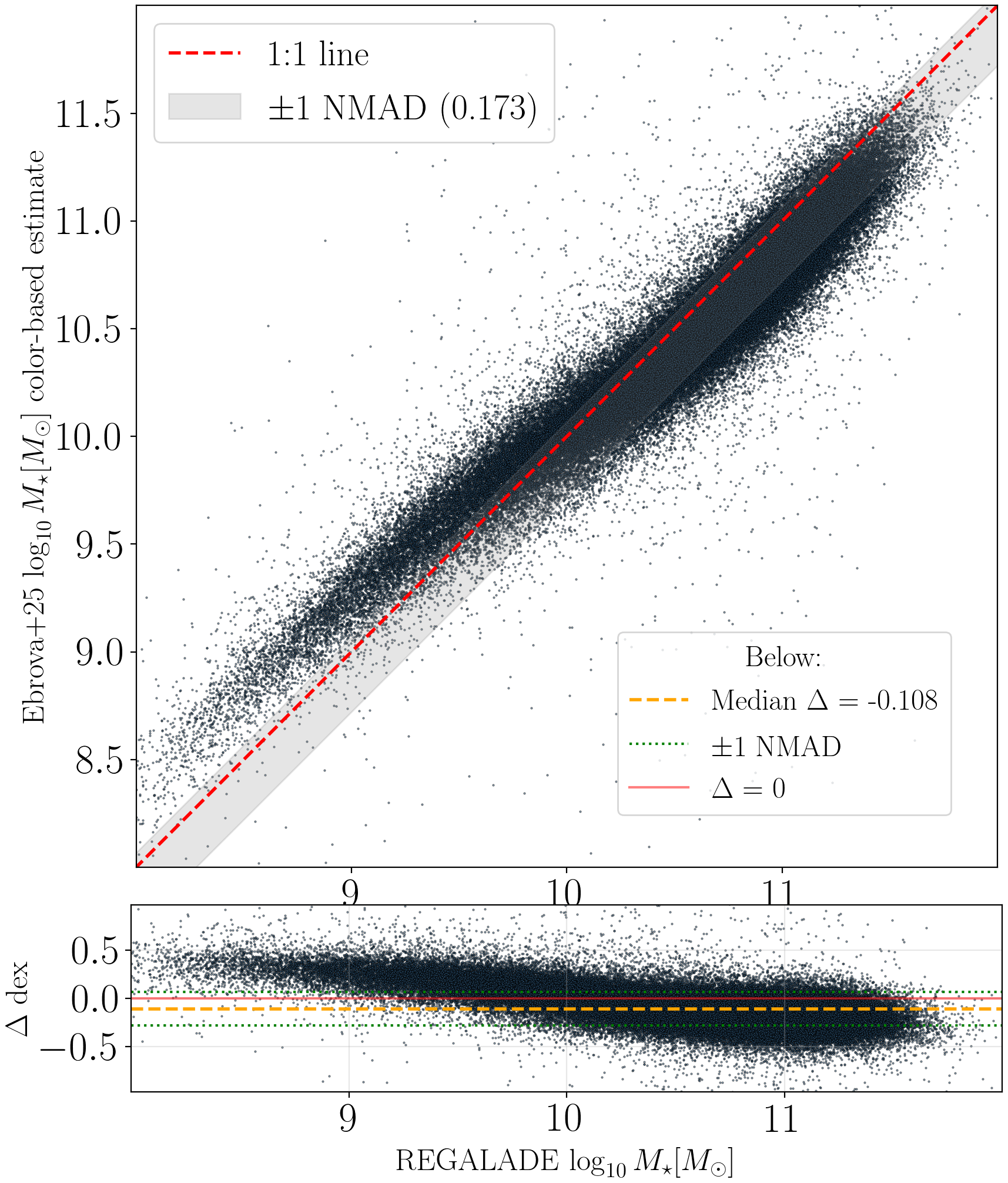}
\caption{Comparison of the total stellar masses $M_{\star}$ for galaxies. \textbf{Left}: the x-axis values are from CIGALE \citep{2019A&A...622A.103B} fitting in DESI DR1 \citep{2025arXiv250314745D}, while the y-axis values are adopted for EGIDE from the colour-related calibrations in \citep{2025A&A...704A.232E}. \textbf{Right}: the x-axis values are from the calibrations in REGALADE, while the vertical axis is the same as on the left.
}
\label{fig:masses_comparison}
\end{figure}

In Figure~\ref{fig:masses_comparison}, we present a comparison of total stellar masses $M_{\star}$ for galaxies in EGIDE obtained in three different ways. In the DESI DR1 EMLines VAC, stellar masses\footnote{Available in column MASS\_CG of the Stellar Mass and Emission Line Catalog.} were derived using Code Investigating GALaxy Emission (CIGALE; \citep{2019A&A...622A.103B}), which combines spectra with broadband photometry from the DESI Legacy Surveys ($g$, $r$, $z$, $W1$, and $W2$ bands). In the REGALADE survey \citep{regalade}, the authors measure stellar masses using Kron magnitudes from Pan-STARRS photometry in the optical $griz$ bands, as well as distance $D$ and photometric redshift $z_{ph}$ estimates, and additional information from two catalogs based on DESI Legacy DR9 and DR10, which include WISE photometry in $W1$ and $W2$. In total, REGALADE-related masses were derived on page 6 of \citep{regalade} using the following calibration ($a,b,c,d,e$ - constant values):$$\log M_{\star} = log D^2 + a \times z_{mag} + b\times r_{mag} + c \times W1_{mag} + d\times z_{ph} + e.$$
On the other hand, \citep{2025A&A...704A.232E} compare optical and IR counterparts from $S4G$ \citep{salo_etal15} and find a simple formula for estimating galaxy masses using absolute stellar magnitudes in the $g$ and $r$ DESI bands. We use available distance measurements and these calibrations to find the stellar mass as (see details in \citep{2025A&A...704A.232E})
$$\log(M_{\star}[M_{\odot}]) = 0.673\times M_g - 1.108\times M_r + 0.996.$$

%On the left side of Figure~\ref{fig:masses_comparison} we compare stellar masses used here from the photometry obtained in Section~\ref{sec:Photometry} and $M_{\star}$ from SED-fitting estimations. We could see, that regardless of visible systematic differences, it lies within $M_{\star}$ relative error ($\approx 30\%$ for CIGALE-based data in this sample) and overall agreement is good. On the right panel we compare two $M_{\star}$ estimations based on photometric calibrations derived in REGALADE and \citep{2025A&A...704A.232E}. Despite the [xfive times] bigger sample and different methods, two parts for Figure~\ref{fig:masses_comparison} looks nearly the same: NMAD is approximately 0.15~dex, the difference interval is about 0.5~dex and method we adopt for EGIDE galaxies overweights massive galaxies and underweights light one, with the transition at the point $\log M_{\star} \approx 10 M_{\odot}$.  In total, we can conclude that \citep{2025A&A...704A.232E} $g-r$ colour-based method for mass estimation works relatively well and could be used as a good proxy of total stellar mass. 

On the left side of Figure~\ref{fig:masses_comparison}, we compare the stellar masses used here from the photometry obtained in Section~\ref{sec:Photometry}, with masses $M_{\star}$ from SED-fitting estimates. We can see that, regardless of visible systematic differences, the NMAD lies within the $M_{\star}$ relative error, which is $\approx 30\%$ for CIGALE-based data in this sample, and the overall agreement is good. On the right panel, we compare the two $M_{\star}$ estimates based on the photometric calibrations derived in REGALADE and \citep{2025A&A...704A.232E}. Despite the five times larger sample than in the left panel and different methods, the two parts of Figure~\ref{fig:masses_comparison} look nearly the same: the NMAD is approximately 0.15 dex, the difference interval is about 0.5 dex, and the method we adopt for EGIDE galaxies overweights massive galaxies and underweights lighter ones, with the transition at $\log M_{\star}/M_{\odot} \approx 10$. In total, we can conclude that the \citep{2025A&A...704A.232E}  $g$ and $r$-based method for mass estimation works relatively well and can be used as a good proxy for total stellar mass.

%Finally in this Appendix we compare apparent Kron magnitudes in $r$ and $z$ bands from REGALADE and measured in this work in Section~\ref{sec:Photometry}. Values from REGALADE are measured in Pan-STARRS system and reduced into DESI bands for proper comparison. In Figure~\ref{fig:mag_comparison} we see that NMAD is $0.10$~mag and the agreement between independent apparent magnitude estimates is decent. For $z$-band we see the sub-line systematics and overall agreement is worse than for $r$-band but the difference is well below 0.5~mag.

\begin{figure}
\centering
\includegraphics[width=0.49\textwidth]{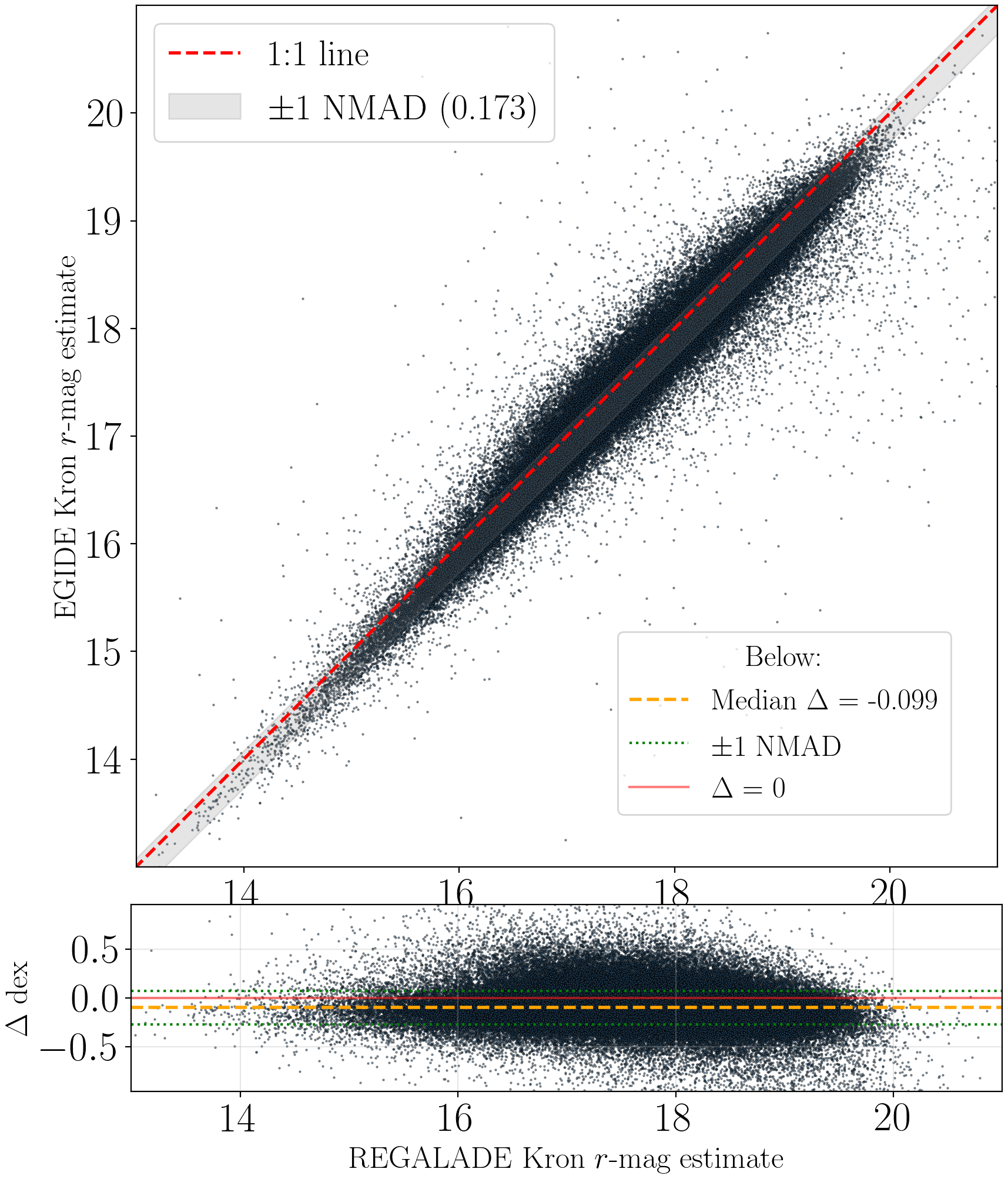}
\includegraphics[width=0.49\textwidth]{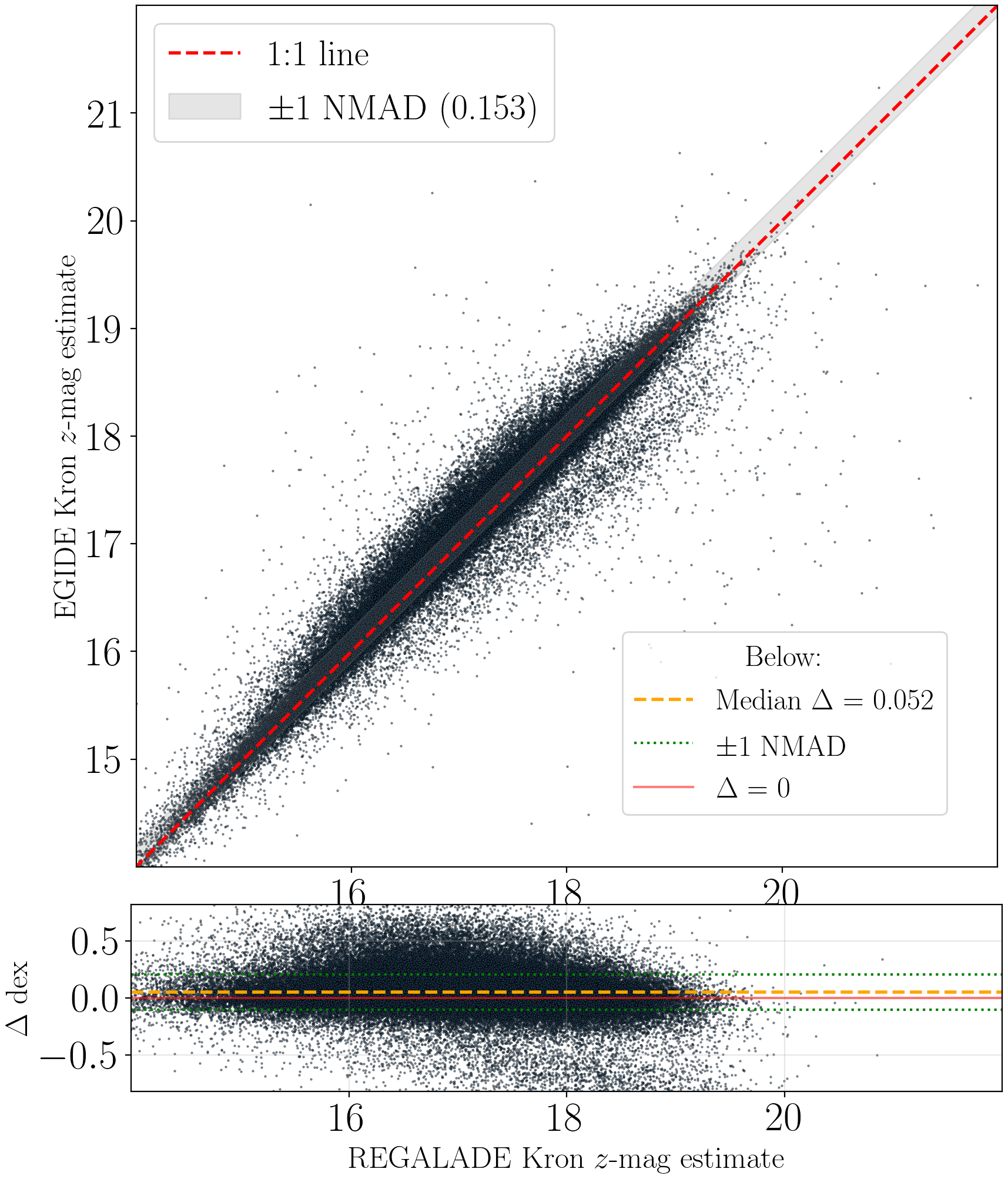}
\caption{Comparison of the apparent Kron magnitude in EGIDE photometry (vertical axis) and in REGALADE (horizontal axis). The left plot is for the $r$ band, and the right plot is for the $z$ band.
}
\label{fig:mag_comparison}
\end{figure}

Finally, in this Appendix we compare the apparent Kron magnitudes in the $r$ and $z$ bands from REGALADE with those measured in this work in Section~\ref{sec:Photometry}. Values from REGALADE are measured in the Pan-STARRS system and reduced to DESI bands for proper comparison. In Figure~\ref{fig:mag_comparison}, we see that the NMAD is $0.10$~mag and the agreement between the independent apparent magnitude estimates is decent. For the $z$-band, we see sub-line systematics, and the overall agreement is worse than for the $r$-band, but the difference is well below 0.5~mag.

\section[\appendixname~\thesection]{Flatness measurements in other surveys}
\label{ap:flatness}
%In this Appendix we provide in Figure~\ref{fig:q_appendix} flatness measurements from other various sources and the details of $b/a$ estimation. We doing this to find the independent evidence that galaxies in EGIDE are indeed edge-on, and to increase the confidence of reader in our work.

\begin{figure}
\centering
\includegraphics[width=0.95\textwidth]{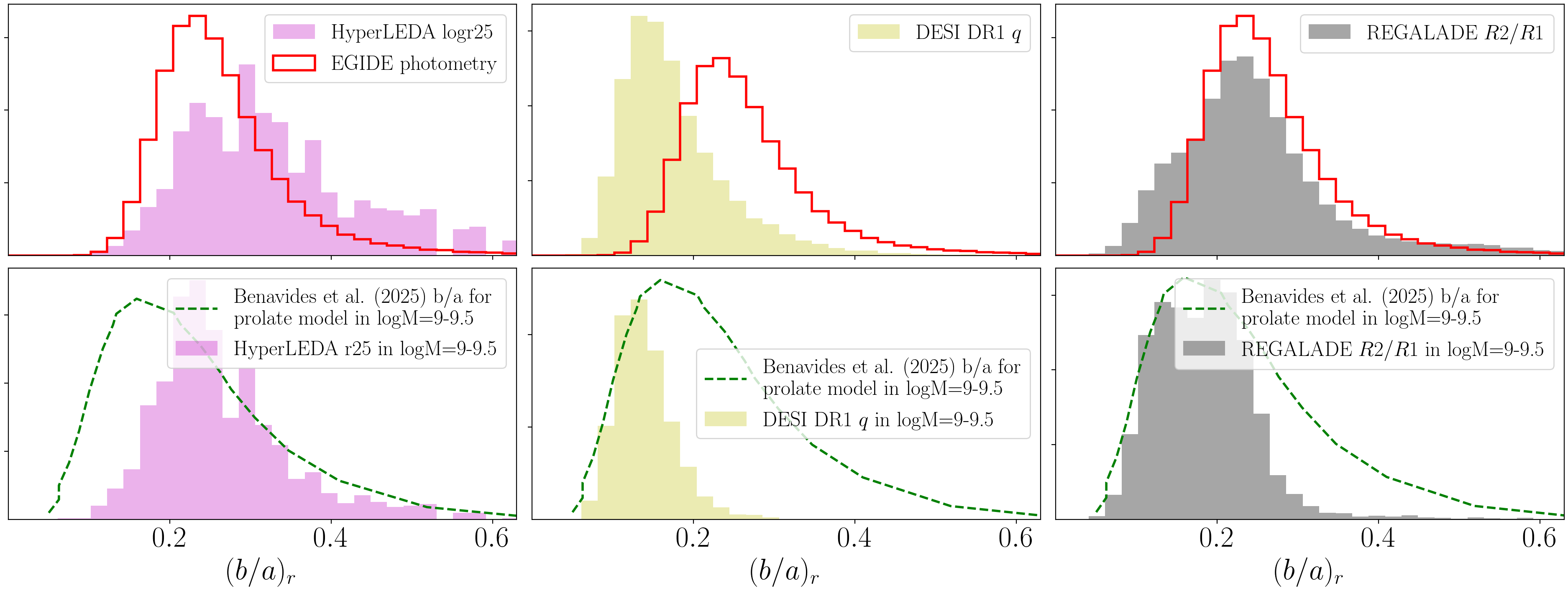}
\caption{
Distributions of the axis ratio from various datasets and comparison with the EGIDE $b/a$. Columns from left to right: $b/a$ from HyperLEDA, $q$ from DESI DR1 ellipticities, and $R2/R1$ ratio from REGALADE. The upper row shows the full distribution for the intersection with each respective survey (see Table~\ref{tab:inters} for details). In the bottom row, we show the distribution only for the $\log M_{\star}/M_{\odot} = 9\textrm{--}9.5$ mass bin, where the green line shows the distribution from figure~6 in \citep{2025arXiv251211035B}.
}
\label{fig:q_appendix}
\end{figure}

In this Appendix, we provide flatness measurements from various other sources and the details of $b/a$ estimation in Figure~\ref{fig:q_appendix}. We do this to find independent evidence that galaxies in EGIDE are indeed edge-on, and to increase the reader's confidence in our work.

%In HyperLEDA \citep{leda} subset we use apparent axis ratio \textsc{logr25}. This parameter is the axis ratio of the isophote 25~mag/arcsec$^2$ in the $B$-band. These values are slightly bigger on average as Figure~\ref{fig:q_appendix} shows.

In the HyperLEDA \citep{leda} subset, we use the apparent axis ratio \textsc{logr25}. This parameter is the axis ratio of the 25 mag/arcsec$^2$ isophote in the $B$-band. These values are slightly larger on average, as Figure~\ref{fig:q_appendix} shows.

%From REGALADE \citep{REGALADE} data we have $R1$ and $R2$ which are semimajor and semiminor axes of ellipse. In this work authors recalibrate sizes from used catalogues to the same notation similar to what is used in the Siena Galaxy Atlas (Moustakas et al. 2023), i.e. till $\mu=26$~mag/arcsec$^2$ isophote in the $r$-band. The $R2/R1$ estimation of flatness is well fitted with $b/a$ measured using the EGIDE photometry in Section~\ref{sec:Photometry}.

From the REGALADE \citep{regalade} data, we have $R1$ and $R2$, which are the semi-major and semi-minor axes of the ellipse. In this work, the authors recalibrate sizes from the used catalogs to the same notation as that used in the Siena Galaxy Atlas \citep{2023ApJS..269....3M}, i.e., to the $\mu = 26$ mag/arcsec$^2$ isophote in the $r$-band. The $R2/R1$ estimate of flattening is in good agreement with the $b/a$ measured using the EGIDE photometry in Section~\ref{sec:Photometry}.

%Finally, in order to compute the shape for DESI DR1 galaxies, we use two ellipticity components $\epsilon_1$ and $\epsilon_2$, which are then converted to $q_{DESI}$ using the formula\footnote{See \url{https://www.legacysurvey.org/dr10/catalogs/\#\#ellipticities} and parameters SHAPE\_E1 and SHAPE\_E2 in Table dr1\_galaxy\_stellarmass\_lineinfo\_v1.0 of the Stellar Mass and Emission Line Catalog.}:

Finally, to compute the shape for DESI DR1 galaxies, we use the two ellipticity components $\epsilon_1$ and $\epsilon_2$, which are then converted to $q_{\text{DESI}}$ using the formula\footnote{See \url{https://www.legacysurvey.org/dr10/catalogs/\#\#ellipticities} and parameters SHAPE\_E1 and SHAPE\_E2 in Table dr1\_galaxy\_stellarmass\_lineinfo\_v1.0 of the Stellar Mass and Emission Line Catalog.}:

$$q_{DESI} = \frac{1-\sqrt{\epsilon_1^2+\epsilon_2^2}}{1+\sqrt{\epsilon_1^2+\epsilon_2^2}}.$$

%The resulted $q_{DESI}$ distribution demonstrate median with standard deviation equals $0.16\pm0.08$. It is hard to say why these values are $q_{DESI}$ is noticeably lower than $b/a$ from our photometry and other surveys, but in any case such small values additionally validate that collected candidates are very elongated and most likely are visible edge-on.

The resulting $q_{\text{DESI}}$ distribution has a median with a standard deviation of $0.16 \pm 0.08$. It is difficult to say why these $q_{\text{DESI}}$ values are noticeably lower than the $b/a$ values from our photometry and other surveys. We notice that in all examined cases probabilistic \textsc{TRACTOR} models results in much bigger $a$ than observed, and thus in smaller $q$. In any case, such small values further validate that the collected candidates are very elongated and most likely visible edge-on.

%\textit{Last but not least important is a comparison with independent oblateness measurements, available in other surveys. We provide such comparison in Appendix~\ref{ap:flatness}. In total, we compare with ellipticity components from DESI, $R2/R1$ or minor-to-major axis ratio measurements in REGALADE, same ratio from HyperLEDA data and $q$ modeling from \citep{2025arXiv251211035B}. As Figure~\ref{fig:q_appendix} demonstrates, all of these measurements have distributions very similar to what measured here and thus validate small $q$ found here. One visible peculiarity is significantly smaller values obtained from DESI DR, which may probably relate to methodology they used [???]. Another interesting feature of Figure~\ref{fig:q_appendix} comes from the comparison with \citep{2025arXiv251211035B} model for galaxies in $\log M_{\star}$ from 9.0 to 9.5 mass bin. We can see that $b/a$ for subsamples, obtained according to different methods and data employed, is actually fit left and right wings of the reconstructed model in this mass bin relatively well. This may have a meaning that oblate model in \citep{2025arXiv251211035B} fit actual thickness of galaxies we obtain directly for edge-on galaxies very well. On the other hand, this may be a coincidence, which needs further testing. }

\section[\appendixname~\thesection]{Gallery of edge-on galaxies from EGIDE}
\label{ap:gallery}
In this Appendix we show preview for other examples of edge-on galaxies from EGIDE. They are arranged in the same way as in Figure~\ref{fig:color_axisRatio}, representing objects of different flatness and colour.

%\begin{figure}
%\centering
%\includegraphics[width=0.95\textwidth]{paper_imgs/grid.png}
%\caption{
%Gallery of galaxies from EGIDE with varying flatness $a/b$ (horizontal axis) and $(g-i)$ Petrosian color (horizontal axis).
%}
%\label{fig:grid}
%\end{figure}

\begin{figure}
\centering
\includegraphics[width=0.99\textwidth]{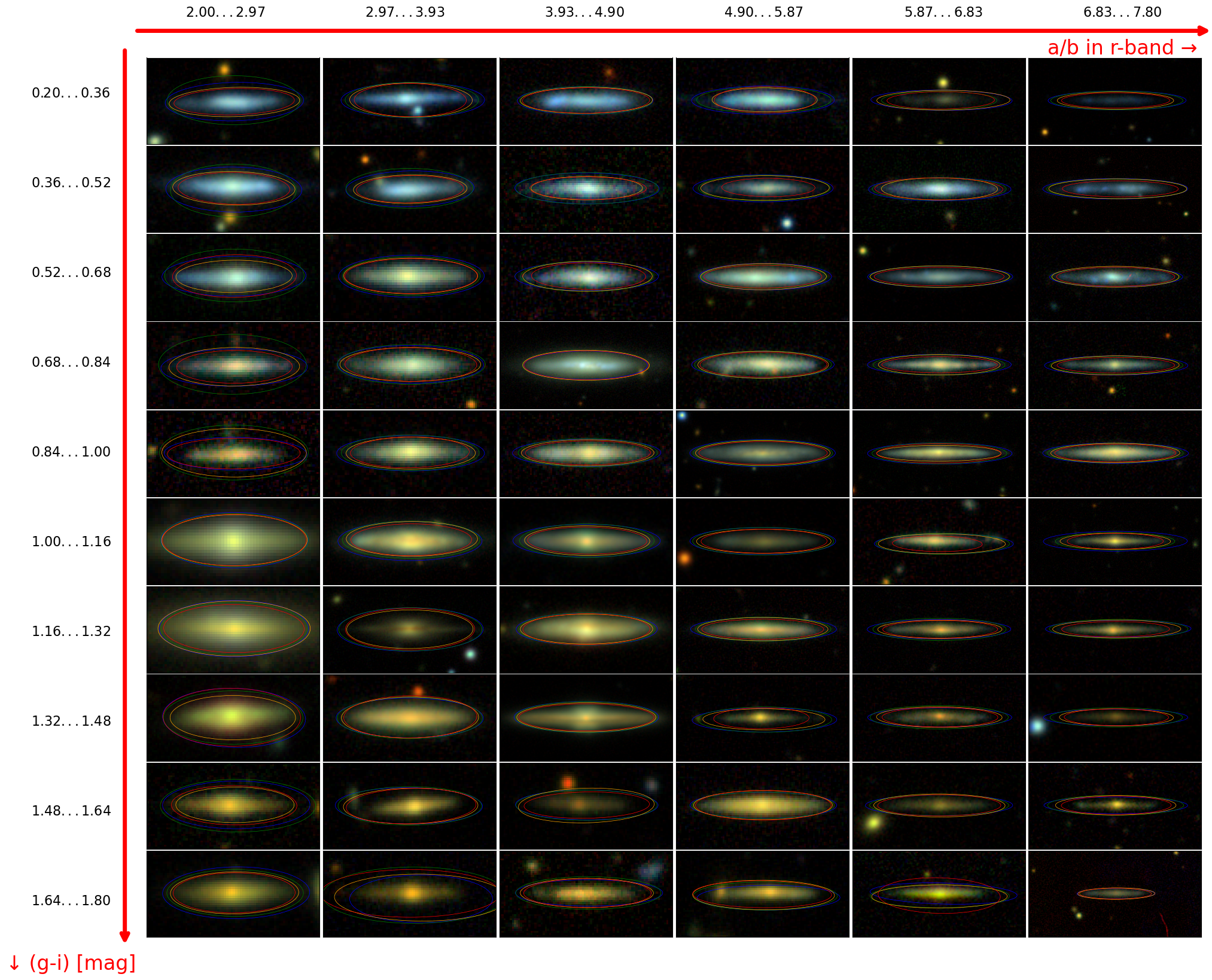}
\caption{
Gallery of galaxies from EGIDE with varying flatness $a/b$ in the $r$-band (horizontal axis) and colour $(g-i)$ using Petrosian magnitudes (vertical axis). Marked ellipses are detected by \textsc{SExtractor} and used in the photometry measurements. The numbers show the range of flatness and colour bin in each column and row, respectively.
}
\label{fig:grid}
\end{figure}

%\appendixtitles{yes} % Leave argument "no" if all appendix headings stay EMPTY (
%then no dot is printed after "Appendix A").
%If the appendix sections contain a heading then change the argument to
%"yes".
%\appendixstart
%\appendix

\bibliography{main}

%} % If the paper is ``preprints'', please uncomment this parenthesis.
\end{document}